\documentclass{article}

\usepackage{arxiv}
\usepackage{titletoc}
\usepackage[utf8]{inputenc} 
\usepackage[T1]{fontenc}    
\usepackage{hyperref}       
\usepackage{url}            
\usepackage{booktabs}       
\usepackage{amsfonts}       
\usepackage{nicefrac}       
\usepackage{microtype}      
\usepackage{lipsum}
\usepackage{graphicx}

\usepackage{caption}
\usepackage{booktabs}
\usepackage{multirow}
\usepackage{graphicx}

\usepackage[utf8]{inputenc} 
\usepackage[T1]{fontenc}    
\usepackage{hyperref}       
\usepackage{url}            
\usepackage{booktabs}       
\usepackage{amsfonts}       
\usepackage{nicefrac}       
\usepackage{microtype}      
\usepackage[table]{xcolor}   
\usepackage{booktabs}
\usepackage{multirow}
\usepackage{pifont}

\usepackage{algorithm}
\usepackage{algpseudocode}
\usepackage{amssymb}
\usepackage{graphicx}   
\usepackage{graphicx}   
\usepackage{amsmath}    
\usepackage{cleveref}

\usepackage{float}
\usepackage{textgreek}
\usepackage{doi}
\usepackage{multirow}
\usepackage{siunitx}
\usepackage{colortbl}
\usepackage{array}
\usepackage{enumitem}

\definecolor{firstbg}{RGB}{255,182,193}
\definecolor{secondbg}{RGB}{255,220,225}
\definecolor{teal}{RGB}{0,128,128}

\newcommand{\std}[1]{\text{\scriptsize(\(\pm\)\,\num{#1})}}

\graphicspath{ {./images/} }

\title{StateXDiff: Cell State-Contextualized Multimodal Diffusion for Single-Cell Perturbation Prediction}

\author{
Peiting Shi$^{1,\dagger}$ \quad
Ningfeng Que$^{1,\dagger}$ \quad
Xianzhe Huang$^{2}$ \quad
Xiaofei Wang$^{3}$ \quad
Jianzhong Jeff Xi$^{1,*}$
\\[0.6em]
$^{1}$Department of Biomedical Engineering, College of Future Technology\\
$^{2}$College of Informatics, Huazhong Agricultural University
\\
$^{3}$Department of Clinical Neurosciences, University of Cambridge, Cambridge
\\[0.5em]
{\small
$^{\dagger}$Equal contribution. \quad
$^{*}$Corresponding author: \texttt{jzxi@pku.edu.cn}.
}
}

\begin{document}
\maketitle
\begin{abstract}
Predicting drug-induced cellular state changes at single-cell resolution remains a central challenge in virtual cell modeling, particularly under out-of-distribution (OOD) conditions. Current approaches predominantly rely on RNA-based assays, which often fail to adequately capture the diverse cellular states underlying drug responses. Moreover, conditional distribution shifts and low signal-to-noise ratios frequently cause models to learn spurious correlations rather than genuine state transitions. To address these limitations, we introduce \textbf{StateXDiff}, a cell State-contextualized multimodal (X) Diffusion framework for predicting single-cell responses to drug perturbations. The framework operates sequentially: first, it learns a disentangled, multimodal representation of cellular state by integrating transcriptomic profiles with inferred protein features; second, it employs a conditional diffusion model to generate perturbation-specific changes. Our approach introduces a \textbf{Virtual Multimodal Cell State}, which augments RNA-based representations with protein-level context, and a \textbf{Mechanism-aware Drug–Gene Template}, which consolidates multi-source biological knowledge for accurate drug representation. Generation is driven by a latent-space diffusion Transformer, regularized through quality-aware triplet constraints, including positive drug-protein pairs or protein-drug mismatched pairs, and explicit protein-reliability weighting.
Extensive evaluation demonstrates that \textbf{StateXDiff} consistently enhances generalization performance across three challenging settings: unseen cell lines, unseen drugs, and combinatorial perturbations.

\end{abstract}


\section{Introduction}

Virtual cells aim to predict the dynamic responses of cells to drug perturbations through computational methods. The goal is not only to accurately predict the effects of known drugs, but also to support novel drug discovery through functional genomics analysis and mechanistic insights\cite{bunne2024build}. Perturbation experiments are costly and data are scarce, with large-scale multi-omics perturbation datasets being particularly limited. This makes predicting transcriptional responses across diverse cellular contexts a significant challenge, especially under out-of-distribution (OOD) conditions\cite{bunne2024build,mccoy2025digital}.

\begin{figure}[H]
  \centering
  \includegraphics[width=0.8\textwidth]{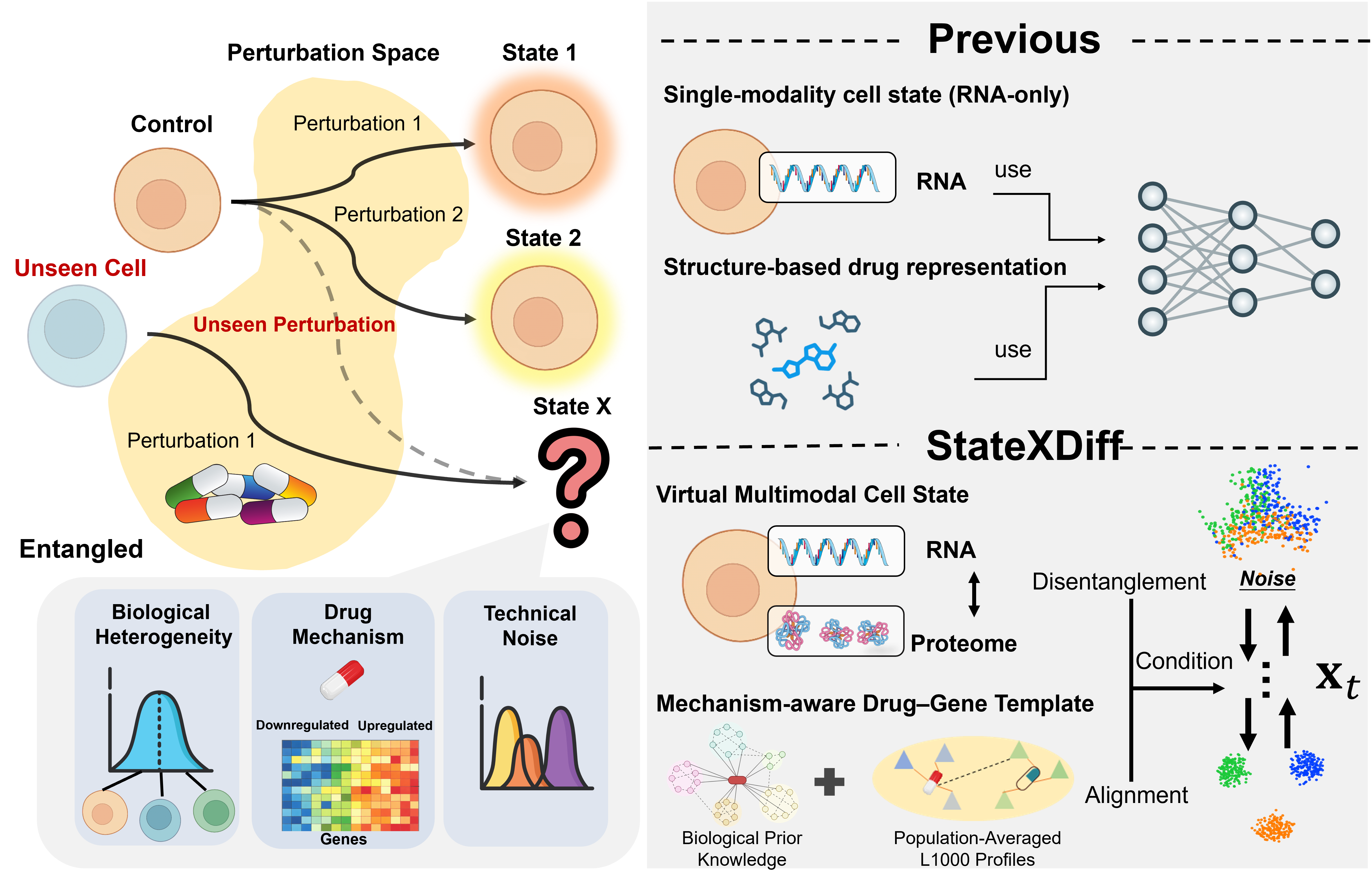}
  \caption{\textbf{Motivation of StateXDiff.}}
  \label{fig_intro}
\end{figure}

Recent systematic benchmarking~\cite{ahlmann2025deep} shows that highly expressive models provide only marginal gains over simple baselines in cross-context or OOD settings. This suggests that the main bottleneck lies not in model capacity, but in the structural noise and distributional shift inherent to perturbation data. In particular, the tight coupling between drug MoA and intrinsic cellular state, together with unobserved biological heterogeneity, substantially lowers the signal-to-noise ratio in gene expression space. Two core limitations characterize existing approaches (Fig.~\ref{fig_intro}): first, drug representations built on chemical structure or generic embeddings fail to resolve shared versus differential MoA at the transcriptional level, whereas resources such as LINCS L1000\cite{Subramanian2017_L1000} combined with drug–target and gene interaction networks can supply stronger mechanistic priors for transferable drug representations; second, perturbation responses are shaped by protein--transcript coupling, but this cross-layer dependency is rarely modeled explicitly. Recent cross-modal translation methods~\cite{liu2025pre,fang2025sclinguist} help bridge this gap by deriving proteome-proximal representations from transcriptomic profiles without requiring matched proteomic measurements. However, incorporating such modalities introduces a structural challenge: the pronounced asymmetry between pseudo-protein and RNA representations makes naive fusion prone to representation collapse, either as complete collapse\cite{chaudhuri2025closer} where pseudo-protein signals are overwhelmed by the dominant RNA modality, or as channel collapse where most feature dimensions degenerate into uninformative representations\cite{liang2023factorized}. These challenges call for a modeling framework that integrates mechanism-aware priors, explicitly models cross-layer interactions, and aligns multimodal representations in a disentangled manner. Such a framework is essential for improving generalization under distributional shift and missing-modality conditions.

We propose \textbf{StateXDiff}, a two-stage State-contextualized multimodal (X) Diffusion framework for predicting single-cell transcriptional responses to drug perturbations. The framework first learns disentangled conditional representations for cells and drugs in its \textbf{Representation Learning Stage}. For cellular context,  we augment transcriptomic profiles with inferred pseudo-protein embeddings and apply a complementary disentangled alignment strategy to untreated cells. This yields a \textbf{Virtual Multimodal Cell State (VMCS)} that preserves shared cellular information and protein-specific signals. For drug context, we integrate chemical structure, L1000 perturbation signatures, and biological prior networks through cross-modal alignment and knowledge distillation to form a unified \textbf{Mechanism-aware Drug–Gene Template (MDT)}. In the subsequent \textbf{Perturbation Generation Stage}, models context-dependent drug responses through latent conditional diffusion. Bidirectional cross-attention first couples cell and drug representations to produce context-aware tokens. These tokens condition a Transformer denoiser that predicts perturbation-response residuals via joint self-attention. Quality-aware triplet constraints and protein reliability modeling further improve robustness by enforcing drug--protein consistency and down-weighting unreliable protein-associated signals.

\textbf{Contributions:}
\begin{itemize}[itemsep=0.5pt, topsep=1pt, leftmargin=*, parsep=0pt]

    \item \textbf{Virtual Multimodal Cell State (VMCS).}
    We introduce pseudo-protein embeddings to complement transcriptomic profiles for perturbation prediction. We further develop a disentangled alignment strategy that aligns RNA and pseudo-protein representations, preserving shared cell-state information while retaining protein-associated signals.

    \item \textbf{Protein-quality aware module (PQM).}
    We introduce a quality-aware protein module that enables the model to selectively leverage protein-associated signals while mitigating errors from low-fidelity cross-modal inferences.

    \item \textbf{Mechanism-aware Drug--Gene Template (MDT).}
    By unifying chemical structure, transcriptomic signatures, and biological graph priors through soft-label contrastive learning and distillation, we learn drug embeddings that capture mechanism-of-action specificity and generalize across environments.
    
    \item \textbf{Perturbation-Aware Conditional Diffusion.}
    We develop a latent-space diffusion Transformer conditioned jointly on VMCS and MDT, enhanced with quality-aware triplet constraints and explicit protein-reliability weighting for robust perturbation prediction.
\end{itemize}

\section{Related Work}

\paragraph{Perturbation Representation Learning.}
Transferable perturbation representations are central to chemical perturbation response prediction. Early methods such as scGen~\cite{Lotfollahi2019_scGen} encode drugs as one-hot vectors, limiting generalization to unseen compounds. Structure-aware methods, such as PRnet~\cite{Qi2024_PRnet}, learn molecular-graph embeddings and improve extrapolation to structurally related drugs, but remain limited in capturing functional similarity beyond structural proximity. Large-scale resources such as L1000~\cite{Subramanian2017_L1000} have enabled perturbation representation pretraining for cross-drug transfer, as exemplified by chemCPA~\cite{Hetzel2022_chemCPA}. Recent methods further incorporate cell-type specificity; for instance, CRISP~\cite{Wang2025_CRISP} combines cell representations learned by single-cell foundation models with cell-type-specific contrastive learning for zero-shot cross-platform prediction. Nevertheless, most existing methods still rely on globally shared perturbation embeddings and treat dosage as a scalar, limiting their ability to model context-dependent drug effects across cellular states.

\paragraph{Cell-State Modeling and Cross-Modality Inference.}
Transcriptomic profiles are widely used to represent cellular states. Strong baselines range from statistical or additive predictors, such as trainMean, baseReg, baseMLP, and baseControl~\cite{Zhang2025_scPerturBench}, to VAE-based models, such as scGen~\cite{Lotfollahi2019_scGen} and trVAE~\cite{Lotfollahi2020_trVAE}, which learn latent spaces for cellular variation and perturbation effects. Recent foundation models, including SCimilarity~\cite{heimberg2025cell} and STATE~\cite{Adduri2025_STATE}, further seek transferable cell representations across datasets. Cross-modality models, such as scTranslator~\cite{Liu2025_scTranslator} and scLinguist~\cite{fang2025sclinguist}, complement RNA-based representations by translating transcriptomes into protein-level readouts. However, RNA-only state representations remain sensitive to measurement noise and biological heterogeneity, and existing state encoders are seldom evaluated for cross-dataset generalization.

\paragraph{Generative Modeling of Perturbation Responses.}
Generative methods predict perturbation responses through latent transitions or conditional generation. Latent-variable models, including scGen~\cite{Lotfollahi2019_scGen}, trVAE~\cite{Lotfollahi2020_trVAE}, CPA~\cite{Lotfollahi2023_CPA}, and chemCPA~\cite{Hetzel2022_chemCPA}, disentangle perturbation effects from cell-state variation in a shared latent space. More recent approaches model distributional shifts using flow matching, as in CellFlow~\cite{CellFlow2025}; or latent diffusion, as in PerturbDiff~\cite{yuan2026perturbdiff} and scDFM~\cite{yu2026scdfm}. Despite these advances, most methods assume complete and reliable conditioning signals and primarily optimize reconstruction fidelity. As a result, they remain vulnerable to noisy or missing conditions and provide limited support for modeling perturbation uncertainty under out-of-distribution settings.

\section{Method}
\label{sec:method}

\begin{figure*}[t]
  \centering
  \includegraphics[width=\textwidth]{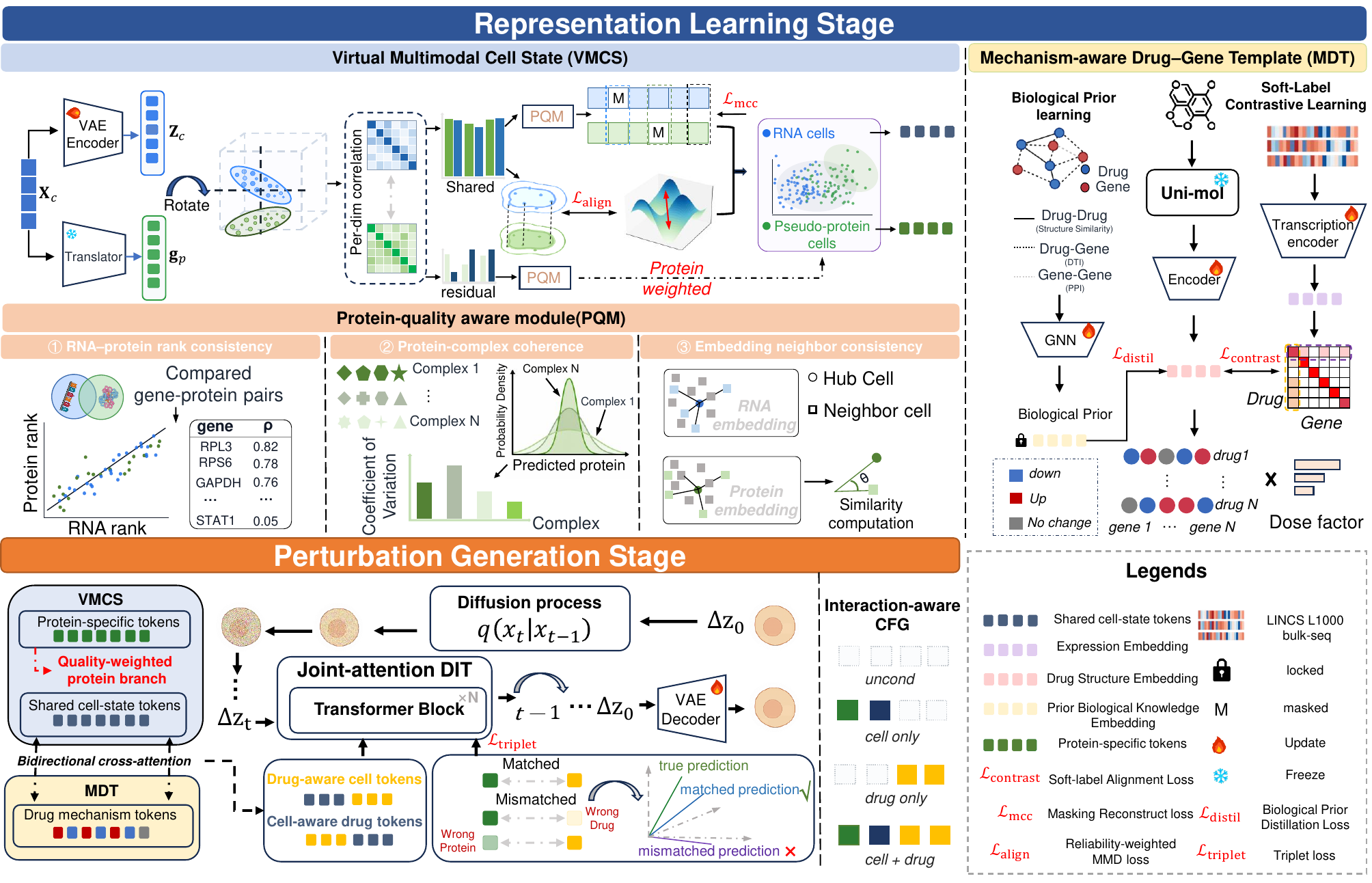}
  \caption{
Overview of \textbf{StateXDiff}. 
Stage~I constructs two complementary condition representations from untreated cells: a \textbf{Virtual Multimodal Cell State (VMCS)} for cellular context modeling and a \textbf{Mechanism-aware Drug Template (MDT)} for drug representation. 
Stage~II uses a \textbf{Perturbation-Aware Conditional Diffusion} module to predict single-cell transcriptional responses to drug perturbations.
}
\label{fig2}

\end{figure*}

Fig.~\ref{fig2} illustrates the proposed \textbf{StateXDiff}, a two-stage framework for predicting single-cell transcriptional responses to drug perturbations.
Stage~I constructs two complementary condition representations from untreated cells:
(i)~a \textbf{Virtual Multimodal Cell State (VMCS)} that integrates transcriptomic and pseudo-protein embeddings, where a \textbf{Protein Quality-Aware Module (PQM)} estimates per-cell pseudo-protein reliability and guides a disentangled decomposition into shared cell-state and protein-associated components (Sec.~\ref{sec:vmcs}), and
(ii)~a \textbf{Mechanism-aware Drug Template (MDT)} that distills priors from chemical structure, transcriptomic responses, and biological graphs into a unified drug representation (Sec.~\ref{sec:mdt}).
Stage~II introduces a \textbf{Perturbation-Aware Conditional Diffusion} module, which conditions a latent diffusion Transformer on the interaction between VMCS and MDT representations to predict perturbation residuals (Sec.~\ref{sec:generation}).

\subsection{Problem Formulation}
\label{sec:problem}
Let $\mathbf{x}_c,\mathbf{x}_p\in\mathbb{R}^G$ denote the control and perturbed transcriptomic profiles, respectively, where $G$ is the number of genes. 
Each perturbation is defined by a drug $d$ and dosage $\delta$. 
We encode transcriptomic profiles into a latent space to obtain $\mathbf{z}_c$ and $\mathbf{z}_p$, and represent the perturbation effect as the latent residual $\Delta\mathbf{z}=\mathbf{z}_p-\mathbf{z}_c$. 
Given the control state $\mathbf{z}_c$ and perturbation condition $(d,\delta)$, our goal is to model the conditional distribution $p(\Delta\mathbf{z}\mid\mathbf{z}_c,d,\delta)$. 
To this end, we construct two condition representations: a virtual multimodal cell state $\mathbf{c}_{\mathrm{cell}}=\mathrm{VMCS}(\mathbf{z}_c)$ and a mechanism-aware drug template $\mathbf{c}_{\mathrm{drug}}=\mathrm{MDT}(d,\delta)$. 
The perturbation residual is then generated by a conditional diffusion model $p_\theta(\Delta\mathbf{z}\mid\mathbf{c}_{\mathrm{cell}},\mathbf{c}_{\mathrm{drug}})$, and the predicted post-perturbation profile $\hat{\mathbf{x}}_p$ is reconstructed from the resulting latent state.


\subsection{VMCS: Virtual Multimodal Cell State}
\label{sec:vmcs}
Transcriptomic measurements do not directly capture post-translational regulation. 
We therefore use a cross-modal translator, scLinguist~\cite{fang2025sclinguist}, to infer a pseudo-protein embedding, denoted as $\mathbf{g}_p$, for each pre-perturbation cell.

\paragraph{Protein-quality aware module (PQM).}
Pseudo-protein inference quality varies across cells. 
We quantify per-cell reliability using three non-parametric scores (Appendix~\ref{app:quality_detail}): 
\textbf{RNA--protein rank consistency} $s_{\mathrm{rc}}^{(i)}$, which measures the Spearman correlation between transcriptomic and inferred protein levels; 
\textbf{protein-complex coherence} $s_{\mathrm{cx}}^{(i)}$, which measures the co-expression tightness of protein complex subunits; and 
\textbf{embedding neighbor consistency} $s_{\mathrm{nb}}^{(i)}$, which measures the cosine similarity to the $k$-NN centroid in transcriptome space. 
After z-score normalization, we average the three scores and map the result through a sigmoid function $\sigma(\cdot)$ to obtain a scalar quality weight:
\begin{equation}
  q_i = \sigma\!\bigl(\tfrac{1}{3}(\tilde{s}_{\mathrm{rc}}^{(i)}+\tilde{s}_{\mathrm{cx}}^{(i)}+\tilde{s}_{\mathrm{nb}}^{(i)})\bigr),
  \label{eq:quality}
\end{equation}
where $q_i\in(0,1)$. 
We validate this design on an independent CITE-seq dataset with paired RNA and surface protein measurements, where $q_i$ shows a strong correlation with measured protein fidelity (Appendix~\ref{app:qi_validation}).

\paragraph{Cross-modal decomposition via learned rotation.}
To disentangle shared transcriptome--proteome variation from modality-specific effects, we first map $\mathbf{z}_c$ and $\mathbf{g}_p$ into a common $D$-dimensional space using projectors $\phi_z$ and $\phi_p$. 
We learn an orthogonal rotation $\mathbf{R}\in O(D)$ parameterized by the Cayley transform:
\begin{equation}
  \mathbf{R}=(\mathbf{I}-\mathbf{S})(\mathbf{I}+\mathbf{S})^{-1},
  \qquad
  \mathbf{S}=\mathbf{A}-\mathbf{A}^{\!\top},
\end{equation}
where $\mathbf{A}\in\mathbb{R}^{D\times D}$ is an unconstrained learnable matrix initialized to zero. 
This gives the rotated features
\begin{equation}
  \tilde{\mathbf{f}}_r = \mathbf{R}\,\phi_z(\mathbf{z}_c),\qquad
  \tilde{\mathbf{f}}_p = \mathbf{R}\,\phi_p(\mathbf{g}_p).
  \label{eq:rotate}
\end{equation}

We use the dimension-wise batch Pearson correlation
$\rho_j=\mathrm{Corr}(\tilde{\mathbf{f}}_{r,j},\,\tilde{\mathbf{f}}_{p,j})$
to estimate cross-modal agreement. 
After rescaling $\rho_j$ to $\hat\rho_j=(\rho_j+1)/2$, we obtain sharedness weights
$w_j^{s}=\sigma\bigl((\hat\rho_j-\tau_s)/\beta\bigr)$, where $\tau_s$ and $\beta$ are learnable. 
These weights decompose the rotated features into a shared cell state and a protein-associated residual:
\begin{equation}
  \mathbf{h}_s = \mathbf{R}^{\!\top}\!\bigl(\mathbf{w}^{s}\odot
    \tfrac{1}{2}(\tilde{\mathbf{f}}_r+\tilde{\mathbf{f}}_p)\bigr),
  \qquad
  \mathbf{r}_p = \mathbf{R}^{\!\top}\!\bigl((\mathbf{1}-\mathbf{w}^{s})\odot\tilde{\mathbf{f}}_p\bigr).
  \label{eq:decomp}
\end{equation}

Finally, we align the shared components with an agreement-weighted MMD loss:
\begin{equation}
  \mathcal{L}_{\mathrm{align}} = \mathrm{MMD}_{k_{\boldsymbol{\rho}}}\!\bigl(\mathbf{w}^{s}\!\odot\!\tilde{\mathbf{f}}_r,\; \mathbf{w}^{s}\!\odot\!\tilde{\mathbf{f}}_p\bigr),
  \label{eq:mmd}
\end{equation}
where $k_{\boldsymbol{\rho}}$ is a mixture-of-Gaussians kernel weighted by $\hat\rho_j$ to emphasize dimensions with stronger transcriptome--proteome agreement (Appendix~\ref{app:mmd}).

\paragraph{Complementary masking.}
To avoid collapse of the shared bottleneck to a single modality, we use complementary masking guided by pseudo-protein reliability. 
For dimension $j$ of cell $i$, the masking probability is proportional to $q_i\hat\rho_j$ and clipped to $[p_{\min},p_{\max}]$. 
A complementary Bernoulli assignment enforces $m_{ij}^{r}m_{ij}^{p}=0$, ensuring that the two modalities are not masked simultaneously at the same dimension. 
We then reconstruct the full shared state from the masked views with a stop-gradient target:
\begin{equation}
  \mathcal{L}_{\mathrm{mcc}}
  = \Bigl\|\mathbf{R}^{\!\top}\!\Bigl(\mathbf{w}^{s}\!\odot
    \tfrac{\tilde{\mathbf{f}}_r^{\,m}+\tilde{\mathbf{f}}_p^{\,m}}{2}\Bigr)
  - \mathrm{sg}\Bigl[\mathbf{R}^{\!\top}\!\Bigl(\mathbf{w}^{s}\!\odot
    \tfrac{\tilde{\mathbf{f}}_r+\tilde{\mathbf{f}}_p}{2}\Bigr)\Bigr]
  \Bigr\|_2^2.
  \label{eq:mcc}
\end{equation}
The VMCS module is trained on untreated cells with
$\mathcal{L}_{\mathrm{vmcs}}=\mathcal{L}_{\mathrm{mcc}}+\lambda_{\mathrm{align}}\mathcal{L}_{\mathrm{align}}$.

\paragraph{Context tokenization.}
The shared state $\mathbf{h}_s$ and reliability-weighted residual $q_i\mathbf{r}_p$ are projected into shared and protein-associated token sets,
$\mathbf{T}_s\in\mathbb{R}^{K_s\times H}$ and $\mathbf{T}_p\in\mathbb{R}^{K_p\times H}$, respectively. 
A lightweight Transformer refines the concatenated tokens $\mathbf{T}_c=[\mathbf{T}_s;\mathbf{T}_p]$, producing the multimodal cell condition $\mathbf{c}_{\mathrm{cell}}$.

\subsection{MDT: Mechanism-Aware Drug--Gene Template}
\label{sec:mdt}

Chemical structure alone incompletely predicts transcriptional response, as structurally similar compounds may engage distinct targets, whereas divergent molecules can converge on the same pathway.
To capture MOA specificity, MDT integrates chemical structure with two external priors, L1000 transcriptomic signatures and a heterogeneous biological graph, into a unified drug context:
\begin{equation}
  \mathbf{T}_d = f_{\mathrm{tok}}(\mathbf{z}_d^s),\qquad
  \mathbf{e}_{\mathrm{drug}} = \psi_d(\mathbf{z}_d^s),
  \label{eq:mdt_assembly}
\end{equation}
where $\mathbf{z}_d^s=f_s(\mathbf{x}_d^{s})$ denotes the UniMol~\cite{zhou2023uni} structure embedding, and $f_{\mathrm{tok}}$ and $\psi_d$ are learned projections.
At inference, MDT only requires $f_s$, $f_{\mathrm{tok}}$, and $\psi_d$, enabling zero-shot generalization to unseen drugs.

~\cite{zhou2023uni}

\paragraph{Biological prior.}
To ensure the integrity of the captured biological context, we freeze the parameters of the graph encoder following its pre-training phase. This yields static yet highly informative embeddings $\mathbf{z}^b_d \in \mathbb{R}^{256}$, effectively decoupling generic biological knowledge from task-specific learning.

\paragraph{Soft-label contrastive learning.}
In a multi-view setting, each drug is represented by one structural view and up to $K$ transcriptomic views selected by Transcriptional Activity Score \cite{Subramanian2017_L1000} (Appendix~\ref{app:multiview}).
Instead of binary labels, we construct a soft similarity matrix $M\in[0,1]^{N\times N}$ that encodes both intra-drug response consistency and inter-drug biological similarity, preventing false negatives among mechanistically related drugs (Appendix~\ref{app:softlabel}).
The weighted contrastive loss is:

\begin{equation}
\mathcal{L}_{\text{contrast}} =
-\frac{1}{|\mathcal{P}|}
\sum_{i \in \mathcal{P}} \sum_{j \neq i}
M_{ij}
\log
\frac{
\exp(\tau\langle\hat{\mathbf{z}}_i,\hat{\mathbf{z}}_j\rangle)
}{
\sum_{k \neq i}
\exp(\tau\langle\hat{\mathbf{z}}_i,\hat{\mathbf{z}}_k\rangle)
}.
\label{eq:contrast_main}
\end{equation}

\paragraph{Knowledge distillation.}
Transcriptomic responses do not explicitly encode target-level semantics.
We distill $\mathbf{z}^b_d$ into the structure encoder via a lightweight projection head $g(\cdot)$:
\begin{equation}
\mathcal{L}_{\text{distill}} =
1 - \mathbb{E}_{d} \cos(g(\mathbf{z}^s_d),\,\mathbf{z}^b_d),
\label{eq:distill_main}
\end{equation}

The total MDT objective is 
$\mathcal{L} = \alpha\cdot\mathcal{L}_{\text{contrast}} + \beta\cdot\mathcal{L}_{\text{distill}}$,
where $\alpha$, $\beta$ controls the distillation weight. The drug context is assembled as 
$\mathbf{c}_{\mathrm{drug}}=(\mathbf{T}_d,\mathbf{e}_{\mathrm{drug}})$,
and dosage is incorporated at the final stage:
\begin{equation}
  \mathbf{c}_{\mathrm{drug}} = \log(1+\delta)\cdot\mathbf{c}_{\mathrm{drug}}.
\end{equation}

After training, $\mathbf{c}_{\mathrm{drug}}$ conditions the downstream diffusion model.
For OOD evaluation, drugs in the test set are excluded from MDT pretraining to prevent data leakage.

\subsection{Perturbation-Aware Conditional Diffusion}
\label{sec:generation}

\paragraph{Bidirectional condition interaction and denoising.}
Before denoising, the cell and drug condition tokens interact through bidirectional cross-attention:
\begin{equation}
  \mathbf{T}_d' = \mathbf{T}_d + \mathrm{XAttn}(\mathbf{T}_d,\,\mathbf{T}_c),
  \qquad
  \mathbf{T}_c' = \mathbf{T}_c + \mathrm{XAttn}(\mathbf{T}_c,\,\mathbf{T}_d).
  \label{eq:preinteract}
\end{equation}
The updated cell tokens $\mathbf{T}_c'$ are segmented into $\mathbf{T}_s'$ and $\mathbf{T}_p'$ according to the original VMCS token allocation.
The noisy residual $\Delta\mathbf{z}_t$ is projected into a noise token $\mathbf{h}_0$, which is concatenated with the condition tokens:
$\mathbf{S}=[\mathbf{h}_0;\,\mathbf{T}_s';\,\mathbf{T}_p';\,\mathbf{T}_d']$.
The denoiser applies token type embeddings and separate QKV projections for noise, shared cell, protein-associated, and drug tokens, and performs joint self-attention over the full sequence with AdaLN-Zero modulation.
It predicts the velocity
$\hat{\mathbf{v}}_t=f_\theta(\Delta\mathbf{z}_t,\,t,\,\mathbf{c}_{\mathrm{cell}},\,\mathbf{c}_{\mathrm{drug}})$,
with target
$\mathbf{v}_t=\sqrt{\bar\alpha_t}\,\boldsymbol{\epsilon}-\sqrt{1\!-\!\bar\alpha_t}\,\Delta\mathbf{z}$.

\paragraph{Quality-aware triplet constraints.}
A drug's effect depends on the match between its targets and the cell's protein state. We therefore construct mismatched triplets by replacing either $d$ with a random batch drug $d'\!\neq\!d$, or $\mathbf{x}_c$ with a cell from another lineage.
A cosine-margin loss pushes these mismatched predictions away from the ground-truth velocity direction:
\begin{equation}
  \mathcal{L}_{\mathrm{triplet}}
  = q_i\;\mathbb{E}_{t}\!\bigl[
    \max\!\bigl(0,\;
      \cos(\hat{\mathbf{v}}_t^{\mathrm{mis}},\,\mathbf{v}_t)
      - \cos(\hat{\mathbf{v}}_t,\,\mathbf{v}_t)
      + m\bigr)\bigr],
  \label{eq:triplet}
\end{equation}
where $\hat{\mathbf{v}}_t^{\mathrm{mis}}$ is the velocity under the mismatched condition, $m$ is a margin, and $q_i$ down-weights cells with unreliable pseudo-protein signals.
The full generation loss is
$\mathcal{L}_{\mathrm{gen}}
= \mathbb{E}_{t,\boldsymbol{\epsilon}}[\|\hat{\mathbf{v}}_t-\mathbf{v}_t\|_2^2]
+ \lambda_{\mathrm{triplet}}\mathcal{L}_{\mathrm{triplet}}$.

\paragraph{Interaction-aware CFG and inference.}
To enable controllable generation, we train with structured dropout: $\mathbf{c}_{\mathrm{cell}}$ or $\mathbf{c}_{\mathrm{drug}}$ (or both) are independently dropped, producing four conditioning modes ($\emptyset$, cell-only, drug-only, joint). At inference, classifier-free guidance composes these modes to amplify the conditioning signal:
\begin{equation}
  \hat{\mathbf{v}}^{\mathrm{guide}}
  = \hat{\mathbf{v}}_{\emptyset}
  + w_c(\hat{\mathbf{v}}_{\mathbf{c}_{\mathrm{cell}}} - \hat{\mathbf{v}}_{\emptyset})
  + w_d(\hat{\mathbf{v}}_{\mathbf{c}_{\mathrm{drug}}} - \hat{\mathbf{v}}_{\emptyset})
  + w_{cd}\bigl(\hat{\mathbf{v}}_{\mathbf{c}_{\mathrm{cell}},\mathbf{c}_{\mathrm{drug}}}
    - \hat{\mathbf{v}}_{\mathbf{c}_{\mathrm{cell}}}
    - \hat{\mathbf{v}}_{\mathbf{c}_{\mathrm{drug}}}
    + \hat{\mathbf{v}}_{\emptyset}\bigr),
  \label{eq:icfg}
\end{equation}
where $w_c,w_d$ control the strength of marginal conditions and $w_{cd}$ amplifies the drug--cell interaction. Sampling uses DDIM\cite{song2020denoising} with this guided velocity (see Appendix~\ref{app:diffusion_detail} for derivation).

\section{Experiments}
\label{headings}
\paragraph{Datasets.}
We evaluate cell-context generalization on Tahoe-100M, KaggleCrossCell, KaggleCrossPatient, CrossPatient, and sciplex3. For perturbation generalization, we use Tahoe-100M and sciplex3 for single-drug prediction, and sciplex3-comb for combinatorial prediction. Dataset details are provided in Appendix~\ref{OOD_Split_Construction}.

\paragraph{Setup.}
We evaluate two OOD settings: \textbf{cell-context generalization}, which tests transfer to held-out cellular contexts, and \textbf{perturbation generalization}, which tests transfer to held-out drugs and drug combinations (Appendix~\ref{Evaluation_Protocol_Details}). We also perform CITE-seq validation, ablations, diversity-controlled scaling, and robustness tests under noise and sparsity. Details are provided in Appendix~\ref{Evaluation_Protocol_Details}.

\paragraph{Metrics.}

We follow the scPerturBench protocol~\cite{Zhang2025_scPerturBench} and report six metrics: MSE, PCC$\Delta$, E-distance (E-dist), Wasserstein distance (WD), KL divergence (KL), and Common-DEGs, covering population-level accuracy, distributional discrepancy, and DEG recovery. Results are reported under the top-100-DEG and top-5000-DEG settings, with WD and Common-DEGs computed only for top-100 DEGs. Details are provided in Appendices~\ref{DEGs_Definition} and~\ref{Evaluation_Metrics}.

\section{Results}
\label{headings}

\subsection{Out-of-Distribution Generalization on Chemical Perturbations}

Table~\ref{tab:unseen_cell_unseen_cell_100degs} reports the top-100-DEG OOD results under cell-context and perturbation generalization. Following the metrics defined above, we evaluate MSE, PCC$\Delta$, E-dist, KL, WD, and Common-DEGs to capture point-wise accuracy, perturbation-response correlation, distributional discrepancy, and DEG recovery. StateXDiff achieves the most consistent gains on distributional discrepancy, attaining the lowest E-dist and WD across all datasets and the best or second-best KL in both settings. This suggests that StateXDiff is particularly effective at recovering the target expression distribution under unseen biological contexts.

The improvements are more nuanced for MSE, PCC$\Delta$, and Common-DEGs. In cell-context generalization, StateXDiff strongly improves E-dist and WD on both Tahoe-100M and KaggleCrossPatient, but individual baselines achieve better MSE, PCC$\Delta$, or Common-DEGs on some datasets, especially under patient-level heterogeneity. In perturbation generalization, StateXDiff shows a more dominant pattern, achieving the best results on five of the six metrics for both Tahoe-100M and sciplex3-MCF7. Evaluations for the 5,000 DEGs are reported in Appendix Tables~\ref{tab:unseen_drug_All_D3_OOD} and~\ref{tab:unseen_cell_allgene_D3_ood}. Overall, these results indicate that StateXDiff provides robust OOD generalization, with the most stable gains in distributional discrepancy and additional improvements in point-wise accuracy and response correlation under unseen perturbations.

\begin{table*}[h]
		\centering
		\renewcommand{\arraystretch}{0.7}
		\sisetup{group-separator={}}
		\caption{Evaluation is based on the top 100 DEGs.}
		\label{tab:unseen_cell_unseen_cell_100degs}
  \vspace{-4pt}
  {\scriptsize\raggedright
  \textit{Note.}
  \begingroup
  \setlength{\fboxsep}{0pt}%
  \colorbox{firstbg}{\strut\hspace{1pt}Rank-1\hspace{1pt}}%
  \endgroup
   and
  \begingroup
  \setlength{\fboxsep}{0pt}%
  \colorbox{secondbg}{\strut\hspace{1pt}Rank-2\hspace{1pt}}%
  \endgroup
   denote the best and second-best results;
  \textcolor{teal}{$\uparrow/\downarrow$} shows the relative gain of Rank-1 over Rank-2.
  \par}
  \vspace{1pt}
		\resizebox{\textwidth}{!}{%
			\begin{tabular}{llcccccc}
				\toprule
				\textbf{Dataset} & \textbf{Method} &
				\textbf{E-dist$\downarrow$} &
				\textbf{MSE$\downarrow$} &
				\textbf{PCC$\Delta\uparrow$} &
				\textbf{KL$\downarrow$} &
				\textbf{WD$\downarrow$} &
				\textbf{ Common-DEGs$\uparrow$} \\
				
				\midrule
				\multicolumn{8}{l}{\textbf{\textit{Cell-context Generalization}}} \\
				\midrule
				
				\multirow{11}{*}{\textbf{Tahoe-100M}}
				& trainMean
				& \num{3.72}\std{1.55}
				& \num{0.0972}\std{0.0652}
				& \num{0.419}\std{0.195}
				& \num{20.8}\std{16.7}
				& \num{11.7}\std{7.1}
				& \num{0.0}\std{0.0}
				\\
				& baseControl
				& \num{2.85}\std{0.019}
				& \cellcolor{secondbg}\num{0.0714}\std{0.001}
				& \num{-0.0142}\std{0.0021}
				& \num{19.9}\std{0.214}
				& \num{9.01}\std{0.106}
				& \num{0.0039}\std{0.0062}
				\\
				& baseMLP
				& \num{4.15523}\std{2.144016}
				& \num{0.258433}\std{0.1132}
				& \num{0.339057}\std{0.253156}
				& \num{34.535425}\std{1.220073}
				& \num{30.33549}\std{11.517}
				& \cellcolor{secondbg}\num{0.09333}\std{0.064216}
				\\
				& baseReg
				& \num{3.18}\std{0.0162}
				& \cellcolor{firstbg}\num{0.0627}\std{0.0005}{\scriptsize\textcolor{teal}{$\downarrow$12\%}}
				& \num{0.65}\std{0.0021}
				& \num{21.0}\std{0.21}
				& \num{7.13}\std{0.0559}
				& \num{0.0096}\std{0.0133}
				\\
				& bioLord
				& \num{5.92}\std{0.0311}
				& \num{0.131}\std{0.0013}
				& \num{0.463}\std{0.0035}
				& \num{48.3}\std{0.0147}
				& \num{13.9}\std{0.131}
				& \num{0.0016}\std{0.0021}
				\\
				& CellFlow
				& \cellcolor{secondbg}\num{1.953486}\std{1.632878}
				& \num{0.094111}\std{0.11381}
				& \num{0.679132}\std{0.213548}
				& \cellcolor{secondbg}\num{17.196754}\std{8.619738}
				& \cellcolor{secondbg}\num{6.27564}\std{3.13848}
				& \num{0.0471}\std{0.00154}
				\\
				& CRISP
				& \num{10.77147}\std{1.287868}
				& \num{0.47921}\std{0.06232}
				& \num{0.411714}\std{0.195301}
				& \num{26.212668}\std{16.658191}
				& \num{24.712}\std{7.1041}
				& \num{0.0269}\std{0.00364}
				\\
				& STATE
				& \num{8.27}\std{2.43}
				& \num{0.316}\std{0.167}
				& \num{0.789}\std{0.214}
				& \num{39.7}\std{8.62}
				& \num{38.2}\std{17.63}
				& \num{0.0147}\std{0.0157}
				\\
        		& trVAE
				& \num{8.13}\std{3.15}
				& \num{0.327}\std{0.367}
				& \cellcolor{firstbg}\num{0.889}\std{0.574}{\scriptsize\textcolor{teal}{$\uparrow$1\%}}
				& \num{41.71}\std{6.62}
				& \num{43.12}\std{8.63}
				& \num{0.051}\std{0.0214}
				\\
				& PerturbDiff
                & \num{4.6720}\std{1.9609}
                & \num{0.1747}\std{0.1075}
                & \num{0.4879}\std{0.2250}
                & \num{39.0554}\std{8.5850}
                & \num{24.3802}\std{12.1636}
				& \num{0.0202}\std{0.0238}
				\\
    			& scDFM
				& \num{3.86}\std{1.31}
				& \num{0.143}\std{0.107}
				& \num{0.771}\std{0.114}
				& \num{21.6}\std{4.12}
				& \num{18.2}\std{9.71}
				& \num{0.0715}\std{0.0157}
				\\
				& \cellcolor{gray!15}\textbf{Ours}
				& \cellcolor{firstbg}\num{1.134}\std{0.737}{\scriptsize\textcolor{teal}{$\downarrow$42\%}}
				& \cellcolor{gray!15}\num{0.0802}\std{0.0244}
				& \cellcolor{secondbg}\num{0.883}\std{0.187}
				& \cellcolor{firstbg}\num{4.787}\std{0.659}{\scriptsize\textcolor{teal}{$\downarrow$72\%}}
				& \cellcolor{firstbg}\num{3.614}\std{1.722}{\scriptsize\textcolor{teal}{$\downarrow$42\%}}
				& \cellcolor{firstbg}\num{0.113}\std{0.259}{\scriptsize\textcolor{teal}{$\uparrow$21\%}}
				\\
				
				\cmidrule{1-8}
				
				\multirow{11}{*}{\textbf{KaggleCrossPatient}}
				& trainMean
				& \num{1.427047}\std{1.067871}
				& \num{0.047287}\std{0.113594}
				& \cellcolor{firstbg}\num{0.72666}\std{0.37376}{\scriptsize\textcolor{teal}{$\uparrow$1\%}}
				& \num{44.958542}\std{2.280192}
				& \num{60.892547}\std{28.449823}
				& \cellcolor{secondbg}\num{0.207667}\std{0.224663}
				\\
				& baseControl
				& \num{4.533147}\std{4.248013}
				& \num{0.441063}\std{0.463321}
				& \num{0.064587}\std{0.112988}
				& \num{45.039951}\std{2.409706}
				& \num{95.959373}\std{70.468781}
				& \num{0.006}\std{0.009322}
				\\
				& baseMLP
				& \num{6.603523}\std{5.334016}
				& \num{0.458433}\std{0.515452}
				& \num{0.219057}\std{0.292156}
				& \num{46.731425}\std{2.220073}
				& \num{65.33549}\std{59.7217}
				& \num{0.087333}\std{0.064216}
				\\
				& baseReg
				& \num{7.078523}\std{5.709588}
				& \num{0.468023}\std{0.522667}
				& \num{0.26649}\std{0.336932}
				& \num{42.440276}\std{3.028718}
				& \num{63.819163}\std{57.247697}
				& \num{0.008}\std{0.010954}
				\\
				& bioLord
				& \num{2.638363}\std{2.394565}
				& \num{0.144107}\std{0.180076}
				& \num{0.608232}\std{0.422056}
				& \num{39.821131}\std{2.853563}
				& \num{31.773277}\std{25.040249}
				& \num{0.001333}\std{0.004342}
				\\
				& CellFlow
				& \num{1.36423}\std{2.17862}
				& \cellcolor{firstbg}\num{0.03265}\std{0.015423}{\scriptsize\textcolor{teal}{$\downarrow$7\%}}
				& \num{0.5206}\std{0.4872}
				& \num{44.65976}\std{5.6437}
				& \num{43.44672}\std{26.8762}
				& \num{0.04232}\std{0.04874}
				\\
				& CRISP
				& \num{3.341212}\std{2.873552}
				& \num{0.1215}\std{0.195291}
				& \num{0.5916}\std{0.417174}
				& \cellcolor{firstbg}\num{33.650811}\std{15.472682}{\scriptsize\textcolor{teal}{$\downarrow$6\%}}
				& \num{33.372127}\std{24.711131}
				& \num{0.157211}\std{0.26527}
				\\
				& STATE
				& \num{0.545847}\std{0.757583}
				& \num{0.03836}\std{0.072986}
				& \num{0.380437}\std{0.575654}
				& \num{41.343833}\std{2.835105}
				& \num{26.499937}\std{16.215588}
				& \cellcolor{firstbg}\num{0.309}\std{0.266}{\scriptsize\textcolor{teal}{$\uparrow$49\%}}
				\\
				& trVAE
				& \num{1.220087}\std{1.17432}
				& \num{0.0434}\std{0.093348}
				& \cellcolor{secondbg}\num{0.716697}\std{0.377697}
				& \num{39.855734}\std{4.08784}
				& \cellcolor{secondbg}\num{18.792547}\std{16.472036}
				& \num{0.078667}\std{0.108556}
				\\
        		& PerturbDiff
				& \num{0.641817}\std{0.578583}
				& \num{0.08836}\std{0.062986}
				& \num{0.410437}\std{0.381654}
				& \num{47.543833}\std{5.835105}
				& \num{28.499937}\std{11.54588}
				& \num{0.189}\std{0.086}
				\\
    			& scDFM
				& \cellcolor{secondbg}\num{0.531847}\std{0.416583}
				& \num{0.04836}\std{0.032986}
				& \num{0.51537}\std{0.411654}
				& \num{38.143833}\std{1.435105}
				& \num{24.812937}\std{16.215588}
				& \num{0.109}\std{0.136}
				\\
				& \cellcolor{gray!15}\textbf{Ours}
				& \cellcolor{firstbg}\num{0.269574}\std{1.15957}{\scriptsize\textcolor{teal}{$\downarrow$49\%}}
				& \cellcolor{secondbg}\num{0.03504}\std{0.11478}
				& \cellcolor{gray!15}\num{0.639927}\std{0.579111}
				& \cellcolor{secondbg}\num{35.869525}\std{2.021119}
				& \cellcolor{firstbg}\num{14.418533}\std{24.449907}{\scriptsize\textcolor{teal}{$\downarrow$23\%}}
				& \cellcolor{gray!15}\num{0.1132}\std{0.23422}
				\\
				
				\midrule
				\multicolumn{8}{l}{\textbf{\textit{Perturbation Generalization}}} \\
				\midrule
				
				\multirow{13}{*}{\textbf{Tahoe-100M}}
				& trainMean
				& \num{4.109}\std{1.797}
				& \num{0.16}\std{0.114}
				& \cellcolor{secondbg}\num{0.7}\std{0.235}
				& \num{25.41}\std{17.99}
				& \num{24.87}\std{13.89}
				& \num{0.0}\std{0.0}
				\\
				& baseControl
				& \num{2.955}\std{0.022}
				& \num{0.082}\std{0.001}
				& \num{-0.003}\std{0.003}
				& \num{22.99}\std{0.279}
				& \num{12.09}\std{0.224}
				& \num{0.004}\std{0.006}
				\\
				& baseMLP
				& \num{14.92}\std{0.078}
				& \num{1.035}\std{0.008}
				& \num{0.079}\std{0.004}
				& \num{26.81}\std{0.279}
				& \num{112.5}\std{0.823}
				& \num{0.0}\std{0.0}
				\\
				& baseReg
				& \num{4.194}\std{0.029}
				& \num{0.164}\std{0.002}
				& \num{0.302}\std{0.004}
				& \num{25.45}\std{0.292}
				& \num{25.26}\std{0.224}
				& \num{0.005}\std{0.009}
				\\
				& bioLord
				& \num{3.88}\std{0.035}
				& \cellcolor{firstbg}\num{0.074}\std{0.001}{\scriptsize\textcolor{teal}{$\downarrow$4\%}}
				& \num{0.535}\std{0.005}
				& \num{47.82}\std{0.021}
				& \cellcolor{secondbg}\num{8.412}\std{0.142}
				& \num{0.0}\std{0.002}
				\\
				& CellFlow
				& \num{2.1649}\std{2.3592}
				& \num{0.0972}\std{0.1267}
				& \num{0.4187}\std{0.3269}
				& \cellcolor{secondbg}\num{13.8104}\std{12.8213}
				& \num{15.5211}\std{14.5143}
				& \cellcolor{secondbg}\num{0.0902}\std{0.0223}
				\\
                & CRISP
                & \num{5.823}\std{2.134}
                & \num{0.312}\std{0.187}
                & \num{0.451}\std{0.264}
                & \num{31.47}\std{14.32}
                & \num{22.84}\std{11.56}
                & \num{0.018}\std{0.024}
				\\
				& STATE
				& \num{3.565}\std{2.359}
				& \num{0.117}\std{0.137}
				& \num{0.489}\std{0.327}
				& \num{23.81}\std{14.84}
				& \num{15.52}\std{14.51}
				& \num{0.029}\std{0.026}
				\\
				& chemCPA
				& \num{7.565}\std{1.359}
				& \num{0.715}\std{0.247}
				& \num{0.694}\std{0.227}
				& \num{38.81}\std{19.44}
				& \num{28.52}\std{13.51}
				& \num{0.089}\std{0.426}
				\\
				& PerturbDiff
                & \num{3.9692}\std{2.7606}
                & \num{0.1582}\std{0.1533}
                & \num{0.5300}\std{0.3093}
                & \num{39.7553}\std{8.4247}
                & \num{23.5712}\std{18.3338}
				& \num{0.0154}\std{0.0186}
				\\
    			& scDFM
				& \cellcolor{secondbg}\num{1.355}\std{1.159}
				& \num{0.108}\std{0.097}
				& \num{0.689}\std{0.317}
				& \num{16.81}\std{10.84}
				& \num{11.12}\std{14.51}
				& \num{0.029}\std{0.026}
				\\
				& \cellcolor{gray!15}\textbf{Ours}
				& \cellcolor{firstbg}\num{0.614}\std{0.746}{\scriptsize\textcolor{teal}{$\downarrow$55\%}}
				& \cellcolor{secondbg}\num{0.077}\std{0.351}
				& \cellcolor{firstbg}\num{0.756}\std{0.283}{\scriptsize\textcolor{teal}{$\uparrow$8\%}}
				& \cellcolor{firstbg}\num{9.744}\std{12.97}{\scriptsize\textcolor{teal}{$\downarrow$29\%}}
				& \cellcolor{firstbg}\num{2.242}\std{14.06}{\scriptsize\textcolor{teal}{$\downarrow$73\%}}
				& \cellcolor{firstbg}\num{0.14}\std{0.148}{\scriptsize\textcolor{teal}{$\uparrow$55\%}}
				\\
				
				\cmidrule{1-8}
				
				\multirow{12}{*}{\textbf{sciplex3\_MCF7}}
				& trainMean
				& \num{0.9656}\std{2.3823}
				& \num{0.1091}\std{0.3181}
				& \num{0.4857}\std{0.1793}
				& \num{42.481}\std{9.9588}
				& \num{65.2709}\std{65.4158}
				& \num{0.0}\std{0.0}
				\\
				& baseControl
				& \num{1.0692}\std{2.4793}
				& \num{0.1208}\std{0.3356}
				& \num{0.0693}\std{0.1023}
				& \num{42.4851}\std{9.95}
				& \num{66.5907}\std{67.3182}
				& \num{0.044}\std{0.022}
				\\
				& baseMLP
				& \num{1.717}\std{4.3479}
				& \num{0.1795}\std{0.5704}
				& \num{0.3036}\std{0.2892}
				& \num{42.7035}\std{10.1807}
				& \num{72.766}\std{72.7848}
				& \cellcolor{secondbg}\num{0.0854}\std{0.088}
				\\
				& baseReg
				& \num{0.9996}\std{1.9384}
				& \num{0.0937}\std{0.2092}
				& \num{0.3555}\std{0.294}
				& \num{42.517}\std{10.0391}
				& \num{63.8486}\std{53.3737}
				& \num{0.0}\std{0.0}
				\\
				& bioLord
				& \num{4.0216}\std{2.9643}
				& \num{0.0865}\std{0.252}
				& \num{0.6013}\std{0.2006}
				& \num{45.8137}\std{2.1646}
				& \num{39.5768}\std{40.3243}
				& \num{0.0}\std{0.0005}
				\\
				& CellFlow
				& \cellcolor{secondbg}\num{0.5749}\std{0.2181}
				& \num{0.0862}\std{0.0173}
				& \num{0.3853}\std{0.2411}
				& \num{39.3265}\std{1.372}
				& \cellcolor{secondbg}\num{26.8942}\std{8.6719}
				& \num{0.0212}\std{0.0023}
				\\
				& CRISP
				& \num{1.038}\std{3.2123}
				& \num{0.1209}\std{0.3215}
				& \num{0.4355}\std{0.214}
				& \num{42.1027}\std{5.3908}
				& \num{41.7036}\std{47.5388}
				& \num{0.0301}\std{0.0158}
				\\
				& STATE
				& \num{0.9195}\std{2.2287}
				& \num{0.1023}\std{0.2942}
				& \num{0.4518}\std{0.2471}
				& \num{41.039}\std{9.6217}
				& \num{54.5977}\std{58.7983}
				& \cellcolor{firstbg}\num{0.0943}\std{0.0592}{\scriptsize\textcolor{teal}{$\uparrow$10\%}}
				\\
				& chemCPA
				& \num{3.5521}\std{2.8851}
				& \num{0.0887}\std{0.2675}
				& \cellcolor{secondbg}\num{0.6575}\std{0.1848}
				& \cellcolor{secondbg}\num{38.7011}\std{7.5788}
				& \num{39.6345}\std{41.7439}
				& \num{0.0102}\std{0.0223}
				\\
            	& PerturbDiff
				& \num{1.21195}\std{1.2384}
				& \num{0.2413}\std{0.2352}
				& \num{0.4116}\std{0.1457}
				& \num{41.539}\std{10.5217}
				& \num{51.8977}\std{48.7983}
				& \num{0.0443}\std{0.0192}
				\\
    			& scDFM
				& \num{0.7595}\std{0.817}
				& \cellcolor{secondbg}\num{0.0823}\std{0.1542}
				& \num{0.5818}\std{0.2441}
				& \num{40.015}\std{6.5247}
				& \num{34.4957}\std{34.7983}
				& \num{0.0743}\std{0.01592}
				\\
				& \cellcolor{gray!15}\textbf{Ours}
				& \cellcolor{firstbg}\num{0.3749}\std{0.3099}{\scriptsize\textcolor{teal}{$\downarrow$35\%}}
				& \cellcolor{firstbg}\num{0.0162}\std{0.0217}{\scriptsize\textcolor{teal}{$\downarrow$80\%}}
				& \cellcolor{firstbg}\num{0.6809}\std{0.3428}{\scriptsize\textcolor{teal}{$\uparrow$4\%}}
				& \cellcolor{firstbg}\num{37.2265}\std{1.435}{\scriptsize\textcolor{teal}{$\downarrow$4\%}}
				& \cellcolor{firstbg}\num{16.8842}\std{8.6318}{\scriptsize\textcolor{teal}{$\downarrow$37\%}}
				& \cellcolor{gray!15}\num{0.0712}\std{0.0037}
				\\
				
				\bottomrule
		\end{tabular}}
	\end{table*}

\subsection{Biologically Consistent Representations}

\begin{figure*}[h]
  \centering
  \includegraphics[width=\textwidth]{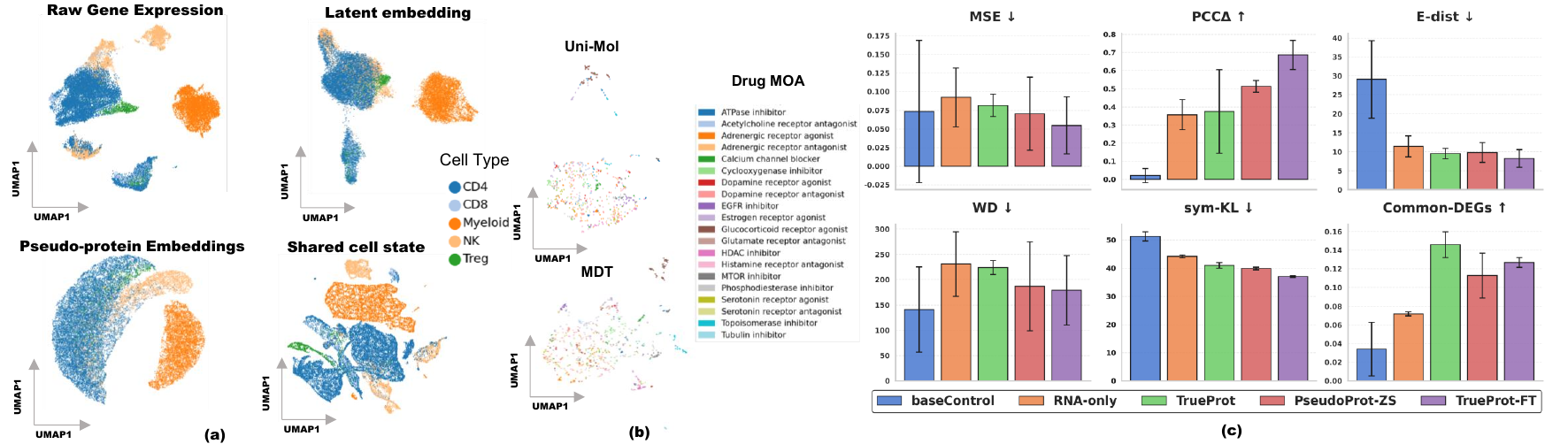}
  \caption[Biological consistency evaluation of StateXDiff.]{
  Biological consistency evaluation of StateXDiff.
  \textbf{a,} Cell-level RNA--protein representation analysis.
  \textbf{b,} Drug-level representation analysis.
  \textbf{c,} Performance comparison of CITE-seq variants.
  }
  \label{fig:cell_drug_citeseq}
\end{figure*}

To assess the biological consistency of the learned representations, we analyze the embedding spaces from both drug and cellular perspectives, and further evaluate the contribution of the protein modality using CITE-seq data. At the drug level, structure-based embeddings fail to capture mechanism-level similarities, whereas representations pretrained on transcriptional responses better reflect functional relatedness aligned with known mechanisms of action (Fig.~\ref{fig:cell_drug_citeseq}a). At the cell level, the aligned RNA–protein representations preserve clear cell-type separation while maintaining continuous transitions between related states (Fig.~\ref{fig:cell_drug_citeseq}b), indicating that the model learns biologically meaningful cell-state structures.

We further evaluate the contribution of the protein modality on CITE-seq data by comparing five methods: \textbf{baseControl}, a baseline without protein modality integration; \textbf{RNA-only}, which uses only RNA; \textbf{PseudoProt-ZS}, which infers pseudo-protein representations from RNA in a zero-shot manner using scLinguist; \textbf{TrueProt}, which incorporates true protein measurements together with RNA; and \textbf{TrueProt-FT}, which further incorporates protein representations derived from fine-tuned scLinguist. As shown in Fig.~\ref{fig:cell_drug_citeseq}c, \textbf{TrueProt-FT} achieves the best overall performance. \textbf{PseudoProt-ZS} ranks second overall and remains competitive with \textbf{TrueProt}, while \textbf{RNA-only} performs worse than the protein-informed variants and \textbf{baseControl} yields the weakest overall results. These findings suggest that incorporating protein information improves representation quality, and that informative protein-level representations can be inferred from RNA through cross-modal learning.

\subsection{Multi-Drug Combination Prediction}

\begin{figure}[htbp]

\centering
\includegraphics[width=0.95\columnwidth]{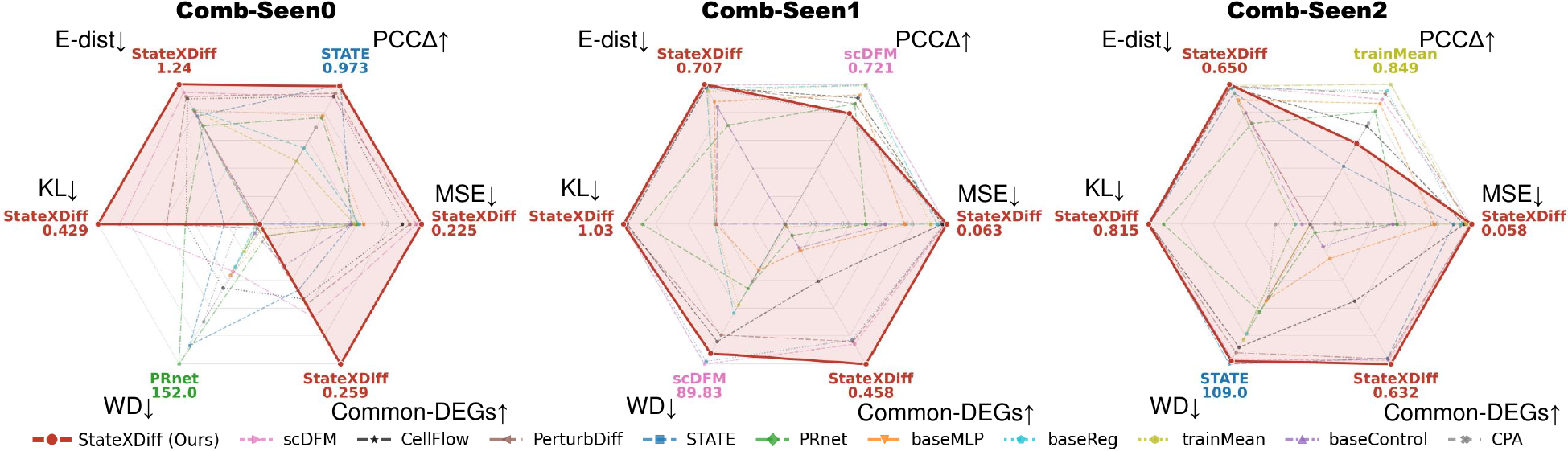}
\caption{Performance comparison on chemical combination prediction evaluated on the top-100 DEGs. Our method achieves the best overall results across most metrics.}
\label{comb}
\end{figure}

We further evaluated StateXDiff on combinatorial chemical perturbation prediction under three visibility-based splits: Comb-Seen0, Comb-Seen1, and Comb-Seen2, where zero, one, or both individual component perturbations of a held-out test combination are present in the training set, respectively. Comb-Seen0 is particularly challenging because neither component drug is observed during training. On the top 100 DEGs, StateXDiff achieved the best overall performance across the three splits, demonstrating strong generalization to unseen drug combinations (Figure~\ref{comb}). Evaluations for the 5,000 DEGs are provided in Figure~\ref{comb_5000}.

\subsection{Ablation and Diversity Scaling}
\label{sec:ablation}

\begin{figure*}[h]
\centering
\begin{minipage}[t]{0.48\textwidth}
\centering
\vspace{0pt}
\captionof{table}{Component ablation (DEG100) on Tahoe.}
\label{tab:tahoe_ablation_deg100}
  \vspace{-4pt}
  {\tiny\raggedright
  \textit{Note.}
  \begingroup
  \setlength{\fboxsep}{0pt}%
  \colorbox{firstbg}{\strut\hspace{1pt}Rank-1\hspace{1pt}}%
  \endgroup
   and
  \begingroup
  \setlength{\fboxsep}{0pt}%
  \colorbox{secondbg}{\strut\hspace{1pt}Rank-2\hspace{1pt}}%
  \endgroup
   denote the best and second-best results;
  \textcolor{teal}{$\uparrow/\downarrow$} shows the relative gain of Rank-1 over Rank-2.
  \par}
  \vspace{1pt}
\scriptsize
\setlength{\tabcolsep}{1.8pt}
\renewcommand{\arraystretch}{1.05}
\resizebox{\linewidth}{!}{
\begin{tabular}{llcccccc}
\toprule
Setting & Method
& MSE $\downarrow$
& PCC$\Delta\uparrow$
& E-dist $\downarrow$
& KL $\downarrow$
& WD $\downarrow$
& Common-DEGs $\uparrow$ \\
\midrule
\multirow{7}{*}{UD}
& w/o VMCS
& 0.0678 & \cellcolor{firstbg}0.3575{\scriptsize\textcolor{teal}{$\uparrow$5\%}} & 1.6231 & 4.8942 & 7.5893 & 0.1421 \\
& w/o pseudo\textsuperscript{$\dagger$}
& 0.0646 & \cellcolor{secondbg}0.3399 & 1.5728 & 4.9566 & 7.4046 & 0.1549 \\
& w/o MDT
& 0.0506 & 0.2543 & 1.3117 & 5.0851 & 6.4767 & 0.1921 \\
& w/o PQM
& 0.0489 & 0.2478 & 1.2651 & 4.7843 & 6.2517 & 0.2068 \\
& w/o triplet
& 0.0498 & 0.2500 & 1.2988 & 5.0680 & 6.4528 & 0.1946 \\
& w/o ICFG
& \cellcolor{secondbg}0.0483 & 0.2451 & \cellcolor{secondbg}1.2478 & \cellcolor{secondbg}4.7215 & \cellcolor{secondbg}6.1926 & \cellcolor{secondbg}0.2098 \\
& \cellcolor{gray!15}\textbf{Ours}
& \cellcolor{firstbg}\textbf{0.0468}{\scriptsize\textcolor{teal}{$\downarrow$3\%}}
& \cellcolor{gray!15}\textbf{0.2387}
& \cellcolor{firstbg}\textbf{1.2184}{\scriptsize\textcolor{teal}{$\downarrow$2\%}}
& \cellcolor{firstbg}\textbf{4.6127}{\scriptsize\textcolor{teal}{$\downarrow$2\%}}
& \cellcolor{firstbg}\textbf{6.0843}{\scriptsize\textcolor{teal}{$\downarrow$2\%}}
& \cellcolor{firstbg}\textbf{0.2189}{\scriptsize\textcolor{teal}{$\uparrow$4\%}} \\
\midrule
\multirow{7}{*}{UC}
& w/o VMCS
& 0.0692 & \cellcolor{firstbg}0.3698{\scriptsize\textcolor{teal}{$\uparrow$5\%}} & 1.6587 & 5.0234 & 7.8723 & 0.1376 \\
& w/o pseudo\textsuperscript{$\dagger$}
& 0.0663 & \cellcolor{secondbg}0.3528 & 1.6045 & 5.0839 & 7.6815 & 0.1498 \\
& w/o MDT
& 0.0521 & 0.2617 & 1.3462 & 5.2146 & 6.7132 & 0.1874 \\
& w/o PQM
& 0.0501 & 0.2553 & 1.2987 & 4.9126 & 6.4873 & 0.2003 \\
& w/o triplet
& 0.0510 & 0.2576 & 1.3314 & 5.1983 & 6.6921 & 0.1902 \\
& w/o ICFG
& \cellcolor{secondbg}0.0496 & 0.2527 & \cellcolor{secondbg}1.2814 & \cellcolor{secondbg}4.8547 & \cellcolor{secondbg}6.4215 & \cellcolor{secondbg}0.2042 \\
& \cellcolor{gray!15}\textbf{Ours}
& \cellcolor{firstbg}\textbf{0.0482}{\scriptsize\textcolor{teal}{$\downarrow$3\%}}
& \cellcolor{gray!15}\textbf{0.2465}
& \cellcolor{firstbg}\textbf{1.2519}{\scriptsize\textcolor{teal}{$\downarrow$2\%}}
& \cellcolor{firstbg}\textbf{4.7385}{\scriptsize\textcolor{teal}{$\downarrow$2\%}}
& \cellcolor{firstbg}\textbf{6.3128}{\scriptsize\textcolor{teal}{$\downarrow$2\%}}
& \cellcolor{firstbg}\textbf{0.2116}{\scriptsize\textcolor{teal}{$\uparrow$4\%}} \\
\bottomrule
\end{tabular}
}
\vspace{3pt}
\raggedright\fontsize{5.5}{6.5}\selectfont
\textsuperscript{$\dagger$}``pseudo'' = inferred pseudo-protein embedding.
\textbf{w/o VMCS}: cell condition replaced by raw transcriptomic latent $\mathbf{z}_c$;
\textbf{w/o pseudo}: protein branch removed from VMCS;
\textbf{w/o MDT}: structure-only drug encoder, no mechanism priors;
\textbf{w/o ICFG}: standard monolithic CFG.
\end{minipage}
\hfill
\begin{minipage}[t]{0.48\textwidth}
\centering
\vspace{0pt}
\includegraphics[width=0.9\linewidth]{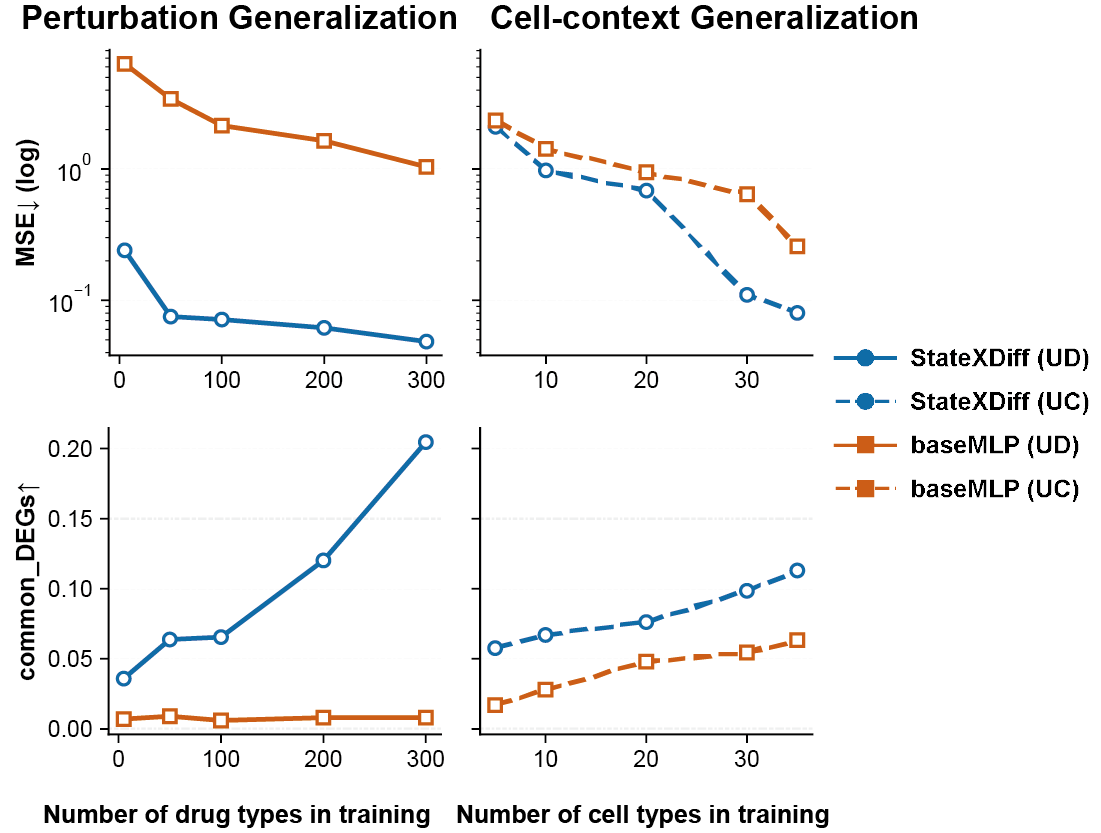}
\captionof{figure}{Performance under varying drug or cell-line diversity.}
\label{fig:diversity}
\end{minipage}
\end{figure*}

We ablate six components on Tahoe under unseen-drug (UD) and unseen-cell-line (UC) settings: the VMCS module, pseudo-protein branch, MDT mechanism priors, protein quality gate PQM, directional triplet loss, and interaction-aware CFG (ICFG). Table~\ref{tab:tahoe_ablation_deg100} reports results on the top-100 DEGs. The pseudo-protein branch and MDT priors contribute the largest gains, particularly under UD, confirming that cross-modal protein knowledge and mechanism-aware drug representations are complementary to transcriptomic features. The reliability gate $q_i$, triplet constraint, and ICFG yield smaller but consistent improvements, with $q_i$ and ICFG especially beneficial under UC where cross-cell-state shifts are more pronounced. Full-gene (DEG5000) ablation results are provided in Appendix~\ref{tab:tahoe_ablation_deg5000}. Figure~\ref{fig:diversity} further evaluates diversity scaling under a fixed sample budget by varying the number of unique drugs or cell lines. StateXDiff scales more effectively than baseMLP, with stronger gains as diversity increases. The improvement is more pronounced under UC, suggesting that cellular heterogeneity, rather than chemical diversity alone, is the primary bottleneck for perturbation generalization.

\subsection{Case Study: Linking Predictions to Resistance Mechanisms and Targeted Inhibition}

\begin{figure}[b]
\centering
\includegraphics[width=0.95\columnwidth]{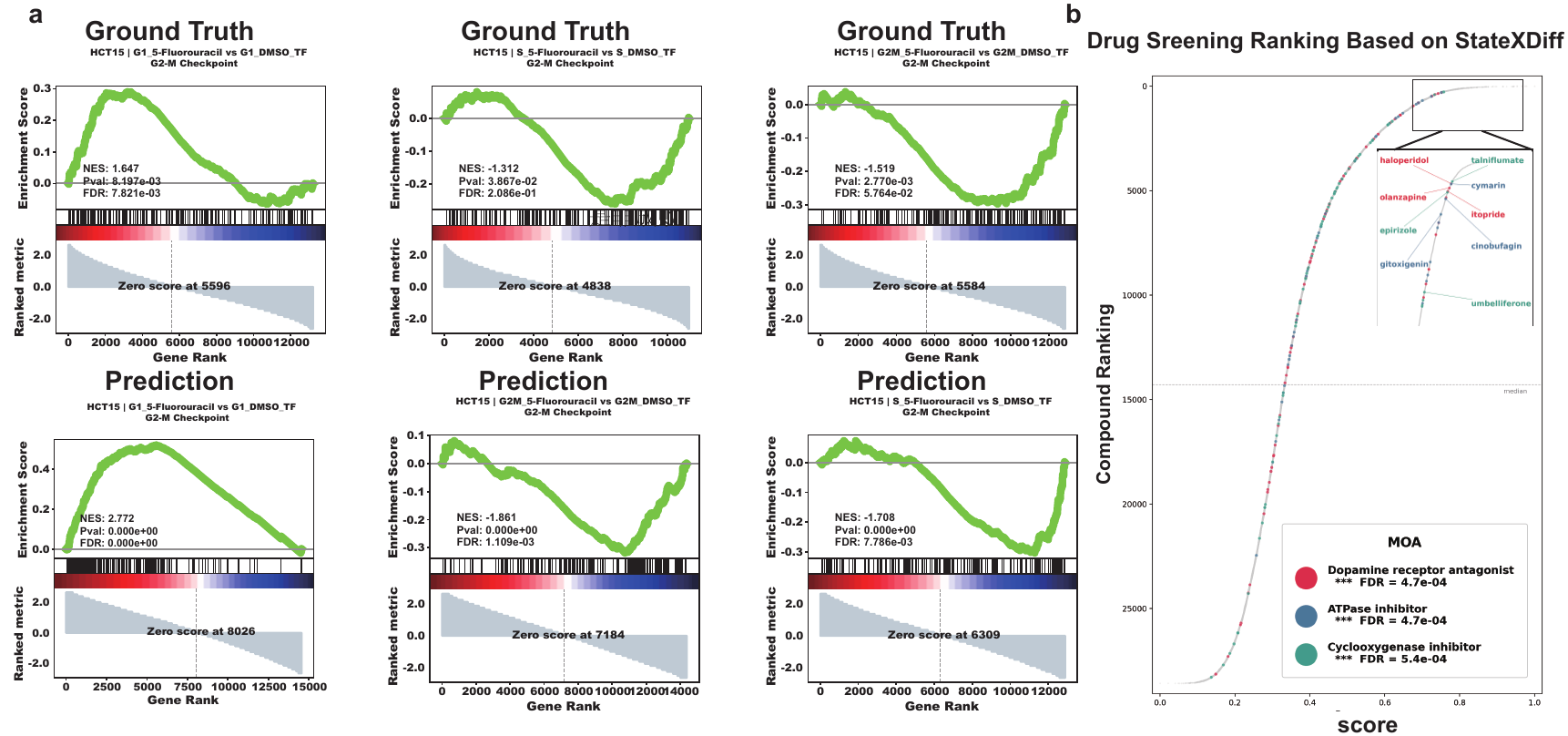}
\caption{Case study on HCT15. a. StateXDiff prediction vs. ground truth for 5-FU response. b. StateXDiff-based drug screening.}
\label{bio_case}
\end{figure}

We applied StateXDiff to the colorectal cancer cell line HCT15, which was not seen during training, as a case study. The model accurately predicted 5-fluorouracil (5-FU)-induced transcriptomic changes and captured heterogeneous responses across cell-cycle subpopulations, showing high concordance with ground-truth expression profiles (Figure~\ref{bio_case}a). Pathway enrichment analysis revealed that apoptosis was activated in the S-phase subpopulation but suppressed in the G1 and G2/M phases, with the strongest suppression observed in G1.

Motivated by this heterogeneity, we further conducted in silico screening of 28,574 compounds in the 5-FU-resistant G1 subpopulation. For each compound, StateXDiff generated post-treatment transcriptomes, which were subsequently analyzed using pathway enrichment to quantify their ability to induce apoptosis relative to the DMSO control. Using rank-based enrichment analysis, we identified 27 significantly enriched mechanisms of action (MOAs) among the top-ranked pro-apoptotic candidates (BH-adjusted FDR < 0.05). Notably, dopamine receptor antagonists, ATPase inhibitors, and cyclooxygenase inhibitors emerged as the most significantly enriched categories (Figure~\ref{bio_case}b). Among them, the top-ranked ATPase inhibitors (e.g., cymarin, cinobufagin, and gitoxigenin) are cardiac glycosides, which have been widely reported to inhibit colorectal cancer cell proliferation \cite{anderson2017cardiac,shah2022chemistry}. These results highlight the potential of our framework to identify mechanism-driven combination therapies targeting resistant subpopulations.

\section{Conclusion}
\label{Conclusion}
We present StateXDiff, a two-stage diffusion framework for single-cell perturbation prediction that integrates mechanism-aware drug representations with reliability-gated multimodal cell-state conditioning. StateXDiff combines RNA-derived pseudo-protein embeddings with cell context and drug mechanism priors, enabling a latent diffusion Transformer to model perturbation responses. On the Tahoe-100M benchmark, StateXDiff consistently improves generalization to unseen drugs, unseen cell lines, and combinatorial perturbations. Further scaling analysis suggests that cell-state heterogeneity, rather than chemical diversity alone, is a key bottleneck for generalization. A current limitation is the use of inferred pseudo-protein embeddings, although the modular VMCS design allows them to be replaced by measured proteomics when paired multi-omic data become available.

\clearpage
\renewcommand{\refname}{References}
\bibliographystyle{unsrt}
\bibliography{references}

\newpage
\appendix
\section*{Appendix}
\startcontents[appendix]
\printcontents[appendix]{}{0}[3]{}
\section{Method Details}
\subsection{Pretrained Backbone Details}
\label{app:backbones}

\subsubsection{Transcriptomic Latent Representation with SCimilarity}
\label{app:scimilarity}

To obtain compact transcriptomic representations, we adopt SCimilarity~\cite{heimberg2025cell}, a pretrained foundation model based on an encoder--decoder architecture.
SCimilarity was pretrained on a large-scale corpus comprising 22.7 million cells from 399 published scRNA-seq studies, enabling the model to learn a unified 128-dimensional latent space that captures shared transcriptional structure across tissues and experimental platforms.

Given log-normalized gene expression vectors $\mathbf{x}_c,\mathbf{x}_p\in\mathbb{R}^{G}$ for control and perturbed cells, respectively, where $G$ denotes genes aligned to the SCimilarity reference ordering, the encoder $\mathcal{E}$ produces latent embeddings
$\mathbf{z}_c=\mathcal{E}(\mathbf{x}_c)$ and $\mathbf{z}_p=\mathcal{E}(\mathbf{x}_p)\in\mathbb{R}^{128}$.
The decoder $\mathcal{D}$ reconstructs the corresponding full expression profiles from the latent representations.

We further fine-tune the pretrained SCimilarity weights on the target dataset for 200{,}000 optimization steps with a learning rate of $1\times10^{-5}$.
After fine-tuning, both the encoder and decoder are frozen for all downstream experiments.

\subsubsection{Pseudo-Proteomic Representation with scLinguist}
\label{app:sclinguist}

We employ scLinguist~\cite{fang2025sclinguist}, a pretrained RNA-to-protein foundation model, to infer pseudo-protein representations from scRNA-seq data in a zero-shot setting.
We use scLinguist to extract two outputs:
\begin{itemize}[nosep,leftmargin=1.5em]
  \item \textbf{Protein predictions} $\hat{\mathbf{p}}\in\mathbb{R}^{6427}$: decoder outputs used solely to compute the reliability score $q_i$ (Eq.~\ref{eq:quality}).
  
  \item \textbf{Protein embedding} $\mathbf{g}_p\in\mathbb{R}^{128}$: a compact cell-level proteomic representation obtained by mean-pooling the encoder's per-gene hidden states across the gene dimension.
  This embedding is used as input to the VMCS modality projection (Eq.~\ref{eq:rotate}).
\end{itemize}

\subsection{VMCS Implementation Details}
\label{app:vmcs_detail}
\subsubsection{Protein-quality aware module (PQM)}
\label{app:quality_detail}

Pseudo-protein quality from scLinguist varies across cells due to transcriptome--proteome discordance.
PQM estimates per-cell reliability $q_i\in(0,1)$ from three non-parametric, complementary sub-scores using only RNA-derived signals.
No learnable parameters are involved; gradients are stopped before $q_i$.

\paragraph{Reference gene set.}
scLinguist predicts 6{,}427 proteins.
For a given dataset, we identify the subset with matching gene symbols in the RNA input ($\sim\!2{,}800$ for Tahoe-100M).
From these, we retain genes whose population-level Spearman correlation between RNA expression and predicted protein exceeds 0.3, yielding $\mathcal{A}$ ($|\mathcal{A}|\!\approx\!80$; Table~\ref{tab:ref_genes}).
Restricting $s_{\mathrm{rc}}$ to $\mathcal{A}$ avoids noisy RNA--protein pairs; $s_{\mathrm{cx}}$ and $s_{\mathrm{nb}}$ are not limited to $\mathcal{A}$.

\paragraph{Sub-scores.}

$\bullet$ \textbf{$s_{\mathrm{rc}}$ (RNA--protein rank consistency):} Spearman correlation between $\{x_g^{(i)}\}_{g\in\mathcal{A}}$ and $\{\hat{p}_g^{(i)}\}_{g\in\mathcal{A}}$:
\begin{equation}
  s_{\mathrm{rc}}^{(i)} = \mathrm{SpearmanR}\bigl(\{x_g^{(i)}\}_{g\in\mathcal{A}},\;\{\hat{p}_g^{(i)}\}_{g\in\mathcal{A}}\bigr).
  \label{eq:src}
\end{equation}

$\bullet$ \textbf{$s_{\mathrm{cx}}$ (Protein-complex coherence):} For $M\!=\!62$ CORUM~\cite{giurgiu2019corum} complexes with $\geq\!3$ subunits in the output panel:
\begin{equation}
  s_{\mathrm{cx}}^{(i)} = 1 - \frac{1}{M}\sum_{m=1}^{M}\min\!\Bigl(\frac{\mathrm{std}(\{\hat{p}_j^{(i)}\}_{j\in\mathcal{P}_m})}{\mathrm{mean}(\{\hat{p}_j^{(i)}\}_{j\in\mathcal{P}_m})+\epsilon},\;1\Bigr).
  \label{eq:scx}
\end{equation}

$\bullet$ \textbf{$s_{\mathrm{nb}}$ (Embedding neighborhood consistency):} Cosine similarity between cell $i$'s protein embedding and the centroid of its $k\!=\!50$ SCimilarity nearest neighbors:
\begin{equation}
  s_{\mathrm{nb}}^{(i)} = \cos\!\bigl(\mathbf{g}_p^{(i)},\;\tfrac{1}{k}\textstyle\sum_{j\in\mathcal{N}_k(i)}\mathbf{g}_p^{(j)}\bigr).
  \label{eq:snb}
\end{equation}

\paragraph{Aggregation.}
Each sub-score is z-normalized: $\tilde{s} = (s - \mu)/(\sigma+\epsilon)$, then averaged and passed through a sigmoid:
\begin{equation}
  q_i = \sigma\!\bigl(\tfrac{1}{3}(\tilde{s}_{\mathrm{rc}}^{(i)}+\tilde{s}_{\mathrm{cx}}^{(i)}+\tilde{s}_{\mathrm{nb}}^{(i)})\bigr).
  \label{eq:quality_app}
\end{equation}
$q_i$ weights the counterfactual triplet loss (Eq.~\ref{eq:triplet}) and modulates the masking probability during VMCS training.

\begin{table}[H]
\centering
\caption{Reference genes for $s_{\mathrm{rc}}$, grouped by functional category.}
\label{tab:ref_genes}
\small
\setlength{\tabcolsep}{8pt}
\begin{tabular}{>{\bfseries}l>{\raggedright\arraybackslash}p{9.5cm}}
\toprule
\textbf{Category} & \textbf{Genes} \\
\midrule
\rowcolor{gray!10} Ribosomal (large) & RPL3, RPL4, RPL5, RPL6, RPL7, RPL8, RPL9, RPL10A, RPL11, RPL12, RPL13, RPL14, RPL15, RPL18, RPL23A, RPL24, RPL27A \\
Ribosomal (small) & RPS2, RPS3, RPS4X, RPS5, RPS6, RPS7, RPS8, RPS9, RPS10, RPS11, RPS12, RPS13, RPS14, RPS15, RPS16, RPS17, RPS18, RPS19, RPS20, RPSA \\
\rowcolor{gray!10} Glycolysis & GAPDH, LDHA, LDHB, PKM, ENO1, ALDOA, PGK1, TPI1 \\
Chaperones & HSP90AA1, HSP90AB1, HSPA8, HSPD1, CALR, CANX \\
\rowcolor{gray!10} Translation initiation & EIF3A, EIF3B, EIF3C, EIF3D, EIF3E, EIF3F, EIF3G, EIF3H, EIF3I, EIF4A1, EIF4G1 \\
Cytoskeletal & ACTB, ACTG1, VIM, TUBA1B, TUBB, B2M \\
\bottomrule
\end{tabular}
\end{table}

\subsubsection{MMD Kernel and Protein Dropout Details}
\label{app:mmd}

The alignment loss (Eq.~\ref{eq:mmd}) uses an agreement-weighted mixture-of-Gaussians kernel:
\begin{equation}
  k_{\boldsymbol{\rho}}(\mathbf{a},\mathbf{b}) = \sum_{l=1}^{L} \gamma_l \,
    \exp\!\Bigl(-\frac{\|\mathbf{a}-\mathbf{b}\|_2^2}{2\sigma_l^2}\Bigr),
  \qquad
  \gamma_l = \frac{\hat\rho^{\top}}{\sum_{m=1}^{L}\hat\rho^{\top}} = \frac{1}{L},
  \label{eq:mmd_kernel}
\end{equation}
with $L=5$ bandwidths $\sigma_l\in\{0.1, 0.5, 1, 2, 5\}$ and $\hat\rho^{\top}=\frac{1}{D}\sum_{j=1}^{D}\hat\rho_j$ as the mean cross-modal agreement over all $D=512$ dimensions.
The agreement weighting focuses MMD on dimensions where RNA and protein genuinely share information, avoiding interference from modality-specific dimensions.

\paragraph{Protein dropout during diffusion training.}
During diffusion training, the protein branch of $\mathbf{c}_{\mathrm{cell}}$ is independently dropped with probability $1-q_i$ per cell.
When dropped, $\mathbf{T}_p$ tokens are replaced by a learned null embedding $\mathbf{e}_{\mathrm{null}}^{p}\in\mathbb{R}^{H}$, and the weighted residual $q_i\mathbf{r}_p$ contributing to the per-cell embedding is zeroed.
This prevents low-quality pseudo-protein signals from corrupting the diffusion condition.

\subsubsection{Empirical Validation of $q_i$}
\label{app:qi_validation}

We validate $q_i$ on the CITE-seq BMMC dataset (10{,}546 cells, 228 surface protein (ADT, antibody-derived tag), 8 cell types), which was \textbf{not} used in scLinguist training.

\paragraph{Protocol.}
(1)~Compute $q_i$ with true ADT vs.\ predicted protein---high correlation indicates $q_i$ captures genuine protein signal rather than translator self-consistency.
(2)~Negative control: compare matched, cross-type, and shuffled RNA--protein pairings.

\begin{figure*}[h]
  \centering
  \includegraphics[width=\textwidth]{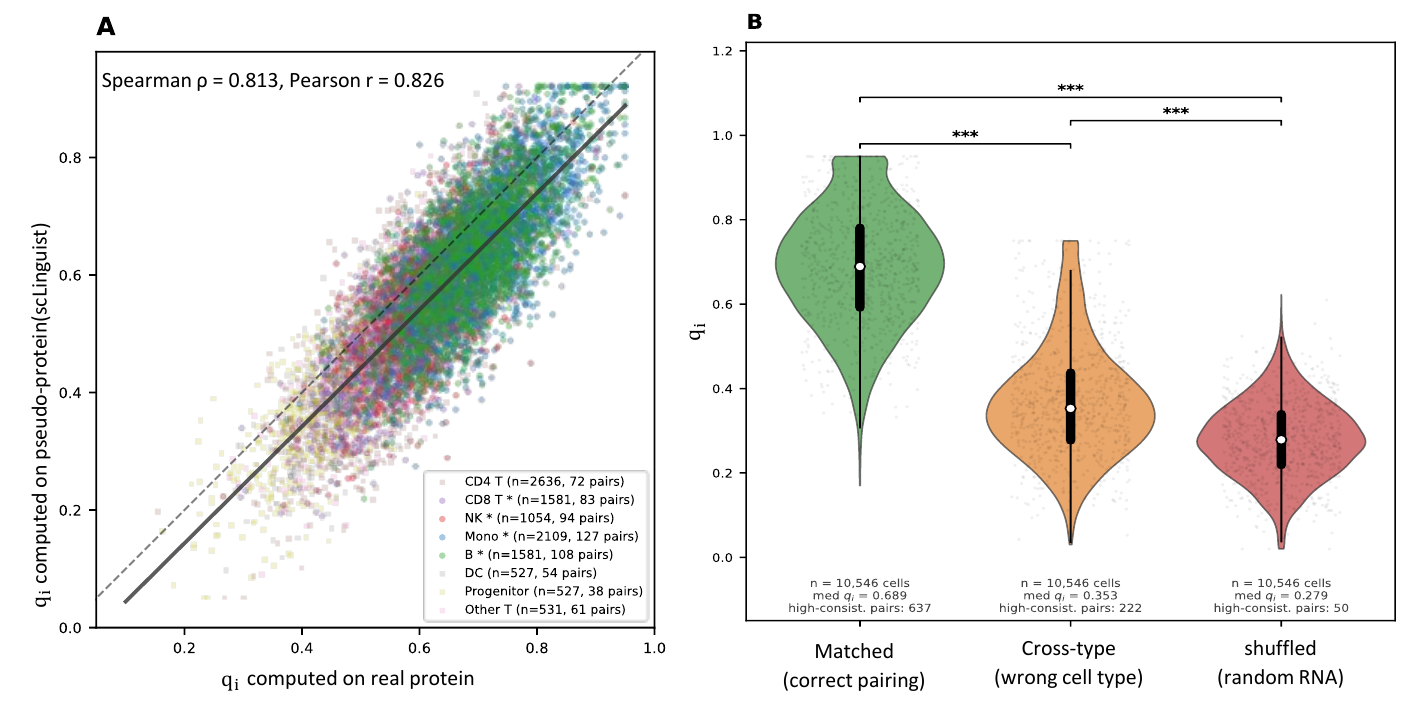}
  \caption[Validation of $q_i$.]{\textbf{(a)} True-ADT vs.\ predicted-protein $q_i$ correlation. \textbf{(b)} Matched vs.\ cross-type vs.\ shuffled controls.}
  \label{fig:qi_validation}
\end{figure*}

\paragraph{Results (Fig.~\ref{fig:qi_validation}).}
True-ADT $q_i$ vs.\ predicted-protein $q_i$: Spearman $\rho\!=\!0.81$, Pearson $r\!=\!0.83$.
Median $q_i$: 0.69 (matched), 0.35 (cross-type), 0.28 (shuffled); all pairwise differences $p\!<\!10^{-300}$ (Mann--Whitney $U$).
High RNA--protein coupling cell types (Mono, B, NK, CD8~T) achieve higher $q_i$, while progenitor cells occupy the low-$q_i$ region.

\subsection{Mechanism-Aware Drug--Gene Template Details}
\label{app:mdt_method}

This section provides the full implementation details for MDT, whose high-level design is described in Sec.~\ref{sec:mdt}.
All pretraining hyperparameters are consolidated in Table~\ref{tab:mdt_params}.

\subsubsection{Biological Prior Encoding via Heterogeneous Graph}
\label{app:graph}

To inject target-level and pathway-level semantics that are not directly accessible from chemical structure, we pretrain a graph neural network on a heterogeneous biological knowledge graph.
After pretraining, the graph encoder parameters are frozen, and the resulting drug embeddings $\mathbf{z}^b_d$ serve as static biological priors for downstream distillation and soft-label construction.

\paragraph{Graph construction.}
We construct a heterogeneous graph $\mathcal{G} = (\mathcal{V}, \mathcal{E})$ comprising drug nodes $\mathcal{V}_d$ and gene/protein nodes $\mathcal{V}_g$.
Drug nodes are initialized with UniMol 3D molecular embeddings, and gene nodes are initialized with ESM-2 protein language model representations:
\begin{equation}
\mathbf{h}_d^{(0)} = W_s \mathbf{x}^s_d, \quad
\mathbf{h}_g^{(0)} = W_g \mathbf{x}^{\text{esm}}_g,
\end{equation}
where $W_s$ and $W_g$ are learnable linear projections that map both node types into a shared 256-dimensional space.
The graph contains four relation types: (i)~drug--target interactions (DTI) from curated databases, (ii)~reverse DTI edges to enable bidirectional message flow, (iii)~protein--protein interactions (PPI) filtered at a STRING evidence score $\ge 700$ to retain high-confidence functional associations, and (iv)~drug--drug similarity (DDS) edges connecting compounds whose UniMol cosine similarity exceeds 0.995.
The inclusion of PPI edges allows target-level information to propagate along biological pathways, while DDS edges encourage structurally near-identical compounds to share representations.

\paragraph{Graph encoder.}
We employ a two-layer HeteroSAGE architecture with relation-specific message passing.
At each layer, node representations are updated by aggregating neighbor features through relation-specific transformations:
\begin{equation}
\mathbf{h}_v^{(l+1)} = \sigma\Bigl(\sum_{r \in \mathcal{R}} W_r \cdot \mathrm{AGG}_r(\mathcal{N}_r(v))\Bigr),
\end{equation}
where $\mathcal{R}$ denotes the set of relation types, $\mathcal{N}_r(v)$ is the set of neighbors of node $v$ under relation $r$, and $\mathrm{AGG}_r$ uses scatter-mean aggregation.
The hidden dimension is 256 throughout, and the final drug node representations are L2-normalized to yield $\mathbf{z}^b_d \in \mathbb{R}^{256}$.

\paragraph{Training objective.}
The graph encoder is jointly optimized with three complementary objectives that encourage biologically coherent representations:
\begin{equation}
\mathcal{L}_{\text{graph}} = \mathcal{L}_{\text{DTI}} + w_{\text{mae}}\mathcal{L}_{\text{MAE}} + w_{\text{er}}\mathcal{L}_{\text{ER}}.
\end{equation}

$\bullet$ \textbf{$\mathcal{L}_{\text{DTI}}$ (Drug--target interaction prediction):} This objective enforces that drug representations are predictive of their known protein targets, formulated as an InfoNCE-style contrastive loss over drug--gene pairs:
\begin{equation}
\mathcal{L}_{\text{DTI}} = -\log\frac{\exp(\mathbf{z}_d^\top \mathbf{z}_g)}{\exp(\mathbf{z}_d^\top \mathbf{z}_g) + \sum_{g'}\exp(\mathbf{z}_d^\top \mathbf{z}_{g'})}.
\end{equation}

$\bullet$ \textbf{$\mathcal{L}_{\text{MAE}}$ (Masked feature reconstruction):} To improve robustness to incomplete annotations and encourage the encoder to learn generalizable node features, we randomly mask 30\% of node input features and reconstruct them from the graph context using mean squared error:
\begin{equation}
\mathcal{L}_{\text{MAE}} = \frac{1}{|\mathcal{M}|}\sum_{v\in\mathcal{M}}\|\mathbf{x}_v - \hat{\mathbf{x}}_v\|^2.
\end{equation}

$\bullet$ \textbf{$\mathcal{L}_{\text{ER}}$ (Edge reconstruction):} This auxiliary objective preserves the topological structure of the graph by predicting the existence of edges from learned node representations:
\begin{equation}
\mathcal{L}_{\text{ER}} = -\sum_{(u,v)} y_{uv}\log\hat{y}_{uv}.
\end{equation}

The loss weights are set to $w_{\text{DTI}}{=}1.0$, $w_{\text{mae}}{=}1.0$, $w_{\text{er}}{=}0.5$.
The graph encoder is trained for 200 epochs with Adam (lr $= 10^{-3}$).
After pretraining, all graph encoder parameters are frozen, and the resulting embeddings $\mathbf{z}^b_d$ are used as fixed biological priors in subsequent stages.

\subsubsection{Multi-view Construction}
\label{app:multiview}

Each drug in the L1000 database is profiled under multiple experimental conditions (cell lines, dosages, time points), yielding a set of transcriptomic response signatures with varying quality and informativeness.
Rather than averaging these into a single representation, we model them as multiple views to capture the inherent variability in perturbation effects across biological contexts.

\paragraph{Transcriptional Activity Score (TAS).}
Not all L1000 profiles are equally informative: some reflect strong transcriptional perturbations while others are dominated by technical noise or weak responses.
We use the Transcriptional Activity Score (TAS) to quantify the overall perturbation strength of each (drug, condition) pair, defined as the average absolute z-score across the 978 landmark genes relative to the plate-matched control.
For each drug, we compute TAS for all available L1000 profiles and retain up to $K{=}3$ views with the highest scores, ensuring that only the most informative response signatures are used for representation learning.

\paragraph{Batch construction.}
Given a batch of $B$ drugs, the total number of views in the contrastive batch is $N = B + \sum_d K_d$, comprising one structural view (from UniMol) and up to $K$ expression views per drug.
Views originating from the same drug form positive pairs, while views from different drugs form negative pairs whose repulsion strength is modulated by the soft similarity matrix $M$ (described below).

\subsubsection{Soft-label Construction}
\label{app:softlabel}

Standard contrastive learning assigns binary labels (positive or negative) to all pairs, which introduces two sources of error in the drug representation setting: it treats all views of the same drug as equally reliable regardless of their quality, and it penalizes mechanistically related drugs as hard negatives.
We address both issues by constructing a continuous similarity matrix $M \in [0,1]^{N \times N}$ that modulates the contrastive objective.

\paragraph{Same-drug similarity.}
For two expression views $i$ and $j$ belonging to the same drug, we measure their consistency using the Pearson correlation $r_{ij}$ between their 978-dimensional response vectors, weighted by their respective TAS values to emphasize high-quality observations:
\begin{equation}
M_{ij} = \max(r_{ij}, \epsilon) \cdot \sqrt{\mathrm{TAS}_i \cdot \mathrm{TAS}_j},
\end{equation}
where $\epsilon{=}0.05$ serves as a floor to prevent zero weights for weakly correlated but still informative views.
This formulation down-weights noisy or inconsistent profiles while preserving their contribution to the learning signal.

\paragraph{Cross-drug similarity.}
For views belonging to different drugs $d_i$ and $d_j$, we incorporate biological similarity derived from the frozen graph embeddings to prevent mechanistically related compounds from being treated as hard negatives:
\begin{equation}
M_{ij} = \max(0, \cos(\mathbf{z}^b_{d_i}, \mathbf{z}^b_{d_j}) - \delta),
\end{equation}
where $\delta{=}0.86$ is a threshold that activates soft positive labels only for highly similar drug pairs in the biological embedding space.
This prevents the contrastive loss from pushing apart drugs that share targets or pathways.

\paragraph{Normalization.}
Each row of $M$ is L1-normalized ($M_{ij} \leftarrow M_{ij} / \sum_j M_{ij}$) so that the soft labels form a proper distribution over positive pairs for each anchor.

\subsubsection{Training and Model Selection}

We train MDT using AdamW ($\beta_1{=}0.9$, $\beta_2{=}0.999$) with cosine learning rate scheduling (8-epoch linear warmup to peak LR $5.48{\times}10^{-4}$, weight decay $10^{-4}$).
The batch size is 512 drugs, training runs for up to 240 epochs with early stopping (patience 50), and gradient clipping is set to 5.0 with bfloat16 mixed precision.

Model selection uses a retrieval-based validation score computed every 5 epochs:
\begin{equation}
\text{score} = \frac{\text{Top1\%}_{e2s} + 2 \cdot \text{Top1\%}_{s2e}}{3},
\end{equation}
where $\text{Top1\%}_{e2s}$ measures top-1\% retrieval accuracy from expression queries to structure targets, and $\text{Top1\%}_{s2e}$ measures the reverse direction.
The asymmetric weighting ($2\times$ for structure-to-expression) reflects that the structure encoder is the component retained at inference, so its retrieval quality is prioritized during selection.

\subsection{Perturbation-Aware Conditional Diffusion}
\label{app:diffusion_detail}

The main text (Sec.~\ref{sec:generation}) describes the conditional diffusion framework at a high level.
Below we provide the explicit mathematical formulation and architectural details omitted from the main body for brevity.

\paragraph{Forward diffusion and velocity parameterization.}
Given a perturbation effect $\Delta\mathbf{z}=\mathbf{z}_p-\mathbf{z}_c$ (Sec.~\ref{sec:problem}), the forward process corrupts it with Gaussian noise over $T=1000$ steps following a cosine schedule:
\begin{equation}
  \Delta\mathbf{z}_t = \sqrt{\bar\alpha_t}\,\Delta\mathbf{z} + \sqrt{1-\bar\alpha_t}\,\boldsymbol{\epsilon},
  \qquad \boldsymbol{\epsilon}\sim\mathcal{N}(\mathbf{0},\mathbf{I}),
  \label{eq:forward}
\end{equation}
where $\bar\alpha_t=\cos(\frac{t/T+s}{1+s}\cdot\frac{\pi}{2})^2/\cos(\frac{s}{1+s}\cdot\frac{\pi}{2})^2$ with $s=0.008$.
The denoiser $f_\theta$ predicts the velocity~\cite{salimans2022progressive}
$\mathbf{v}_t=\sqrt{\bar\alpha_t}\,\boldsymbol{\epsilon}-\sqrt{1-\bar\alpha_t}\,\Delta\mathbf{z}$,
from which $\Delta\mathbf{z}$ is recovered via $\Delta\hat{\mathbf{z}} = \sqrt{\bar\alpha_t}\,\Delta\mathbf{z}_t - \sqrt{1-\bar\alpha_t}\,\hat{\mathbf{v}}_t$.
The training objective is $\mathcal{L}_{\mathrm{gen}} = \mathbb{E}_{t,\boldsymbol{\epsilon}}[\|\hat{\mathbf{v}}_t-\mathbf{v}_t\|_2^2] + \lambda_{\mathrm{triplet}}\mathcal{L}_{\mathrm{triplet}}$ (Eq.~\ref{eq:triplet}), with $\lambda_{\mathrm{triplet}}=0.1$, margin $m=0.1$.

\paragraph{Triplet loss design.}
A perturbation's direction depends not only on the drug but also on whether the cell's protein state provides the necessary targets.
We construct mismatched triplets to enforce this dependency: for each training pair $(i,d)$, we form a mismatched condition by replacing either the drug $d$ with a random $d'\neq d$ from the batch, or the cell context $\mathbf{c}_{\mathrm{cell}}^{(i)}$ with one from a different cell line, each with probability 0.5.
The denoiser is run under this mismatched condition to produce $\hat{\mathbf{v}}_t^{\mathrm{mis}}$.
The cosine-margin loss (Eq.~\ref{eq:triplet}) then penalizes cases where $\hat{\mathbf{v}}_t^{\mathrm{mis}}$ is closer to the ground-truth $\mathbf{v}_t$ than $\hat{\mathbf{v}}_t$ is (by a margin $m=0.1$).
Using cosine rather than Euclidean distance makes the loss invariant to velocity magnitude, focusing purely on directional accuracy.
The per-cell weight $q_i$ down-weights cells with unreliable pseudo-protein signals: for a low-$q_i$ cell, the protein-derived condition is noisy, so the counterfactual signal is less informative and the triplet loss contributes less to training.

\paragraph{Inference.}
At inference, we sample from $\Delta\mathbf{z}_T\sim\mathcal{N}(\mathbf{0},\mathbf{I})$ and run DDIM~\cite{song2020denoising} for $50$ steps with $\eta=0$ (deterministic).
Each step applies interaction-aware CFG (Eq.~\ref{eq:icfg}), which requires four forward passes: one unconditional ($\hat{\mathbf{v}}_{\emptyset}$), one with cell-only ($\hat{\mathbf{v}}_{\mathbf{c}_{\mathrm{cell}}}$), one with drug-only ($\hat{\mathbf{v}}_{\mathbf{c}_{\mathrm{drug}}}$), and one fully conditioned ($\hat{\mathbf{v}}_{\mathbf{c}_{\mathrm{cell}},\mathbf{c}_{\mathrm{drug}}}$).
These are composed with guidance weights $w_c=1.0$, $w_d=1.5$, $w_{cd}=2.0$.
The interaction term (coefficient $w_{cd}$) amplifies non-additive drug--cell synergy, while $w_d>w_c$ reflects that drug identity is the dominant factor in perturbation response.
On a single A100, each step takes $\sim\!8$\,ms, yielding $\sim\!0.4$\,s per cell.

\paragraph{Denoiser architecture.}
The denoiser is a 12-layer Transformer with hidden dimension $H=768$ and 12 attention heads (head dim 64).
We detail the key components described verbally in Sec.~\ref{sec:generation}.

\textbf{Token sequence.}
The noise token $\mathbf{h}_0=\mathbf{W}_{\mathrm{in}}\Delta\mathbf{z}_t\in\mathbb{R}^{H}$ is concatenated with the condition tokens after bidirectional cross-attention (Eq.~\ref{eq:preinteract}), forming
$\mathbf{S}=[\mathbf{h}_0;\,\mathbf{T}_s';\,\mathbf{T}_p';\,\mathbf{T}_d']\in\mathbb{R}^{71\times H}$
(1 noise, 25 shared-cell, 10 protein, 35 drug tokens).
Learnable token-type embeddings $\mathbf{e}_n,\mathbf{e}_s,\mathbf{e}_p,\mathbf{e}_d\in\mathbb{R}^{H}$ are added.

\textbf{Global condition.}
$\mathbf{t}_{\mathrm{emb}}$ is a 256-dim sinusoidal PE followed by a 2-layer MLP (SiLU).
$\bar{\mathbf{e}}_{\mathrm{cell}},\bar{\mathbf{e}}_{\mathrm{drug}}\in\mathbb{R}^{H}$ are mean-pooled from $\mathbf{T}_c',\mathbf{T}_d'$.
The global AdaLN condition is $\mathbf{c}_{\mathrm{global}}=\mathbf{t}_{\mathrm{emb}}+\bar{\mathbf{e}}_{\mathrm{drug}}+\bar{\mathbf{e}}_{\mathrm{cell}}\in\mathbb{R}^{H}$.

\textbf{AdaLN-Zero.}
For each block $\ell$, a linear projection of $\mathbf{c}_{\mathrm{global}}$ produces modulation parameters:
\begin{equation}
  \mathrm{AdaLN}(\mathbf{x},\mathbf{c}_{\mathrm{global}})
  = \gamma_{\ell}\,\bigl(\boldsymbol{\alpha}_{\ell}\odot\mathrm{LayerNorm}(\mathbf{x})+\boldsymbol{\delta}_{\ell}\bigr),
  \qquad
  (\boldsymbol{\alpha}_{\ell},\boldsymbol{\delta}_{\ell},\gamma_{\ell}) = \mathrm{Linear}_\ell(\mathbf{c}_{\mathrm{global}}),
  \label{eq:adaln}
\end{equation}
where $\boldsymbol{\alpha}_{\ell}\!\in\!\mathbb{R}^{H}$ (scale), $\boldsymbol{\delta}_{\ell}\!\in\!\mathbb{R}^{H}$ (shift), and $\gamma_{\ell}\!\in\!\mathbb{R}$ (gate, initialized to zero).
Applied before both self-attention and FFN; zero-initialized gates ensure identity mapping at the start of training.

\textbf{Four-type joint self-attention.}
Let $\mathbf{S}^{(m)}$ be the subsequence of $\mathbf{S}$ for type $m\!\in\!\{n,s,p,d\}$ (noise, shared-cell, protein, drug).
Dedicated QKV projections per type:
\begin{equation}
  \mathbf{Q}^{(m)} = \mathbf{S}^{(m)}\mathbf{W}_Q^{(m)},\quad
  \mathbf{K}^{(m)} = \mathbf{S}^{(m)}\mathbf{W}_K^{(m)},\quad
  \mathbf{V}^{(m)} = \mathbf{S}^{(m)}\mathbf{W}_V^{(m)},
  \label{eq:type_qkv}
\end{equation}
where $\mathbf{W}_Q^{(m)},\mathbf{W}_K^{(m)},\mathbf{W}_V^{(m)}\!\in\!\mathbb{R}^{H\times d_k}$ ($d_k\!=\!64$ per head).
All types are concatenated (e.g., $\mathbf{Q}=[\mathbf{Q}^{(n)};\mathbf{Q}^{(s)};\mathbf{Q}^{(p)};\mathbf{Q}^{(d)}]$) and standard multi-head attention is applied jointly over the full 71-token sequence.

\textbf{SwiGLU FFN and gated residual.}
The FFN uses a gated linear unit $\mathrm{FFN}(\mathbf{x}) = \mathbf{W}_2\bigl(\mathrm{SiLU}(\mathbf{W}_1\mathbf{x})\odot\mathbf{W}_3\mathbf{x}\bigr)$ with $\mathbf{W}_1,\mathbf{W}_3\!\in\!\mathbb{R}^{H\times 4H}$, $\mathbf{W}_2\!\in\!\mathbb{R}^{4H\times H}$.
The block output is $\mathbf{x}_{\ell+1} = \mathbf{x}_\ell + \alpha_{\ell}\!\cdot\!\mathrm{FFN}\bigl(\mathrm{Attn}(\mathrm{AdaLN}(\mathbf{x}_\ell))\bigr)$, where the per-block gate $\alpha_{\ell}$ is learned and initialized near zero.
\textbf{Output head.} After 12 blocks, only the noise token $\mathbf{h}_0$ is read out: $\hat{\mathbf{v}}_t = \mathbf{W}_{\mathrm{out}}\,\mathrm{AdaLN}(\mathbf{h}_0,\mathbf{c}_{\mathrm{global}})\in\mathbb{R}^{128}$.

\paragraph{Structured condition dropout.}
During training, conditions are dropped with probabilities $p_{\emptyset}\!=\!0.05$ (unconditional), $p_c\!=\!0.05$ (cell-only), $p_d\!=\!0.05$ (drug-only), and $0.85$ (joint).
Within the cell branch, protein tokens $\mathbf{T}_p$ are independently dropped with probability $1\!-\!q_i$, replaced by a learned null embedding $\mathbf{e}_{\mathrm{null}}^{p}\!\in\!\mathbb{R}^{H}$, and $q_i\mathbf{r}_p$ is zeroed in $\bar{\mathbf{e}}_{\mathrm{cell}}$.
When a full condition branch is dropped, its tokens are replaced by null embeddings and its contribution to $\mathbf{c}_{\mathrm{global}}$ is zeroed.
This yields four conditioning modes ($\emptyset$, cell-only, drug-only, joint) required for interaction-aware CFG (Eq.~\ref{eq:icfg}) at inference.

\subsection{Training and Optimization Details}
\label{app:training_detail}

Training proceeds in two stages, consistent with the main text: (1)~representation learning, comprising VMCS training on pre-perturbation cells and MDT drug representation pretraining on L1000 data, and (2)~diffusion model training on perturbation pairs.
All pretrained encoders ($\mathcal{E}$, scLinguist) remain frozen throughout.
VMCS modules are frozen after Stage~1 (or optionally fine-tuned at lr $10^{-6}$ during Stage~2).
The MDT structure encoder is frozen during Stage~2; drug condition assembly layers are fine-tuned.

\begin{table}[t]
\centering
\caption{VMCS training hyperparameters.}
\label{tab:vmcs_params}
\small
\renewcommand{\arraystretch}{1.05}
\begin{tabular}{@{}llr@{\hskip 8pt}l@{}}
\toprule
\textbf{Category} & \textbf{Parameter} & \textbf{Value} & \textbf{Description} \\
\midrule
Data
  & Batch size     & 512  & Pre-perturbation cells \\
  & Epochs         & 100  & -- \\
\midrule
Architecture
  & Projection dim $D$       & 512  & Common embedding space \\
  & Shared tokens $K_s$      & 25   & Cell-state condition tokens \\
  & Protein tokens $K_p$     & 10   & Protein-residual tokens \\
  & Condition mixer layers   & 2    & Transformer encoder \\
\midrule
VMCS
  & EMA momentum $\gamma$                  & 0.1  & For $\rho_j$ statistics \\
  & Initial threshold $\tau_s$             & 0.5  & Learnable sigmoid threshold \\
  & Initial temperature $\beta$            & 0.1  & Learnable sigmoid temperature \\
  & MMD bandwidths                         & \multicolumn{2}{l@{}}{\{0.1, 0.5, 1, 2, 5\}} \\
\midrule
Masking
  & $p_{\min}$ / $p_{\max}$               & 0.05 / 0.30  & Mask probability range \\
\midrule
Loss
  & $\lambda_{\mathrm{align}}$ & 1.0  & MMD alignment weight \\
\midrule
Optimization
  & Optimizer              & AdamW & $\beta_1{=}0.9$, $\beta_2{=}0.999$ \\
  & Peak learning rate     & $10^{-4}$ & -- \\
  & Weight decay           & 0.01  & -- \\
  & LR schedule            & Cosine & 5\% linear warmup \\
  & Gradient clipping      & 1.0   & Max gradient norm \\
  & Precision              & bfloat16 & Mixed precision \\
\bottomrule
\end{tabular}
\end{table}

\subsubsection{MDT Pretraining}

MDT (Multi-modal Drug representation with Template) pretraining constructs a drug embedding space from L1000 transcriptional profiles and molecular structures.
Hyperparameters are listed in Table~\ref{tab:mdt_params}.

\begin{table}[t]
\centering
\caption{MDT pretraining hyperparameters.}
\label{tab:mdt_params}
\small
\renewcommand{\arraystretch}{1.05}
\begin{tabular}{@{}llr@{\hskip 8pt}l@{}}
\toprule
\textbf{Category} & \textbf{Parameter} & \textbf{Value} & \textbf{Description} \\
\midrule
Data
  & Training observations & 28{,}177 & After QC filtering \\
  & Number of drugs      & 5{,}684  & Unique compounds \\
  & Gene features        & 978      & L1000 landmark genes \\
  & Structure features   & 512      & UniMol 3D embedding \\
\midrule
Architecture
  & Struct encoder    & 3-layer MLP & 512$\to$1024$\to$512$\to$512 \\
  & Expr encoder     & 3-layer MLP & 978$\to$1024$\to$512$\to$512 \\
  & Shared embedding dim & 512     & L2-normalized output \\
  & Projection head     & 2-layer MLP & 512$\to$256$\to$256 \\
  & Gene decoder        & 2-layer MLP & 512$\to$1024$\to$978 \\
  & Activation / norm   & GELU / LayerNorm & Per layer \\
  & Dropout             & 0.223   & All encoders \\
\midrule
Soft Mask
  & $\epsilon$ (same-drug floor)      & 0.05  & Minimum positive weight \\
  & $\delta$ (cross-drug threshold)   & 0.86  & Bio-similarity cutoff \\
  & TAS weighting                     & $\sqrt{\text{TAS}_i\!\cdot\!\text{TAS}_j}$ & Quality-aware \\
  & Views per drug ($k$)              & 3     & Multi-cell-line sampling \\
\midrule
Loss
  & Temperature init ($\tau_\text{struct},\tau_\text{expr}$) & $1/0.07{\approx}14.3$ & Learnable \\
  & $\alpha$ ($\mathcal{L}_\text{distill}$) & 0.014 & HG knowledge distillation \\
  & Distillation target                    & 256d  & Frozen Stage~1 bio embedding \\
\midrule
Optimization
  & Optimizer        & AdamW & $\beta_1{=}0.9$, $\beta_2{=}0.999$ \\
  & Learning rate    & $5.48{\times}10^{-4}$ & Peak LR \\
  & Weight decay     & $10^{-4}$ & -- \\
  & Scheduler        & Cosine warmup & 8-epoch linear warmup \\
  & Gradient clipping& 5.0    & Max norm \\
  & Batch size       & 512    & Drugs per batch \\
  & Epochs           & 240    & Early stop patience 50 \\
\midrule
HG Pretrain
  & GNN architecture          & HeteroSAGE & R-GCN style, scatter-mean \\
  & GNN layers / hidden dim   & 2 / 256   & Output: 256d per drug \\
  & Edge types                & DTI, PPI, DDS & PPI$\geq$700, DDS$\geq$0.995 \\
  & HetMAE mask rate          & 0.3       & SCE loss ($\alpha{=}2.0$) \\
  & Loss weights (DTI/feat/edge) & 1.0/1.0/0.5 & -- \\
  & Epochs / LR               & 200 / $10^{-3}$ & Adam optimizer \\
\bottomrule
\end{tabular}
\end{table}

\subsubsection{Diffusion Model Training}

The diffusion model is trained on perturbation pairs $(\mathbf{x}_c,\mathbf{x}_p,d)$ with frozen VMCS and MDT modules.
Hyperparameters are listed in Table~\ref{tab:diffusion_params}.

\begin{table}[t]
\centering
\caption{Diffusion training hyperparameters.}
\label{tab:diffusion_params}
\small
\renewcommand{\arraystretch}{1.05}
\begin{tabular}{@{}llr@{\hskip 8pt}l@{}}
\toprule
\textbf{Category} & \textbf{Parameter} & \textbf{Value} & \textbf{Description} \\
\midrule
Data
  & Batch size   & 512 & Perturbation pairs \\
  & Epochs       & 200 & -- \\
\midrule
Architecture
  & Hidden dim $H$        & 768  & Transformer hidden size \\
  & Depth $L$             & 12   & DiT blocks \\
  & Attention heads       & 12   & Head dim $H/12{=}64$ \\
  & Drug tokens $K_d$     & 35   & -- \\
  & Sequence length       & 71   & $1{+}25{+}10{+}35$ \\
\midrule
Diffusion
  & Timesteps $T$     & 1{,}000 & Cosine schedule ($s{=}0.008$) \\
  & Prediction target & $v$-prediction & Velocity parameterization \\
\midrule
Loss
  & $\lambda_{\mathrm{triplet}}$ (triplet) & 0.1  & Directional triplet \\
  & Triplet margin $m$                      & 0.1  & Cosine margin \\
\midrule
Cond.\ dropout
  & $p_{\emptyset}$ (unconditional)  & 0.05  & Both conditions dropped \\
  & $p_c$ (cell only)               & 0.05  & Drug dropped \\
  & $p_d$ (drug only)               & 0.05  & Cell dropped \\
  & Protein dropout                 & $1{-}q_i$ & Reliability-adaptive \\
\midrule
Optimization
  & Optimizer         & AdamW & $\beta_1{=}0.9$, $\beta_2{=}0.999$ \\
  & Learning rate     & $10^{-4}$ & Peak LR \\
  & Weight decay      & 0.01  & -- \\
  & LR schedule       & Cosine & 5\% linear warmup \\
  & Gradient clipping & 1.0   & Max gradient norm \\
  & EMA decay         & 0.9999 & Parameter EMA \\
  & Precision         & bfloat16 & Mixed precision \\
\bottomrule
\end{tabular}
\end{table}

\subsubsection{Inference}

Inference uses DDIM sampling with interaction-aware CFG.
Parameters are listed in Table~\ref{tab:inference_params}.

\begin{table}[t]
\centering
\caption{Inference parameters.}
\label{tab:inference_params}
\small
\renewcommand{\arraystretch}{1.05}
\begin{tabular}{@{}lr@{\hskip 8pt}l@{}}
\toprule
\textbf{Parameter} & \textbf{Value} & \textbf{Description} \\
\midrule
Sampler               & DDIM      & Deterministic ($\eta{=}0$) \\
Sampling steps        & 50        & Uniformly spaced \\
Cell guidance $w_c$   & 1.0       & -- \\
Drug guidance $w_d$   & 1.5       & -- \\
Interaction guidance $w_{cd}$ & 2.0 & -- \\
Batch size            & 64        & Cells per batch \\
\bottomrule
\end{tabular}
\end{table}

\subsubsection{Model Scale and Dimensionality Summary}
\label{app:model_scale}

Table~\ref{tab:model_breakdown} details the parameter count per module.
The full StateXDiff has ${\sim}$85M trainable parameters, with the diffusion Transformer dominating at 72M (85\% of the total).
Pretrained frozen backbones (SCimilarity: 12M, scLinguist: 45M) are excluded from training.

\begin{table}[t]
\centering
\caption{Parameter count and status per module.}
\label{tab:model_breakdown}
\small
\renewcommand{\arraystretch}{1.05}
\begin{tabular}{@{}lrr@{}}
\toprule
\textbf{Module} & \textbf{Params} & \textbf{Status} \\
\midrule
\rowcolor{gray!10} VMCS projectors $\phi_z,\phi_p$ + rotation $\mathbf{R}$  & 1.6M & Trainable \\
Condition mixer (2-layer Transformer)                                     & 4.2M & Trainable \\
\rowcolor{gray!10} MDT structure encoder (HeteroSAGE + MLP)                & 2.3M & Trainable \\
Drug condition assembly (tokenizer + projection)                           & 4.8M & Trainable \\
\rowcolor{gray!10} Diffusion Transformer ($L{=}12$, $H{=}768$)             & 72M  & Trainable \\
\midrule
\textbf{Total trainable}                                                   & \textbf{85M} & \\
\midrule
\rowcolor{gray!10} SCimilarity encoder $\mathcal{E}$                       & 12M  & Frozen \\
scLinguist (encoder + decoder)                                             & 45M  & Frozen \\
\bottomrule
\end{tabular}
\end{table}

Table~\ref{tab:all_dims} consolidates all key dimensional constants referenced throughout the method.

\begin{table}[t]
\centering
\caption{Key dimensional constants.}
\label{tab:all_dims}
\small
\renewcommand{\arraystretch}{1.05}
\begin{tabular}{@{}lll@{}}
\toprule
\textbf{Symbol} & \textbf{Value} & \textbf{Description} \\
\midrule
\rowcolor{gray!10} $G$ & Dataset-dependent & RNA genes (18{,}961 for Tahoe-100M) \\
$|\mathcal{A}|$ & ${\approx}80$ & Reference genes for $s_{\mathrm{rc}}$ \\
\rowcolor{gray!10} $\mathbf{z}_c,\mathbf{z}_p,\Delta\mathbf{z},\hat{\mathbf{v}}_t$ & $\mathbb{R}^{128}$ & SCimilarity latent / velocity output \\
$\mathbf{g}_p$ & $\mathbb{R}^{128}$ & scLinguist protein embedding \\
\rowcolor{gray!10} $\hat{\mathbf{p}}$ & $\mathbb{R}^{6{,}427}$ & scLinguist protein predictions \\
$D$ & 512 & Common projection / shared state dim \\
\rowcolor{gray!10} $H$ & 768 & Transformer hidden dimension \\
$L$ & 12 & DiT blocks \\
\rowcolor{gray!10} Heads & 12 & Per-head dim $d_k=64$ \\
FFN ratio & 4 & SwiGLU expansion ($4H=3{,}072$) \\
\rowcolor{gray!10} $K_s$ & 25 & Shared cell tokens \\
$K_p$ & 10 & Protein-associated tokens \\
\rowcolor{gray!10} $K_d$ & 35 & Drug tokens \\
$|\mathbf{S}|$ & 71 & Total denoiser sequence $(1{+}25{+}10{+}35)$ \\
\rowcolor{gray!10} $T$ & 1{,}000 & Diffusion timesteps (cosine, $s{=}0.008$) \\
DDIM steps & 50 & $\eta{=}0$ \\
\bottomrule
\end{tabular}
\end{table}

\paragraph{Hardware.}
All experiments run on a single NVIDIA A800 80\,GB GPU with bfloat16 mixed precision.
Stage~1 (VMCS) trains in ${\sim}$40 minutes; MDT pretraining in ${\sim}$1 hour (both batch size 512).
Stage~2 (diffusion) trains in ${\sim}$10 hours (batch size 512).
Inference processes ${\sim}$2{,}500 cells/min on a single A800 (batch size 64, 50 DDIM steps with 4 forward passes per step).


\section{Evaluation Protocol Details}
\label{Evaluation_Protocol_Details}
We focus on chemical perturbation prediction under two representative out-of-distribution (OOD) generalization challenges: cell-context generalization and perturbation generalization. Our evaluation protocol adopts the framework of scPerturBench~\cite{Zhang2025_scPerturBench}, including its curated datasets, OOD split definitions, and six complementary metrics (MSE, PCC-$\Delta$, E-dist, WD, KL,  Common-DEGs). Under the cell-context generalization setting, we consider datasets spanning cross-cell-type and cross-patient transfer protocols. Under the perturbation generalization setting, we evaluate datasets covering both single-compound and combinatorial perturbation tasks, including three single-drug datasets and one drug-combination dataset, with perturbation scales ranging from tens to thousands. This design enables systematic assessment across varying cellular contexts, perturbation sizes, and combinatorial complexities.

Our work directly builds on the findings of Zhang et al.~\cite{Zhang2025_scPerturBench}, who showed that complex models often fail to outperform simple baselines under limited data or large distribution shifts, and identified cellular context embeddings as a key direction for improving generalization. First, we retain the recommended simple baselines---\texttt{baseControl}, \texttt{baseReg}, \texttt{baseMLP}, and \texttt{trainMean}---and show that StateXDiff consistently surpasses them across OOD settings, confirming that the proposed multimodal priors provide a genuine signal beyond memorization of training statistics. Second, we extend the benchmark's method coverage by adding recent generative models, including CellFlow, CRISP, and STATE, and provide head-to-head comparisons under identical splits and metrics. Third, we go beyond the original benchmark by evaluating on the large-scale Tahoe-100M dataset, combinatorial perturbation prediction, synthetic noise and sparsity stress tests, and diversity-controlled scaling experiments, thereby probing generalization limits not covered by Zhang et al.

Beyond real data, we further establish a controlled synthetic benchmark by progressively injecting technical noise and sparsity into real datasets and scaling up the dataset size, allowing us to rigorously validate model robustness and scalability under controlled conditions. We note that while StateXDiff achieves the largest margins on Tahoe-100M, where training diversity is high, the performance gaps narrow on smaller datasets such as CrossPatient and sciplex3 subsets. This trend is consistent with the observation of Zhang et al.~\cite{Zhang2025_scPerturBench} that data scale and train-test similarity remain dominant factors in OOD generalization.
\subsection{Definition of 100 DEGs and 5000  DEGs}
For each perturbation condition, differentially expressed genes are identified by comparing 
perturbed cells against control cells using \texttt{sc.tl.rank\_genes\_groups} from Scanpy, 
and subsequently ranked in descending order of absolute z-score.

Top-100 DEGs are defined as the 100 genes with the highest absolute z-scores. Top-5000 
DEGs are selected by the same criterion, but their precise interpretation depends on the 
preprocessing protocol applied to each dataset: for datasets in which the gene space was 
constrained to 5,000 highly variable genes (HVGs) during preprocessing, the top-5000 DEG 
setting is equivalent to evaluating across all retained genes; for datasets in which the 
full gene complement was preserved, it constitutes a genuine selection of the 5,000 most 
differentially expressed genes.
\label{DEGs_Definition}

\subsection{Evaluation Metrics Definitions}
\label{Evaluation_Metrics}
To rigorously evaluate perturbation prediction performance, we assess model fidelity from two complementary dimensions: population-average agreement, which quantifies the accuracy of recovering mean gene expression levels, and population-distribution agreement, which evaluates the model’s ability to reproduce full single-cell expression distributions, including cellular heterogeneity and higher-order structural features. To capture error magnitude, consistency of perturbation-induced shifts, and distributional divergence, we employ a complementary suite of six metrics that characterize model performance from three perspectives: numerical fitting accuracy, distributional consistency, and biological signal recovery.

Specifically, we use Mean Squared Error (MSE) to quantify the average gene-wise deviation between predicted and ground-truth expression, reflecting overall pointwise accuracy. We use a metric termed PCC$\Delta$, defined as the Pearson correlation of expression changes before and after perturbation, to assess how well a model captures the direction and relative magnitude of perturbation effects. From a distributional perspective, we adopt Energy distance and Wasserstein distance to measure discrepancies between predicted and real cell population distributions, thereby evaluating whether models can reproduce global structural shifts induced by perturbations. In addition, we include KL divergence to quantify distribution mismatch in terms of probabilistic inconsistency. Finally, to assess biological interpretability, we report Common-DEGs, which measures the overlap between differentially expressed genes identified from model predictions and those derived from ground-truth data. By jointly summarizing these population-average and population-distribution metrics, we systematically compare our approach against baseline methods and analyze the trade-offs between mean-level accuracy and distribution-level fidelity.

\subsubsection{Population-average metric}
\paragraph{Mean Squared Error (MSE)}

The Mean Squared Error (MSE) quantifies the global magnitude of prediction error by 
penalizing significant deviations between predicted and observed values. It is typically 
calculated based on group averages, where lower values indicate superior accuracy.
\[
\mathrm{MSE} = \frac{1}{n}\sum_{i=1}^{n}(y_i-\hat{y}_i)^2
\]
where $y_i$ and $\hat{y}_i$ denote the true and predicted values for observation $i$, 
respectively, and $n$ represents the total number of observations.

\paragraph{Energy Distance (E-distance)}

Energy distance assesses the discrepancy between two statistical populations by contrasting between-group distances with within-group distances. This metric provides a robust evaluation of mismatch that extends beyond simple mean differences, with larger values indicating poorer agreement.
\[
E\text{-distance}(X,Y) = 2\cdot\frac{1}{nm}\sum_{i=1}^{n}\sum_{j=1}^{m}\lvert x_i-y_j\rvert
-\frac{1}{n^2}\sum_{i=1}^{n}\sum_{j=1}^{n}\lvert x_i-x_j\rvert
-\frac{1}{m^2}\sum_{i=1}^{m}\sum_{j=1}^{m}\lvert y_i-y_j\rvert
\]
where $X=\{x_i\}_{i=1}^{n}$ and $Y=\{y_j\}_{j=1}^{m}$ represent samples from the true and predicted groups, respectively.

\paragraph{PCC$\Delta$}

To evaluate concordance in the directionality and relative patterns of perturbation effects, we calculate the Pearson correlation coefficient on expression deltas relative to an unperturbed control. Values approaching 1 indicate strong consistency in the predicted perturbation-induced shifts.

Pearson correlation is defined as:
\[
\mathrm{PCC} = \frac{\sum_{i=1}^{n}(x_i-\bar{x})(y_i-\bar{y})}
{\sqrt{\sum_{i=1}^{n}(x_i-\bar{x})^2}\sqrt{\sum_{i=1}^{n}(y_i-\bar{y})^2}}
\]
Accordingly, PCC$\Delta$ is computed on the expression changes:
\[
\mathrm{PCC\text{-}delta} =
\mathrm{PCC}\bigl(Y_{\text{true}}-Y_{\text{control}},\, Y_{\text{pred}}-Y_{\text{control}}\bigr)
\]
\subsubsection{Population-distribution metric}
\paragraph{Wasserstein Distance}

Distribution-level agreement is evaluated using the 1-Wasserstein distance, which measures the minimal transport cost required to transform the predicted distribution into the true distribution. Lower values reflect closer distributional alignment.
\[
W_1(P,Q) = \inf_{\gamma\in \Gamma(P,Q)} \mathbb{E}_{(x,y)\sim\gamma}\bigl[\lvert x-y\rvert\bigr]
\]
where $P$ and $Q$ denote the true and predicted distributions, and $\Gamma(P,Q)$ represents the set of all joint distributions (couplings) with marginals $P$ and $Q$.

\paragraph{KL-divergence}

We further quantify distributional discrepancy using the Kullback-Leibler (KL) divergence, an information-theoretic measure of the information loss when the predicted distribution approximates the true distribution. Smaller values indicate closer agreement (noting that this metric is asymmetric and depends on the overlap of support).
\[
D_{KL}(P\parallel Q) = \int p(x)\log\frac{p(x)}{q(x)}\, dx
\]
where $p(x)$ and $q(x)$ are the probability densities of $P$ (true) and $Q$ (predicted), respectively.

\paragraph{Common-DEGs}

To assess concordance in perturbation-driven differential expression, we compute the ``Common-DEGs'' metric, defined as the overlap between the top-ranked differentially expressed genes (DEGs) identified in the predicted versus true perturbation outcomes. Higher values signify better recovery of biologically relevant perturbation signals.
\[
\mathrm{Common\text{-}DEGs} = \frac{\left\lvert \mathrm{TopDEGs}_{\text{pred}}\cap \mathrm{TopDEGs}_{\text{true}}\right\rvert}
{\left\lvert \mathrm{TopDEGs}_{\text{true}}\right\rvert}
\]
where $\mathrm{TopDEGs}_{\text{pred}}$ and $\mathrm{TopDEGs}_{\text{true}}$ denote the sets of top-ranked DEGs selected from the predicted and true datasets, respectively.

\subsection{OOD Split Construction}
\label{OOD_Split_Construction}

We evaluate generalization along three axes: cellular context, perturbation identity, and combinatorial perturbation. All preprocessing statistics, highly variable gene selection, normalization parameters, and hyperparameter selection are computed using the training split only and then applied unchanged to validation and test data.

For \textbf{cellular context generalization}, we perform leave-one-context-out evaluation. The held-out context is defined as patient donor for CrossPatient and KaggleCrossPatient, cell type for KaggleCrossCell, and cell line for Sciplex3. No cells from the held-out context are used for model training or validation. For Tahoe-100M, which contains a substantially larger and more diverse set of cell lines, we reserve a subset of cell lines as unseen contexts and remove all conditions involving these cell lines from training and validation.

For \textbf{perturbation generalization}, splits are constructed at the drug-identity level. All samples involving held-out drugs, across all doses and cellular contexts, are removed from training and validation. Thus, the model is evaluated on drug identities that are not observed in the single-cell perturbation training data. For Sciplex3 subsets, including A549, MCF7, and K562, the split is performed independently within each cell line. Our unseen-drug setting refers to single-cell unseen-drug generalization. Held-out drugs are excluded from the single-cell perturbation training data, but may be covered by external L1000 perturbational transcriptomic signatures used to learn mechanism-aware drug priors. L1000 provides population-level drug-induced transcriptional response signatures rather than single-cell-resolution perturbation measurements. This setting reflects a practical scenario in which many compounds have L1000-level mechanistic response profiles, whereas matched single-cell perturbation measurements remain scarce. StateXDiff is designed to transfer such population-level perturbational priors to single-cell response prediction.

For \textbf{combinatorial perturbation generalization}, test combinations are grouped according to whether their constituent single drugs appear in the training set: \textit{combo-seen2}, where both drugs are observed individually during training; \textit{combo-seen1}, where only one constituent drug is observed; and \textit{combo-seen0}, where neither drug is observed. The exact drug pair in each test combination is never included in training.

\paragraph{Gene selection and evaluation.}
HVG selection and all preprocessing statistics are fit using the training split only and then applied unchanged to validation and test data. We evaluate each model under two DEG-resolution settings. The first setting uses the top 100 differentially expressed genes (top-100 DEGs), focusing on the most strongly responsive genes. The second setting uses the top 5,000 differentially expressed genes (top-5,000 DEGs), providing a broader evaluation over a substantially larger response-associated gene set. In both settings, DEGs are selected per test condition from the ground-truth perturbation response and are used only for evaluation, not for model training, validation, or hyperparameter selection.

\subsection{CITE-seq evaluation details.}
\label{CITE-seq_evaluation}
We evaluate the contribution of the protein modality using the GSE152469 CITE-seq dataset~\cite{kurtulus2019checkpoint}, a longitudinal single-cell multi-omic dataset profiling peripheral blood mononuclear cells (PBMCs) from a chronic lymphocytic leukemia (CLL) patient undergoing ibrutinib treatment. The dataset provides paired RNA expression profiles and surface protein (ADT) measurements across three time points: before treatment (M0), clinical response after 3 months (M3), and disease progression after 27 months (M27). In this study, we select pre-treatment and post-treatment samples to characterize cellular state transitions induced by drug exposure.

For data splitting, we adopt a leave-one-cell-type-out strategy, holding out one cell type as the validation set to assess generalization to unseen cellular contexts. To ensure a fair comparison, all methods share the same RNA encoder, latent representation space, and diffusion backbone. The only differences across variants lie in the source and utilization of the protein modality, enabling a clear isolation of the effect of protein information on OOD perturbation prediction.

\begin{itemize}[nosep,leftmargin=*]
  \item \textbf{TrueProt}: Measured surface proteins from the CITE-seq antibody panel, 
    projected to $D$ dimensions by a learned linear layer.
  \item \textbf{PseudoProt-ZS}: Zero-shot protein representation predicted by scLinguist, 
    where the 128-dimensional bottleneck embedding $\mathbf{g}_p$ is extracted from the 
    frozen encoder and then passed through $\phi_p$.
  \item \textbf{TrueProt-FT}: scLinguist fine-tuned on the \emph{training split only} 
    of the GSE152469 CITE-seq data. The resulting 128-dimensional bottleneck embedding is 
    extracted and used in the same way as PseudoProt-ZS.
  \item \textbf{RNA-only}: No protein features are used; the model relies only on 
    $\mathbf{z}_c$ from the RNA encoder.
\end{itemize}

\subsection{Robustness Analysis under Noise and Sparsity}
Using the same experimental protocol and evaluation metrics as in the cell-context generalization setting, we systematically evaluate the robustness of the proposed method against baseline approaches. In the noise and sparsity stress tests, we vary sparsity levels and progressively inject noise with increasing intensity on real data. The top row reports mean squared error (MSE), while the bottom row reports PCC$\Delta$. The results indicate that under low-noise and low-sparsity conditions, StateXDiff achieves the best performance on both metrics. As noise intensity or sparsity increases, performance consistently degrades across all methods.
(Table~\ref{tab:Noise and Sparsity})

\begin{table}[htbp]
  \centering
  \caption{Dataset Configuration for Robustness Analysis under Noise and Sparsity}
    \begin{tabular}{ccccc}
    \toprule
    \textbf{Original dataset} & \textbf{DataSet} & \textbf{Experiment} & \textbf{Noise} & \textbf{Sparsity} \\
    \midrule
    \multirow{10}[2]{*}{\textbf{Sciplex3-A549}} & Simulated dataset 1 & Noise 1 & 0.1   & 0.963  \\
          & Simulated dataset 2 & Noise 3 & 0.3   & 0.963  \\
          & Simulated dataset 3 & Noise 5 & 0.5   & 0.963  \\
          & Simulated dataset 4 & Noise 7 & 0.7   & 0.963  \\
          & Simulated dataset 5 & Noise 9 & 0.9   & 0.963  \\
          & Simulated dataset 6 & Sparsity 1 & 0     & 0.969  \\
          & Simulated dataset 7 & Sparsity 3 & 0     & 0.977  \\
          & Simulated dataset 8 & Sparsity 5 & 0     & 0.983  \\
          & Simulated dataset 9 & Sparsity 7 & 0     & 0.990  \\
          & Simulated dataset 10 & Sparsity 9 & 0     & 0.997  \\
    \bottomrule
    \end{tabular}%
  \label{tab:Noise and Sparsity}%
\end{table}%

\subsection{Experimental Setup and Diversity Scaling Protocol}

\paragraph{Dataset and splits.}
All diversity scaling experiments are conducted on the \textbf{Tahoe-100M} dataset, which is pre-partitioned into three mutually exclusive subsets: \textbf{train}, \textbf{test} (in-distribution evaluation), and \textbf{ood} (out-of-distribution evaluation).
The test and ood splits are kept fixed across all experimental conditions to ensure fair comparability; only the composition of the training subset varies.

\paragraph{Fixed-sample-budget protocol.}
A naive approach to studying diversity would simply add more data as diversity increases, confounding the effect of identity diversity with that of sample count.
To isolate diversity as the sole variable, we fix the total training set size at $N_{\mathrm{fix}} = 0.7 \times |\mathrm{Train}|$ for every diversity condition, i.e., $|\mathcal{T}_{\mathrm{fix}}(K)| = N_{\mathrm{fix}}$ regardless of the diversity level $K$.
Within each condition, samples are allocated via identity-balanced subsampling: each selected identity contributes approximately $\lfloor N_{\mathrm{fix}} / K \rfloor$ samples, with the remainder filled by random completion to exactly match $N_{\mathrm{fix}}$.
This design ensures that any observed performance difference is attributable to the diversity of drug or cell-line identities rather than to differences in training set size.

\paragraph{Drug diversity test.}
We evaluate sensitivity to chemical diversity by varying the number of unique drug identities $K_d \in \{5, 50, 150, 300\}$ included in the training subset.
For each $K_d$, we randomly sample $K_d$ drugs from the full training pool $D_{\mathrm{train}}$ and construct the fixed-size subset $\mathcal{T}_{\mathrm{fix}}(K_d)$ following the balanced protocol above.
This range spans from a highly restricted chemical space (5 drugs) to a moderately diverse one (300 drugs), allowing us to characterize how perturbation generalization scales with the breadth of training-time chemical exposure.

\paragraph{Cell-line diversity test.}
Analogously, we vary the number of unique cell lines $K_c \in \{5, 15, 25, 35\}$ to assess how cellular context diversity affects generalization.
For each $K_c$, we sample $K_c$ cell-line identities from $C_{\mathrm{train}}$ and construct $\mathcal{T}_{\mathrm{fix}}(K_c)$ with balanced allocation.
Because Tahoe-100M contains a substantially larger number of drugs than cell lines, the cell-line diversity range is correspondingly narrower but still covers a meaningful spectrum from low to high heterogeneity.

\paragraph{Training and evaluation.}
All models (StateXDiff and baseMLP) are trained with identical architectures, parameter budgets, and optimization hyperparameters across all diversity conditions.
For each diversity level, the model is trained once on the corresponding subset and then evaluated on both the fixed test split and the fixed ood split.
We report the same suite of metrics used throughout this work: MSE, PCC$\Delta$, E-distance, KL divergence, Common-DEGs, and Wasserstein distance.

\section{More Experimental Results}

\subsection{Superior Robustness on Synthetic Data}

We evaluated the resilience of StateXDiff against varying levels of data degradation, as illustrated in Figure~\ref{bio_case_s}. Under low-to-moderate perturbation regimes (up to \texttt{noise3} and \texttt{sparsity3}), StateXDiff consistently achieves the lowest $MSE$ and highest $PCC\Delta$, demonstrating superior robustness in distilling biological signals from imperfect data. However, as the perturbation intensity scales to extreme levels (\textit{e.g.}, \texttt{noise9} or \texttt{sparsity9}), where the underlying biological manifold is substantially obliterated, the performance gap between our model and baselines effectively vanishes. In these high-stochasticity regimes, the metrics for StateXDiff converge toward or are surpassed by simpler baselines like baseMLP, indicating an intrinsic information bottleneck where the loss of critical biological features limits the efficacy of complex generative modeling. 

\begin{figure}[H]
\centering
\includegraphics[width=1\columnwidth]{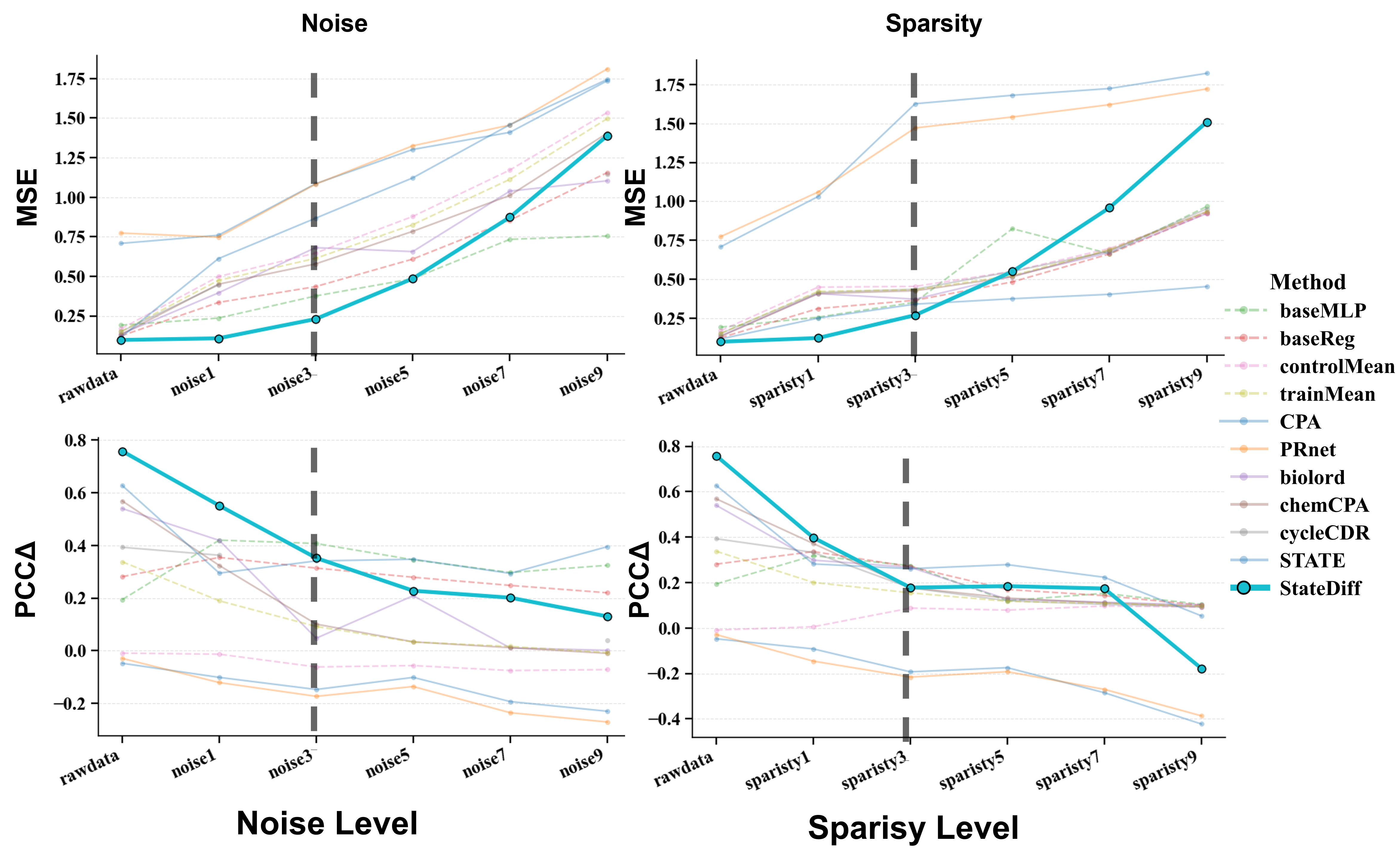}
\caption{Robustness evaluation under noise and sparsity.
Noise intensity (left) and sparsity levels (right) are progressively increased on real data. The top row reports mean squared error (MSE), and the bottom row reports PCC$\Delta$.}
\label{fig_sy}
\end{figure}

\subsection{Scaling Behaviors under Training Diversity}

\begin{figure}[h]
\centering
\includegraphics[width=1\columnwidth]{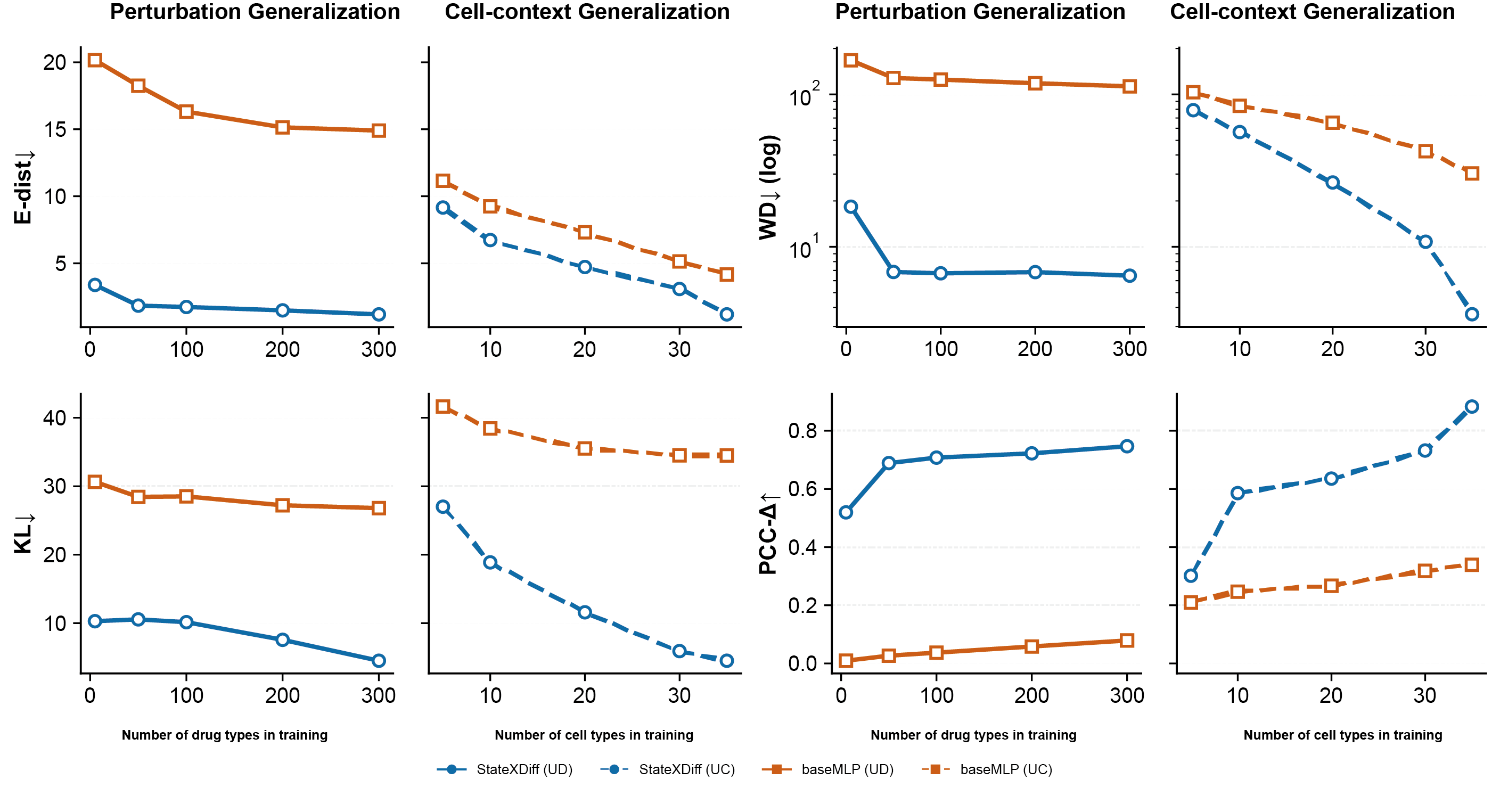}
\caption{Scaling behavior under increasing training diversity.}
\label{div_scaling_appendix}
\end{figure}

\begin{figure}[H]
\centering
\includegraphics[width=1\columnwidth]{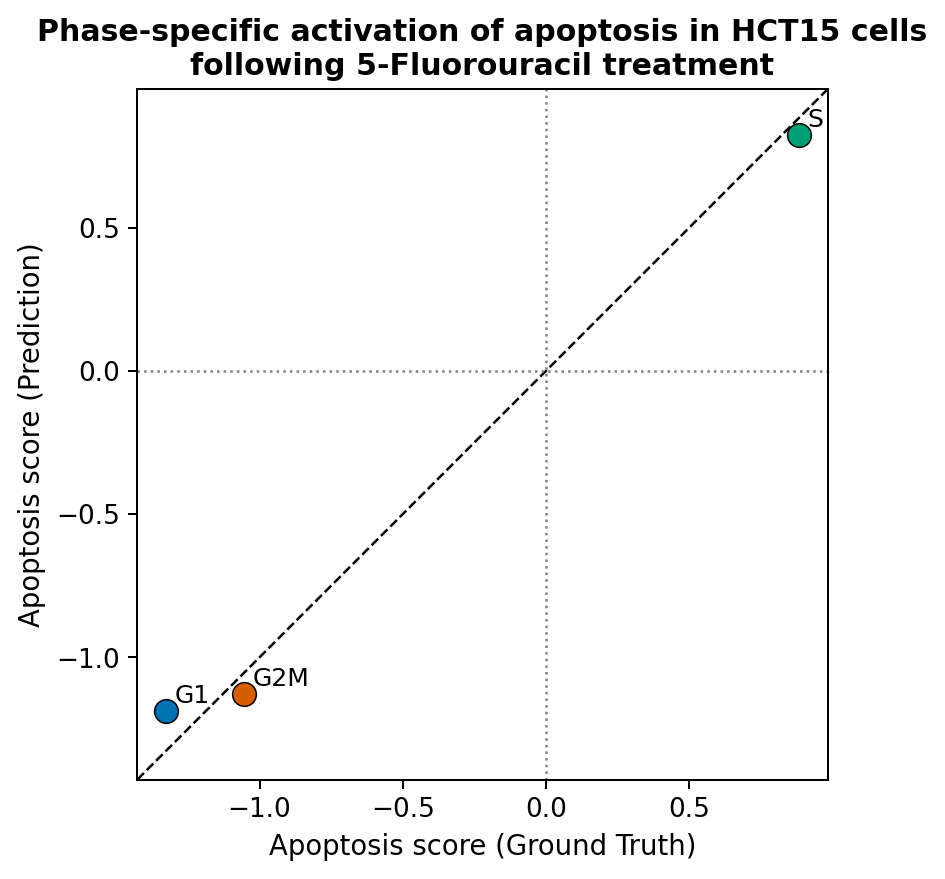}
\caption{Phase-specific prediction of apoptosis pathway responses in HCT15 cells following 5-Fluorouracil treatment. The scatter plot compares predicted and ground-truth changes in apoptosis scores across cell-cycle phases (G1, G2M, and S), defined as treatment relative to control. Positive values indicate pathway activation, while negative values indicate suppression. The dashed line denotes the identity line (y = x), and the dotted lines indicate zero change.}
\label{bio_case_s}
\end{figure}

We evaluate the scaling behavior under increasing training diversity while keeping the computational budget fixed, considering both perturbation generalization (UD) and cell-context generalization (UC). Figure~\ref{div_scaling_appendix} reports results across multiple metrics, including E-distance, KL divergence, Wasserstein distance (WD), and Pearson correlation (PCC).

In the UD setting, as the number of drug types in training increases, StateXDiff shows consistent improvements across all discrepancy metrics. WD decreases rapidly at smaller scales and continues to decline more gradually as diversity increases. Similar trends are observed for KL divergence and E-distance. Meanwhile, PCC increases steadily and stabilizes at a relatively high level. In contrast, baseMLP exhibits limited improvement: WD remains nearly constant on a log scale, while KL and E-distance plateau early. The increase in PCC is also comparatively modest.

In the UC setting, increasing the number of cell types leads to more pronounced improvements for StateXDiff. WD and KL decrease consistently with training scale, without clear signs of saturation at the largest scale considered. PCC also shows a steady upward trend. For baseMLP, although some improvement is observed, the overall magnitude of change is smaller, and a persistent gap remains across all metrics.

Comparing the two settings, the UC task exhibits higher discrepancy levels (e.g., WD and KL) and lower PCC values overall, suggesting a more challenging generalization regime.

Overall, StateXDiff demonstrates consistent gains with increasing data diversity in both settings, whereas baseMLP shows early saturation across multiple metrics.

\subsection{Phase-Specific Activation of Apoptosis in 5-FU Treated HCT15 Cells: Prediction vs. Ground Truth}

\subsection{Additional OOD Generalization Results}
\begin{figure}[htbp]

\centering
\includegraphics[width=1\columnwidth]{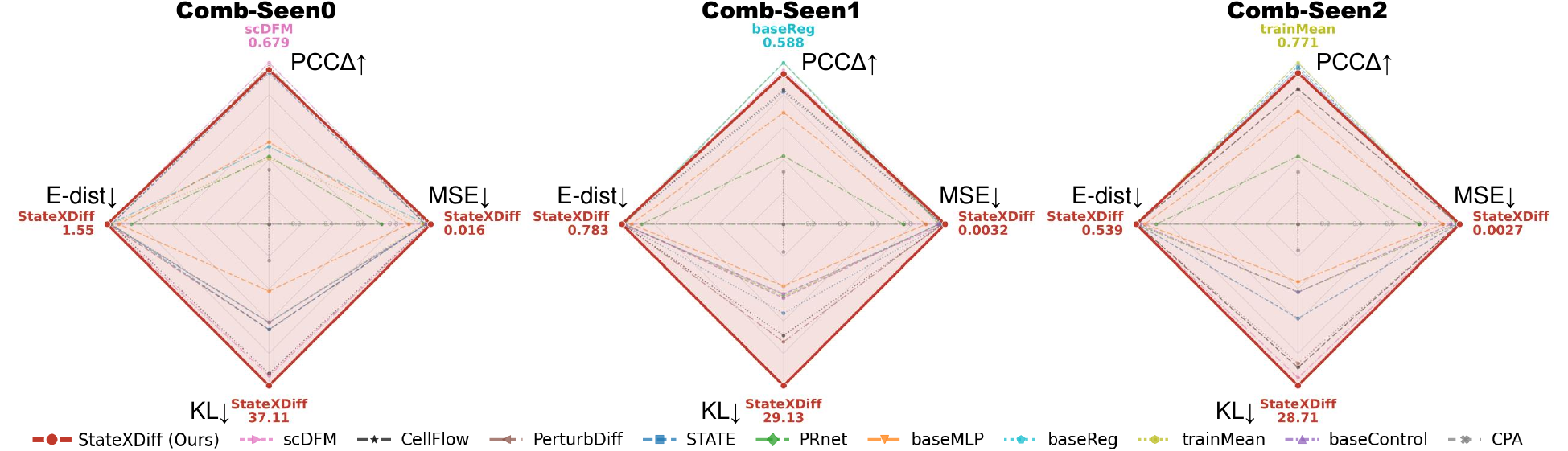}
\caption{Performance comparison on chemical combination prediction evaluated on the top-5000 DEGs. Our method achieves the best overall results across most metrics.}
\label{comb_5000}
\end{figure}

The supplementary material includes additional experimental results under different generalization settings.
Table~\ref{tab:unseen_drug_100degs_D3_OOD} reports the results on unseen drugs evaluated with the top-100 DEG genes.
Table~\ref{tab:unseen_drug_All_D3_OOD} presents the results on unseen drugs evaluated on the full gene set.
Table~\ref{tab:unseen_cell_100degs_D3_ood} shows the results on unseen cells with the top-100 DEG genes, while
Table~\ref{tab:unseen_cell_allgene_D3_ood} reports the corresponding results on unseen cells using all genes.
Figure~\ref{comb} and ~\ref{comb_5000} present the results for the drug combination setting.

\begin{table}[H]
\centering
\sisetup{group-separator={}}
\renewcommand{\arraystretch}{1.3}
\caption{Performance comparison of different methods on top 100 DEGs under the Unseen Drug (UD) setting. Arrows indicate whether lower ($\downarrow$) or higher ($\uparrow$) values are better.}
\label{tab:unseen_drug_100degs_D3_OOD}

  \vspace{2pt}
  {\scriptsize\raggedright
  \textit{Note.}
  \begingroup
  \setlength{\fboxsep}{0pt}%
  \colorbox{firstbg}{\strut\hspace{1pt}Rank-1\hspace{1pt}}%
  \endgroup
   and
  \begingroup
  \setlength{\fboxsep}{0pt}%
  \colorbox{secondbg}{\strut\hspace{1pt}Rank-2\hspace{1pt}}%
  \endgroup
   denote the best and second-best results;
  \textcolor{teal}{$\uparrow/\downarrow$} shows the relative gain of Rank-1 over Rank-2.
  \par}

\setlength{\tabcolsep}{4pt}
\resizebox{\textwidth}{!}{%
\begin{tabular}{llcccccc}
\toprule
\textbf{Dataset} & \textbf{Method} &
\textbf{E-dist$\downarrow$} &
\textbf{MSE$\downarrow$} &
\textbf{PCC$\Delta\uparrow$} &
\textbf{KL$\downarrow$} &
\textbf{WD$\downarrow$} &
\textbf{ Common-DEGs$\uparrow$} \\
\midrule
\multirow{12}{*}{\textbf{sciplex3\_A549}}
& trainMean
& \num{1.24}\std{2.02}
& \num{0.127}\std{0.315}
& \num{0.395}\std{0.159}
& \num{47.8}\std{0.933}
& \num{82.6}\std{71.4}
& \num{0.0}\std{0.0}
\\
& baseControl
& \num{1.39}\std{2.09}
& \num{0.144}\std{0.33}
& \num{0.0594}\std{0.0955}
& \num{47.8}\std{0.931}
& \num{84.4}\std{73.1}
& \num{0.0384}\std{0.0187}
\\
& baseMLP
& \num{1.59}\std{2.62}
& \num{0.156}\std{0.364}
& \num{0.282}\std{0.259}
& \num{47.9}\std{1.13}
& \num{85.8}\std{68.0}
& \num{0.0709}\std{0.0778}
\\
& baseReg
& \num{1.2499}\std{1.4173}
& \num{0.1115}\std{0.1888}
& \num{0.3495}\std{0.2657}
& \num{47.8566}\std{0.925}
& \num{81.159}\std{59.3689}
& \num{0.0003}\std{0.0019}
\\
& bioLord
& \num{4.57}\std{3.1}
& \num{0.114}\std{0.273}
& \num{0.603}\std{0.176}
& \num{47.3}\std{1.43}
& \num{45.2}\std{44.0}
& \num{0.0002}\std{0.0013}
\\
& CellFlow
& \num{1.2189}\std{1.4788}
& \num{0.1046}\std{0.6646}
& \num{0.6867}\std{0.1615}
& \num{45.8516}\std{1.5344}
& \num{67.431}\std{64.469}
& \num{0.0037}\std{0.0013}
\\
& CRISP
& \num{1.3476}\std{1.0376}
& \cellcolor{secondbg}\num{0.0963}\std{0.3436}
& \num{0.4646}\std{0.169}
& \num{45.2233}\std{1.4261}
& \cellcolor{firstbg}\num{40.9902}\std{45.3529}{\scriptsize\textcolor{teal}{$\downarrow$9\%}}
& \num{0.0027}\std{0.0029}
\\
& STATE
& \cellcolor{secondbg}\num{1.12}\std{1.75}
& \num{0.116}\std{0.272}
& \num{0.373}\std{0.223}
& \num{47.0}\std{1.23}
& \num{71.3}\std{66.1}
& \cellcolor{secondbg}\num{0.0902}\std{0.01}
\\
& chemCPA
& \num{4.06}\std{2.92}
& \num{0.111}\std{0.277}
& \num{0.627}\std{0.162}
& \num{42.9}\std{1.95}
& \cellcolor{secondbg}\num{44.9}\std{44.3}
& \num{0.011}\std{0.0147}
\\
& PerturbDiff
& \num{1.31}\std{1.41}
& \num{0.136}\std{0.244}
& \num{0.413}\std{0.313}
& \num{51.3}\std{1.57}
& \num{57.7}\std{46.3}
& \num{0.0905}\std{0.05}
\\
& scDFM
& \num{1.14}\std{1.31}
& \num{0.113}\std{0.132}
& \cellcolor{secondbg}\num{0.693}\std{0.343}
& \cellcolor{secondbg}\num{42.1}\std{2.13}
& \num{51.7}\std{42.6}
& \cellcolor{firstbg}\num{0.1018}\std{0.0815}{\scriptsize\textcolor{teal}{$\uparrow$12\%}}
\\
& \cellcolor{gray!15}\textbf{Ours}
& \cellcolor{firstbg}\num{1.03}\std{1.1}{\scriptsize\textcolor{teal}{$\downarrow$8\%}}
& \cellcolor{firstbg}\num{0.0743}\std{0.137}{\scriptsize\textcolor{teal}{$\downarrow$23\%}}
& \cellcolor{firstbg}\num{0.759}\std{0.213}{\scriptsize\textcolor{teal}{$\uparrow$10\%}}
& \cellcolor{firstbg}\num{37.1}\std{9.48}{\scriptsize\textcolor{teal}{$\downarrow$12\%}}
& \cellcolor{gray!15}\num{54.8}\std{33.2}
& \cellcolor{gray!15}\num{0.0413}\std{0.0034}
\\

\midrule

\multirow{12}{*}{\textbf{sciplex3\_K562}}
& trainMean
& \num{0.744}\std{1.15}
& \num{0.0648}\std{0.128}
& \num{0.39}\std{0.179}
& \num{46.8}\std{0.717}
& \num{76.1}\std{50.9}
& \num{0.0001}\std{0.0018}
\\
& baseControl
& \num{0.821}\std{1.17}
& \num{0.0731}\std{0.135}
& \num{0.0558}\std{0.0839}
& \num{46.9}\std{0.876}
& \num{77.1}\std{51.9}
& \num{0.0414}\std{0.0207}
\\
& baseMLP
& \num{1.24}\std{3.15}
& \num{0.115}\std{0.398}
& \num{0.301}\std{0.254}
& \num{47.1}\std{1.29}
& \num{81.4}\std{57.9}
& \cellcolor{secondbg}\num{0.0723}\std{0.0729}
\\
& baseReg
& \num{0.739}\std{0.852}
& \num{0.0566}\std{0.0782}
& \num{0.325}\std{0.257}
& \num{46.1}\std{0.845}
& \num{75.4}\std{46.2}
& \num{0.0001}\std{0.0021}
\\
& bioLord
& \num{4.14}\std{1.99}
& \num{0.0605}\std{0.121}
& \num{0.564}\std{0.207}
& \num{47.4}\std{1.58}
& \cellcolor{firstbg}\num{41.7}\std{27.9}
& \num{0.0}\std{0.0}
\\
& CellFlow
& \num{0.6453}\std{1.1233}
& \num{0.0571}\std{0.3232}
& \num{0.4839}\std{0.1382}
& \num{36.2337}\std{1.277}
& \num{58.325}\std{32.3534}
& \num{0.0118}\std{0.0207}
\\
& CRISP
& \num{0.7031}\std{0.8521}
& \num{0.0632}\std{0.1243}
& \num{0.5322}\std{0.167}
& \num{42.7369}\std{1.946}
& \num{42.6857}\std{27.9339}
& \num{0.0093}\std{0.0347}
\\
& STATE
& \num{0.666}\std{1.02}
& \num{0.0603}\std{0.115}
& \num{0.404}\std{0.225}
& \num{45.8}\std{1.28}
& \num{61.4}\std{42.3}
& \num{0.06}\std{0.0306}
\\
& chemCPA
& \num{3.82}\std{1.91}
& \num{0.0612}\std{0.124}
& \cellcolor{secondbg}\num{0.613}\std{0.154}
& \num{41.7}\std{1.96}
& \cellcolor{secondbg}\num{41.7}\std{28.2}
& \num{0.005}\std{0.011}
\\
& PerturbDiff
& \num{0.694}\std{1.34}
& \num{0.0743}\std{0.132}
& \num{0.413}\std{0.325}
& \num{43.1}\std{1.38}
& \num{59.4}\std{41.3}
& \num{0.031}\std{0.0126}
\\
& scDFM
& \cellcolor{secondbg}\num{0.622}\std{0.902}
& \cellcolor{secondbg}\num{0.0543}\std{0.125}
& \num{0.534}\std{0.315}
& \cellcolor{secondbg}\num{35.1}\std{11.28}
& \num{54.3}\std{41.9}
& \cellcolor{firstbg}\num{0.08}\std{0.0206}{\scriptsize\textcolor{teal}{$\uparrow$11\%}}
\\
& \cellcolor{gray!15}\textbf{Ours}
& \cellcolor{firstbg}\num{0.619}\std{0.406}
& \cellcolor{firstbg}\num{0.0455}\std{0.0563}{\scriptsize\textcolor{teal}{$\downarrow$16\%}}
& \cellcolor{firstbg}\num{0.674}\std{0.241}{\scriptsize\textcolor{teal}{$\uparrow$10\%}}
& \cellcolor{firstbg}\num{33.2}\std{12.4}{\scriptsize\textcolor{teal}{$\downarrow$5\%}}
& \cellcolor{gray!15}\num{55.6}\std{33.5}
& \cellcolor{gray!15}\num{0.034}\std{0.0073}
\\

\bottomrule
\end{tabular}

} 

\end{table}

\begin{table}[H]
\centering
\sisetup{group-separator={}}
\renewcommand{\arraystretch}{0.9}
\caption{Performance comparison of different methods on top 5000 DEGs under the Unseen Drug (UD) setting. Arrows indicate whether lower ($\downarrow$) or higher ($\uparrow$) values are better.}
\label{tab:unseen_drug_All_D3_OOD}

  \vspace{2pt}
  {\scriptsize\raggedright
  \textit{Note.}
  \begingroup
  \setlength{\fboxsep}{0pt}%
  \colorbox{firstbg}{\strut\hspace{1pt}Rank-1\hspace{1pt}}%
  \endgroup
   and
  \begingroup
  \setlength{\fboxsep}{0pt}%
  \colorbox{secondbg}{\strut\hspace{1pt}Rank-2\hspace{1pt}}%
  \endgroup
   denote the best and second-best results;
  \textcolor{teal}{$\uparrow/\downarrow$} shows the relative gain of Rank-1 over Rank-2.
  \par}

\setlength{\tabcolsep}{4pt}
\resizebox{\textwidth}{!}{%
\begin{tabular}{llcccc}
\toprule
\textbf{Dataset} &
\textbf{Method} &
\textbf{E-dist$\downarrow$} &
\textbf{MSE$\downarrow$} &
\textbf{PCC$\Delta\uparrow$} &
\textbf{KL$\downarrow$} \\
\midrule

\multirow{12}{*}{\textbf{Tahoe-100M}}
& trainMean
& \num{13.7858}\std{3.2401}
& \num{0.0556}\std{0.0168}
& \num{0.7821}\std{0.1367}
& \num{39.8135}\std{2.5065}
\\
& baseControl
& \num{4.7022}\std{0.0397}
& \num{0.0157}\std{0.0002}
& \num{-0.0009}\std{0.0004}
& \num{35.0542}\std{0.0649}
\\
& baseMLP
& \num{59.1208}\std{0.2037}
& \num{0.3815}\std{0.0015}
& \num{0.0937}\std{0.0023}
& \num{41.9105}\std{0.0235}
\\
& baseReg
& \num{14.161}\std{0.0569}
& \num{0.0575}\std{0.0003}
& \num{0.2214}\std{0.0025}
& \num{39.8253}\std{0.0407}
\\
& bioLord
& \num{12.3827}\std{0.1095}
& \num{0.0190}\std{0.00034}
& \num{0.3674}\std{0.006}
& \num{46.6376}\std{0.0195}
\\
& CellFlow
& \num{8.2048}\std{1.214}
& \num{0.0421}\std{0.0131}
& \num{0.5436}\std{0.2046}
& \num{41.5013}\std{0.1373}
\\
& CRISP
& \num{9.3838}\std{2.1431}
& \num{0.0564}\std{0.0134}
& \cellcolor{firstbg}\num{0.9221}\std{0.1467}{\scriptsize\textcolor{teal}{$\uparrow$11\%}}
& \num{49.8235}\std{1.5085}
\\
& STATE
& \num{7.2038}\std{2.714}
& \num{0.016}\std{0.0091}
& \num{0.7836}\std{0.2046}
& \num{42.5013}\std{0.4573}
\\
& CPA
& \num{7.6735}\std{0.1545}
& \num{0.112}\std{0.003}
& \num{0.1642}\std{0.0017}
& \num{32.6682}\std{0.071}
\\
& PRnet
& \num{13.9473}\std{0.0529}
& \num{0.0573}\std{0.0003}
& \num{0.3699}\std{0.004}
& \num{40.0226}\std{0.0395}
\\
& chemCPA
& \num{13.3827}\std{1.8151}
& \num{0.0176}\std{0.0003}
& \num{0.7891}\std{0.3401}
& \num{14.6376}\std{2.0195}
\\
& PerturbDiff
& \num{13.1361}\std{9.7881}
& \num{0.0514}\std{0.0417}
& \num{0.6191}\std{0.2038}
& \num{42.6989}\std{1.0650}
\\
& scDFM
& \cellcolor{secondbg}\num{4.2038}\std{2.714}
& \cellcolor{secondbg}\num{0.0091}\std{0.0081}
& \num{0.7936}\std{0.1146}
& \cellcolor{secondbg}\num{12.9013}\std{0.1273}
\\
& \cellcolor{gray!15}\textbf{Ours}
& \cellcolor{firstbg}\num{0.2347}\std{0.4907}{\scriptsize\textcolor{teal}{$\downarrow$94\%}}
& \cellcolor{firstbg}\num{0.0028}\std{0.0065}{\scriptsize\textcolor{teal}{$\downarrow$69\%}}
& \cellcolor{secondbg}\num{0.8295}\std{0.1199}
& \cellcolor{firstbg}\num{1.9952}\std{1.8758}{\scriptsize\textcolor{teal}{$\downarrow$85\%}}
\\

\midrule

\multirow{12}{*}{\textbf{sciplex3\_A549}}
& trainMean
& \num{1.2863}\std{1.3244}
& \num{0.0067}\std{0.0095}
& \num{0.2061}\std{0.0586}
& \num{44.4856}\std{0.6117}
\\
& baseControl
& \num{1.3635}\std{1.3746}
& \num{0.0073}\std{0.0099}
& \num{0.0143}\std{0.0211}
& \num{44.4878}\std{0.6111}
\\
& baseMLP
& \num{9.1284}\std{39.3792}
& \num{0.1474}\std{0.762}
& \num{0.1582}\std{0.1653}
& \num{44.8539}\std{1.9739}
\\
& baseReg
& \num{2.0468}\std{1.5386}
& \num{0.0128}\std{0.0113}
& \num{0.1845}\std{0.1743}
& \num{44.566}\std{0.6021}
\\
& bioLord
& \num{17.154}\std{1.6542}
& \num{0.0057}\std{0.01}
& \num{0.3663}\std{0.1115}
& \num{46.6639}\std{0.7077}
\\
& CellFlow
& \cellcolor{firstbg}\num{1.1189}\std{1.4788}{\scriptsize\textcolor{teal}{$\downarrow$8\%}}
& \num{0.016}\std{0.6646}
& \num{0.2887}\std{0.1715}
& \num{40.8516}\std{1.5344}
\\
& CRISP
& \num{1.9476}\std{1.0376}
& \num{0.1263}\std{0.3736}
& \cellcolor{firstbg}\num{0.4375}\std{0.717}{\scriptsize\textcolor{teal}{$\uparrow$4\%}}
& \num{41.228}\std{1.4661}
\\
& STATE
& \num{1.3012}\std{1.1848}
& \num{0.0064}\std{0.0083}
& \num{0.2128}\std{0.1142}
& \num{43.5536}\std{0.9058}
\\
& chemCPA
& \num{15.3158}\std{1.3935}
& \cellcolor{secondbg}\num{0.005}\std{0.0083}
& \num{0.3948}\std{0.1081}
& \num{44.6327}\std{0.424}
\\
& PerturbDiff
& \num{1.4012}\std{1.1848}
& \num{0.0083}\std{0.0103}
& \num{0.3128}\std{0.2142}
& \num{41.0536}\std{1.058}
\\
& scDFM
& \num{1.3112}\std{1.2848}
& \num{0.0051}\std{0.0083}
& \num{0.4128}\std{0.3142}
& \cellcolor{secondbg}\num{39.9436}\std{0.9458}
\\
& \cellcolor{gray!15}\textbf{Ours}
& \cellcolor{secondbg}\num{1.22}\std{1.0916}
& \cellcolor{firstbg}\num{0.0049}\std{0.0055}{\scriptsize\textcolor{teal}{$\downarrow$2\%}}
& \cellcolor{secondbg}\num{0.4486}\std{0.1744}
& \cellcolor{firstbg}\num{39.2673}\std{1.0364}{\scriptsize\textcolor{teal}{$\downarrow$2\%}}
\\

\midrule

\multirow{12}{*}{\textbf{sciplex3\_K562}}
& trainMean
& \num{0.8105}\std{0.8909}
& \num{0.0046}\std{0.006}
& \num{0.2173}\std{0.0601}
& \num{43.9614}\std{0.6572}
\\
& baseControl
& \num{0.8603}\std{0.9264}
& \num{0.005}\std{0.0063}
& \num{0.0116}\std{0.0181}
& \num{43.9618}\std{0.6532}
\\
& baseMLP
& \num{6.7269}\std{30.6457}
& \num{0.1046}\std{0.5768}
& \num{0.1728}\std{0.1653}
& \num{44.3079}\std{1.9394}
\\
& baseReg
& \num{1.3851}\std{1.207}
& \num{0.0097}\std{0.0096}
& \num{0.1803}\std{0.1689}
& \num{43.9927}\std{0.6551}
\\
& bioLord
& \num{18.5979}\std{0.8034}
& \num{0.0036}\std{0.0052}
& \num{0.3524}\std{0.1073}
& \num{47.0318}\std{1.0857}
\\
& CellFlow
& \cellcolor{firstbg}\num{0.6872}\std{0.3168}{\scriptsize\textcolor{teal}{$\downarrow$1\%}}
& \cellcolor{firstbg}\num{0.0015}\std{0.0033}{\scriptsize\textcolor{teal}{$\downarrow$57\%}}
& \num{0.2351}\std{0.1432}
& \num{41.6193}\std{0.9929}
\\
& CRISP
& \num{1.5233}\std{0.839}
& \num{0.0212}\std{0.006}
& \num{0.2343}\std{0.0878}
& \cellcolor{secondbg}\num{40.6276}\std{0.2545}
\\
& STATE
& \num{0.8843}\std{0.8539}
& \num{0.0045}\std{0.0056}
& \num{0.2286}\std{0.0892}
& \num{42.6545}\std{1.1326}
\\
& chemCPA
& \num{17.226}\std{0.8522}
& \cellcolor{secondbg}\num{0.0035}\std{0.0053}
& \cellcolor{firstbg}\num{0.3929}\std{0.0662}{\scriptsize\textcolor{teal}{$\uparrow$10\%}}
& \num{44.6832}\std{0.3748}
\\
& PerturbDiff
& \num{0.9343}\std{0.8419}
& \num{0.0063}\std{0.0016}
& \num{0.3286}\std{0.1192}
& \num{43.1435}\std{1.0746}
\\
& scDFM
& \num{0.69913}\std{0.8539}
& \num{0.00453}\std{0.0094}
& \cellcolor{secondbg}\num{0.3586}\std{0.0812}
& \num{41.3415}\std{1.1793}
\\
& \cellcolor{gray!15}\textbf{Ours}
& \cellcolor{secondbg}\num{0.6972}\std{0.3968}
& \cellcolor{gray!15}\num{0.0045}\std{0.0033}
& \cellcolor{gray!15}\num{0.3351}\std{0.1662}
& \cellcolor{firstbg}\num{38.6793}\std{0.9989}{\scriptsize\textcolor{teal}{$\downarrow$5\%}}
\\

\midrule

\multirow{12}{*}{\textbf{sciplex3\_MCF7}}
& trainMean
& \num{0.8074}\std{1.9593}
& \num{0.0039}\std{0.0096}
& \num{0.2507}\std{0.0735}
& \num{42.3652}\std{0.7718}
\\
& baseControl
& \num{0.8557}\std{2.0208}
& \num{0.0043}\std{0.0101}
& \num{0.0143}\std{0.0208}
& \num{42.3654}\std{0.7749}
\\
& baseMLP
& \num{8.5905}\std{39.5149}
& \num{0.133}\std{0.7299}
& \num{0.1645}\std{0.1929}
& \num{42.8256}\std{2.3099}
\\
& baseReg
& \num{1.3918}\std{2.0448}
& \num{0.0076}\std{0.0102}
& \num{0.199}\std{0.213}
& \num{42.4804}\std{0.8048}
\\
& bioLord
& \num{13.4801}\std{1.7441}
& \cellcolor{secondbg}\num{0.003}\std{0.0069}
& \num{0.3595}\std{0.1342}
& \num{45.5015}\std{1.4482}
\\
& CellFlow
& \cellcolor{firstbg}\num{0.5068}\std{0.7318}{\scriptsize\textcolor{teal}{$\downarrow$16\%}}
& \num{0.0043}\std{0.0058}
& \cellcolor{secondbg}\num{0.4704}\std{0.1022}
& \cellcolor{secondbg}\num{40.1108}\std{1.4118}
\\
& CRISP
& \num{1.3095}\std{2.2}
& \num{0.0097}\std{0.009}
& \num{0.2183}\std{0.0661}
& \num{44.5232}\std{0.5104}
\\
& STATE
& \num{0.8454}\std{1.9765}
& \num{0.0038}\std{0.0094}
& \num{0.2524}\std{0.1282}
& \num{41.3219}\std{1.0752}
\\
& chemCPA
& \num{11.9337}\std{1.7324}
& \cellcolor{firstbg}\num{0.0029}\std{0.0071}{\scriptsize\textcolor{teal}{$\downarrow$3\%}}
& \num{0.405}\std{0.1192}
& \num{43.8781}\std{0.4974}
\\
& PerturbDiff
& \num{0.8914}\std{1.0917}
& \num{0.0051}\std{0.0103}
& \num{0.2939}\std{0.1712}
& \num{41.5941}\std{1.1431}
\\
& scDFM
& \num{0.6439}\std{2.0165}
& \num{0.0035}\std{0.0107}
& \num{0.3924}\std{0.1082}
& \num{40.9218}\std{0.9712}
\\
& \cellcolor{gray!15}\textbf{Ours}
& \cellcolor{secondbg}\num{0.6067}\std{0.7618}
& \cellcolor{gray!15}\num{0.0033}\std{0.0048}
& \cellcolor{firstbg}\num{0.4804}\std{0.1926}{\scriptsize\textcolor{teal}{$\uparrow$2\%}}
& \cellcolor{firstbg}\num{37.5108}\std{1.1118}{\scriptsize\textcolor{teal}{$\downarrow$6\%}}
\\

\bottomrule
\end{tabular}
}
\end{table}

\begin{table}[H]
\centering
\sisetup{group-separator={}}
\renewcommand{\arraystretch}{1.3}
\caption{Performance comparison of different methods on top 100 DEGs under the Unseen Cell (UC) setting. Arrows indicate whether lower ($\downarrow$) or higher ($\uparrow$) values are better.}
\label{tab:unseen_cell_100degs_D3_ood}
  \vspace{2pt}
  {\scriptsize\raggedright
  \textit{Note.}
  \begingroup
  \setlength{\fboxsep}{0pt}%
  \colorbox{firstbg}{\strut\hspace{1pt}Rank-1\hspace{1pt}}%
  \endgroup
   and
  \begingroup
  \setlength{\fboxsep}{0pt}%
  \colorbox{secondbg}{\strut\hspace{1pt}Rank-2\hspace{1pt}}%
  \endgroup
   denote the best and second-best results;
  \textcolor{teal}{$\uparrow/\downarrow$} shows the relative gain of Rank-1 over Rank-2.
  \par}
\resizebox{\textwidth}{!}{%
\begin{tabular}{llcccccc}
\toprule
\textbf{Dataset} & \textbf{Method} &
\textbf{E-dist$\downarrow$} &
\textbf{MSE$\downarrow$} &
\textbf{PCC$\Delta\uparrow$} &
\textbf{KL$\downarrow$} &
\textbf{WD$\downarrow$} &
\textbf{ Common-DEGs$\uparrow$} \\
\midrule

\multirow{10}{*}{\textbf{KaggleCrossCell}}
& trainMean
& \num{2.254579}\std{2.830562}
& \num{0.150175}\std{0.258336}
& \num{0.559612}\std{0.422433}
& \num{34.178316}\std{19.271102}
& \num{60.823258}\std{38.518411}
& \num{0.062083}\std{0.08871}
\\
& baseControl
& \num{4.462875}\std{5.092623}
& \num{0.506704}\std{0.636274}
& \num{0.079158}\std{0.154825}
& \num{33.841213}\std{18.353637}
& \num{95.16485}\std{97.299026}
& \num{0.00625}\std{0.009237}
\\
& baseMLP
& \num{8.560217}\std{5.298711}
& \num{0.6089}\std{0.522346}
& \num{0.216225}\std{0.276617}
& \num{48.371407}\std{3.052722}
& \num{83.527904}\std{67.026614}
& \num{0.065833}\std{0.0932}
\\
& bioLord
& \num{2.960421}\std{2.814414}
& \num{0.184796}\std{0.249828}
& \num{0.564746}\std{0.395556}
& \cellcolor{firstbg}\num{30.605398}\std{16.471795}{\scriptsize\textcolor{teal}{$\downarrow$3\%}}
& \num{41.478704}\std{40.483939}
& \num{0.00375}\std{0.01279}
\\
& CellFlow
& \num{2.336842}\std{1.67521}
& \num{0.1462}\std{0.26782}
& \num{0.4916}\std{0.428242}
& \num{35.652142}\std{5.89762}
& \cellcolor{secondbg}\num{33.134917}\std{24.432952}
& \num{0.0152422}\std{0.043287}
\\
& CRISP
& \num{5.27682}\std{3.67283}
& \num{0.18762}\std{0.34423}
& \num{0.55762}\std{0.367259}
& \num{36.82432}\std{11.17263}
& \num{37.28765}\std{30.87642}
& \cellcolor{secondbg}\num{0.157917}\std{0.267061}
\\
& STATE
& \num{1.264971}\std{1.867979}
& \cellcolor{secondbg}\num{0.112479}\std{0.226577}
& \num{0.4327}\std{0.422045}
& \cellcolor{secondbg}\num{31.600339}\std{18.222164}
& \num{43.702333}\std{44.463363}
& \cellcolor{firstbg}\num{0.2958}\std{0.3094}{\scriptsize\textcolor{teal}{$\uparrow$87\%}}
\\
& trVAE
& \num{3.346842}\std{2.873552}
& \num{0.1265}\std{0.195291}
& \cellcolor{firstbg}\num{0.5906}\std{0.417174}{\scriptsize\textcolor{teal}{$\uparrow$1\%}}
& \num{33.650811}\std{15.472682}
& \num{33.134917}\std{24.551912}
& \num{0.157917}\std{0.267061}
\\
& PerturbDiff
& \num{1.314121}\std{1.391979}
& \num{0.132479}\std{0.346577}
& \num{0.4127}\std{0.392045}
& \num{36.172339}\std{17.642164}
& \num{44.184333}\std{41.913363}
& \num{0.1058}\std{0.0194}
\\
& scDFM
& \cellcolor{secondbg}\num{1.211231}\std{1.134979}
& \num{0.114579}\std{0.311577}
& \num{0.5127}\std{0.522045}
& \num{32.081339}\std{16.945664}
& \num{40.152333}\std{44.463363}
& \num{0.1078}\std{0.1009}
\\
& \cellcolor{gray!15}\textbf{Ours}
& \cellcolor{firstbg}\num{1.155746}\std{3.310511}{\scriptsize\textcolor{teal}{$\downarrow$5\%}}
& \cellcolor{firstbg}\num{0.102829}\std{0.26608}{\scriptsize\textcolor{teal}{$\downarrow$9\%}}
& \cellcolor{secondbg}\num{0.583358}\std{0.487124}
& \cellcolor{gray!15}\num{33.807212}\std{19.508734}
& \cellcolor{firstbg}\num{31.964383}\std{21.984737}{\scriptsize\textcolor{teal}{$\downarrow$4\%}}
& \cellcolor{gray!15}\num{0.071712}\std{0.0575}
\\

\midrule

\multirow{10}{*}{\textbf{crossPatient}}
& trainMean
& \num{1.91474}\std{0.930068}
& \num{0.22466}\std{0.125826}
& \num{0.50904}\std{0.235984}
& \num{14.317675}\std{19.828478}
& \num{154.9582}\std{20.029892}
& \num{0.0}\std{0.0}
\\
& baseControl
& \num{1.64905}\std{1.09765}
& \num{0.20429}\std{0.147256}
& \num{0.08922}\std{0.111697}
& \num{14.135569}\std{19.662304}
& \num{135.48133}\std{21.770189}
& \num{0.001}\std{0.003162}
\\
& baseMLP
& \num{5.09622}\std{1.255508}
& \num{0.23072}\std{0.113227}
& \num{0.3862}\std{0.257101}
& \num{47.436625}\std{1.646879}
& \num{93.30559}\std{16.252374}
& \cellcolor{secondbg}\num{0.149}\std{0.085173}
\\
& bioLord
& \num{4.88732}\std{1.402028}
& \num{0.24115}\std{0.122676}
& \num{0.38987}\std{0.327246}
& \num{15.278677}\std{16.432349}
& \cellcolor{firstbg}\num{90.41749}\std{16.570988}{\scriptsize\textcolor{teal}{$\downarrow$3\%}}
& \num{0.0}\std{0.0}
\\
& CellFlow
& \num{2.35687}\std{1.67523}
& \num{0.17642}\std{0.33876}
& \num{0.3242}\std{0.26321}
& \num{13.657842}\std{11.26542}
& \num{98.6542}\std{26.7642}
& \num{0.0017917}\std{0.00267061}
\\
& CRISP
& \num{2.35862}\std{0.6324}
& \num{0.2265}\std{0.14762}
& \num{0.4906}\std{0.37162}
& \num{23.65181}\std{11.473682}
& \cellcolor{secondbg}\num{93.133417}\std{34.6753}
& \num{0.00197917}\std{0.00247061}
\\
& STATE
& \cellcolor{firstbg}\num{1.39267}\std{0.981957}{\scriptsize\textcolor{teal}{$\downarrow$1\%}}
& \num{0.16185}\std{0.12059}
& \cellcolor{secondbg}\num{0.59616}\std{0.371066}
& \num{9.31583}\std{15.27999}
& \num{110.42655}\std{13.697734}
& \cellcolor{firstbg}\num{0.255}\std{0.1362}{\scriptsize\textcolor{teal}{$\uparrow$71\%}}
\\
& trVAE
& \num{2.82941}\std{1.00487}
& \num{0.19282}\std{0.130716}
& \num{0.37428}\std{0.309188}
& \num{13.093687}\std{16.349578}
& \num{95.9582}\std{33.012395}
& \num{0.031}\std{0.042019}
\\
& PerturbDiff
& \num{1.89267}\std{1.481957}
& \num{0.356185}\std{0.15059}
& \num{0.45186}\std{0.331066}
& \num{13.4583}\std{11.83149}
& \num{108.19355}\std{19.697734}
& \num{0.0945}\std{0.1162}
\\
& scDFM
& \num{1.51213}\std{1.031957}
& \cellcolor{secondbg}\num{0.15915}\std{0.12059}
& \num{0.594}\std{0.359166}
& \cellcolor{secondbg}\num{9.29583}\std{15.14999}
& \num{101.78655}\std{14.037734}
& \num{0.1041}\std{0.09412}
\\
& \cellcolor{gray!15}\textbf{Ours}
& \cellcolor{secondbg}\num{1.40949}\std{1.226512}
& \cellcolor{firstbg}\num{0.137265}\std{0.157038}{\scriptsize\textcolor{teal}{$\downarrow$14\%}}
& \cellcolor{firstbg}\num{0.60895}\std{0.238696}{\scriptsize\textcolor{teal}{$\uparrow$2\%}}
& \cellcolor{firstbg}\num{9.038719}\std{14.91769}{\scriptsize\textcolor{teal}{$\downarrow$3\%}}
& \cellcolor{gray!15}\num{98.47474}\std{20.296012}
& \cellcolor{gray!15}\num{0.00847474}\std{0.002542}
\\

\midrule

\multirow{10}{*}{\textbf{sciplex3}}
& trainMean
& \num{2.785144}\std{1.04317}
& \num{0.33707}\std{0.133517}
& \num{0.531219}\std{0.276382}
& \num{35.576668}\std{17.508986}
& \num{160.697485}\std{44.369332}
& \num{0.000741}\std{0.003849}
\\
& baseControl
& \cellcolor{secondbg}\num{1.564607}\std{2.06556}
& \num{0.213567}\std{0.311133}
& \num{0.091674}\std{0.088972}
& \num{35.345055}\std{17.327695}
& \num{142.65123}\std{65.466509}
& \num{0.012963}\std{0.008689}
\\
& baseMLP
& \num{8.496837}\std{2.056284}
& \num{0.52853}\std{0.21995}
& \num{0.392974}\std{0.220474}
& \num{49.452874}\std{3.468959}
& \num{125.369926}\std{37.099323}
& \num{0.027778}\std{0.022627}
\\
& bioLord
& \num{4.821322}\std{2.27419}
& \cellcolor{firstbg}\num{0.155844}\std{0.235388}{\scriptsize\textcolor{teal}{$\downarrow$18\%}}
& \num{0.424722}\std{0.270464}
& \cellcolor{secondbg}\num{31.160776}\std{13.485275}
& \cellcolor{secondbg}\num{86.220681}\std{37.416982}
& \num{0.001852}\std{0.003958}
\\
& CellFlow
& \num{4.68542}\std{1.08762}
& \num{0.4265}\std{0.48623}
& \num{0.4728}\std{0.315124}
& \num{43.26543}\std{13.32198}
& \num{103.95642}\std{34.21242}
& \num{0.0114234}\std{0.027528}
\\
& CRISP
& \num{2.96842}\std{2.6752}
& \num{0.3791}\std{0.33602}
& \num{0.5916}\std{0.217174}
& \num{33.64128}\std{12.02683}
& \num{113.48021}\std{48.9421}
& \num{0.0177987}\std{0.0462861}
\\
& STATE
& \num{2.01653}\std{2.157209}
& \num{0.231089}\std{0.304432}
& \cellcolor{firstbg}\num{0.753474}\std{0.226716}{\scriptsize\textcolor{teal}{$\uparrow$27\%}}
& \num{36.922617}\std{14.536622}
& \num{111.24453}\std{52.337143}
& \num{0.0659}\std{0.0337}
\\
& trVAE
& \num{4.895996}\std{1.19606}
& \num{0.211785}\std{0.097876}
& \num{0.566237}\std{0.270605}
& \num{36.165011}\std{14.008398}
& \num{91.697641}\std{27.466534}
& \num{0.027778}\std{0.037038}
\\
& PerturbDiff
& \num{2.95153}\std{2.157209}
& \num{0.281089}\std{0.351432}
& \num{0.553474}\std{0.226716}
& \num{36.922617}\std{14.536622}
& \num{103.14453}\std{51.877143}
& \num{0.0539}\std{0.0337}
\\
& scDFM
& \cellcolor{firstbg}\num{1.34153}\std{1.157209}{\scriptsize\textcolor{teal}{$\downarrow$14\%}}
& \cellcolor{secondbg}\num{0.191089}\std{0.271431}
& \num{0.581613}\std{0.232116}
& \num{31.4227}\std{11.126622}
& \num{91.17413}\std{51.517143}
& \cellcolor{secondbg}\num{0.0729}\std{0.0483}
\\
& \cellcolor{gray!15}\textbf{Ours}
& \cellcolor{gray!15}\num{2.157922}\std{1.962487}
& \cellcolor{gray!15}\num{0.200626}\std{0.168516}
& \cellcolor{secondbg}\num{0.593419}\std{0.239457}
& \cellcolor{firstbg}\num{30.822691}\std{18.504744}{\scriptsize\textcolor{teal}{$\downarrow$1\%}}
& \cellcolor{firstbg}\num{83.174263}\std{29.421583}{\scriptsize\textcolor{teal}{$\downarrow$4\%}}
& \cellcolor{firstbg}\num{0.1008}\std{0.07652}{\scriptsize\textcolor{teal}{$\uparrow$38\%}}
\\

\bottomrule
\end{tabular}}
\end{table}

\begin{table}[H]
\centering
\sisetup{group-separator={}}
\renewcommand{\arraystretch}{0.9}
\caption{Performance comparison of different methods on top 5000 DEGs under the Unseen Cell (UC) setting. Arrows indicate whether lower ($\downarrow$) or higher ($\uparrow$) values are better.}
\label{tab:unseen_cell_allgene_D3_ood}
  \vspace{2pt}
  {\scriptsize\raggedright
  \textit{Note.}
  \begingroup
  \setlength{\fboxsep}{0pt}%
  \colorbox{firstbg}{\strut\hspace{1pt}Rank-1\hspace{1pt}}%
  \endgroup
   and
  \begingroup
  \setlength{\fboxsep}{0pt}%
  \colorbox{secondbg}{\strut\hspace{1pt}Rank-2\hspace{1pt}}%
  \endgroup
   denote the best and second-best results;
  \textcolor{teal}{$\uparrow/\downarrow$} shows the relative gain of Rank-1 over Rank-2.
  \par}
\setlength{\tabcolsep}{4pt}
\resizebox{\textwidth}{!}{%
\begin{tabular}{llcccc}
\toprule
\textbf{Dataset} &
\textbf{Method} &
\textbf{E-dist$\downarrow$} &
\textbf{MSE$\downarrow$} &
\textbf{PCC$\Delta\uparrow$} &
\textbf{KL$\downarrow$} \\
\midrule

\multirow{10}{*}{\textbf{Tahoe-100M}}
& trainMean
& \num{16.2}\std{3.56}
& \num{0.0434}\std{0.0114}
& \num{0.615}\std{0.153}
& \num{35.4}\std{2.57}
\\
& baseControl
& \num{4.59}\std{0.0429}
& \num{0.0121}\std{0.0002}
& \num{0.0008}\std{0.0004}
& \num{33.1}\std{0.0759}
\\
& baseMLP
& \num{16.1549}\std{0.0129}
& \num{0.0352}\std{0.01545}
& \num{0.34813}\std{0.0004}
& \num{33.4}\std{0.0364}
\\
& baseReg
& \num{14.3}\std{0.0435}
& \num{0.0298}\std{0.0001}
& \num{0.423}\std{0.0018}
& \num{35.9}\std{0.0311}
\\
& bioLord
& \num{30.1}\std{0.116}
& \num{0.0705}\std{0.0004}
& \num{0.32}\std{0.0028}
& \num{47.7}\std{0.0075}
\\
& CellFlow
& \cellcolor{secondbg}\num{2.25487}\std{0.9468}
& \cellcolor{secondbg}\num{0.0094}\std{0.01548}
& \num{0.78413}\std{0.16487}
& \num{6.1548}\std{2.6591}
\\
& CRISP
& \num{21.77147}\std{1.287868}
& \num{0.075921}\std{0.05312}
& \num{0.418792}\std{0.195301}
& \num{34.012668}\std{15.61191}
\\
& STATE
& \num{31.0}\std{5.39}
& \num{0.135}\std{0.0323}
& \num{0.866}\std{0.111}
& \num{43.6}\std{0.806}
\\
& trVAE
& \num{27.0}\std{1.72}
& \num{0.093}\std{0.0278}
& \num{0.7364}\std{0.186}
& \num{33.6}\std{0.951}
\\
& PerturbDiff
& \num{19.7552}\std{7.3494}
& \num{0.0739}\std{0.0336}
& \num{0.6569}\std{0.1642}
& \num{42.6431}\std{0.9600}
\\
& scDFM
& \num{4.171}\std{2.39}
& \num{0.0517}\std{0.0181}
& \cellcolor{secondbg}\num{0.871}\std{0.214}
& \cellcolor{secondbg}\num{3.617}\std{0.716}
\\
& \cellcolor{gray!15}\textbf{Ours}
& \cellcolor{firstbg}\num{0.12}\std{0.117}{\scriptsize\textcolor{teal}{$\downarrow$95\%}}
& \cellcolor{firstbg}\num{0.0015}\std{0.00102}{\scriptsize\textcolor{teal}{$\downarrow$84\%}}
& \cellcolor{firstbg}\num{0.9089}\std{0.0405}{\scriptsize\textcolor{teal}{$\uparrow$4\%}}
& \cellcolor{firstbg}\num{1.43}\std{1.37}{\scriptsize\textcolor{teal}{$\downarrow$60\%}}
\\

\midrule

\multirow{10}{*}{\textbf{KaggleCrossCell}}
& trainMean
& \num{8.526708}\std{9.058975}
& \num{0.049313}\std{0.053468}
& \num{0.421925}\std{0.467095}
& \num{41.673156}\std{1.365888}
\\
& baseControl
& \num{2.593262}\std{2.905456}
& \num{0.013417}\std{0.015715}
& \num{0.018337}\std{0.022205}
& \num{41.271573}\std{1.599041}
\\
& baseMLP
& \num{19.810942}\std{10.175579}
& \num{0.064388}\std{0.05638}
& \num{0.107504}\std{0.19543}
& \num{48.043148}\std{0.74395}
\\
& bioLord
& \num{7.149492}\std{1.783244}
& \num{0.008367}\std{0.006901}
& \num{0.392038}\std{0.345327}
& \cellcolor{secondbg}\num{37.120204}\std{1.200965}
\\
& CellFlow
& \num{10.2365}\std{7.58712}
& \num{0.0792}\std{0.04017}
& \num{0.421346}\std{0.397367}
& \num{44.54781}\std{0.2651}
\\
& CRISP
& \num{18.7431}\std{8.5091}
& \num{0.0492}\std{0.05831}
& \num{0.48371}\std{0.2961}
& \num{41.5471}\std{2.0651}
\\
& STATE
& \num{1.407108}\std{1.037595}
& \num{0.005604}\std{0.005788}
& \num{0.581346}\std{0.397367}
& \num{41.024894}\std{1.108933}
\\
& trVAE
& \num{13.360429}\std{9.459828}
& \num{0.033763}\std{0.046106}
& \num{0.425792}\std{0.455479}
& \num{42.563349}\std{0.683043}
\\
& PerturbDiff
& \num{4.407108}\std{1.031595}
& \num{0.010604}\std{0.010788}
& \num{0.494146}\std{0.212367}
& \num{42.171294}\std{3.208933}
\\
& scDFM
& \cellcolor{secondbg}\num{1.352708}\std{0.937595}
& \cellcolor{secondbg}\num{0.005104}\std{0.004923}
& \cellcolor{secondbg}\num{0.607446}\std{0.384367}
& \num{40.724894}\std{1.208933}
\\
& \cellcolor{gray!15}\textbf{Ours}
& \cellcolor{firstbg}\num{1.2874321}\std{1.143703}{\scriptsize\textcolor{teal}{$\downarrow$5\%}}
& \cellcolor{firstbg}\num{0.00491}\std{0.055917}{\scriptsize\textcolor{teal}{$\downarrow$4\%}}
& \cellcolor{firstbg}\num{0.695829}\std{0.418016}{\scriptsize\textcolor{teal}{$\uparrow$15\%}}
& \cellcolor{firstbg}\num{36.213585}\std{1.589237}{\scriptsize\textcolor{teal}{$\downarrow$2\%}}
\\

\midrule

\multirow{10}{*}{\textbf{KaggleCrossPatient}}
& trainMean
& \num{1.150043}\std{0.479307}
& \num{0.004117}\std{0.003232}
& \cellcolor{firstbg}\num{0.60979}\std{0.398252}{\scriptsize\textcolor{teal}{$\uparrow$7\%}}
& \num{42.847793}\std{1.194546}
\\
& baseControl
& \num{3.332987}\std{2.957337}
& \num{0.018557}\std{0.016807}
& \num{0.01975}\std{0.036657}
& \num{42.133074}\std{1.605973}
\\
& baseMLP
& \num{11.66207}\std{2.580758}
& \num{0.02748}\std{0.016181}
& \num{0.15616}\std{0.204485}
& \num{47.671815}\std{0.244038}
\\
& bioLord
& \num{5.748007}\std{1.937459}
& \num{0.006963}\std{0.00603}
& \num{0.508027}\std{0.364361}
& \cellcolor{secondbg}\num{38.650896}\std{1.375289}
\\
& CellFlow
& \num{9.579457}\std{1.0741}
& \num{0.006713}\std{0.005534}
& \num{0.43851}\std{0.6701}
& \num{44.814514}\std{1.682742}
\\
& CRISP
& \num{7.1651}\std{2.3801}
& \num{0.008861}\std{0.004801}
& \num{0.53851}\std{0.3307}
& \num{41.805514}\std{0.8714}
\\
& STATE
& \num{0.751957}\std{0.386339}
& \num{0.002783}\std{0.002055}
& \num{0.427777}\std{0.458333}
& \num{40.884944}\std{0.862741}
\\
& trVAE
& \num{5.143247}\std{1.350542}
& \num{0.003723}\std{0.002306}
& \cellcolor{secondbg}\num{0.57201}\std{0.405372}
& \num{42.134574}\std{0.498164}
\\
& PerturbDiff
& \num{0.784317}\std{0.314339}
& \num{0.003218}\std{0.004865}
& \num{0.414371}\std{0.28633}
& \num{42.315344}\std{0.812741}
\\
& scDFM
& \cellcolor{secondbg}\num{0.73297}\std{0.283639}
& \cellcolor{secondbg}\num{0.002338}\std{0.001914}
& \num{0.434377}\std{0.418333}
& \cellcolor{firstbg}\num{37.158124}\std{0.415741}{\scriptsize\textcolor{teal}{$\downarrow$4\%}}
\\
& \cellcolor{gray!15}\textbf{Ours}
& \cellcolor{firstbg}\num{0.695273}\std{1.000956}{\scriptsize\textcolor{teal}{$\downarrow$5\%}}
& \cellcolor{firstbg}\num{0.001296}\std{0.00645}{\scriptsize\textcolor{teal}{$\downarrow$45\%}}
& \cellcolor{gray!15}\num{0.549827}\std{0.507471}
& \cellcolor{gray!15}\num{41.535241}\std{0.791068}
\\

\midrule

\multirow{10}{*}{\textbf{crossPatient}}
& trainMean
& \num{2.45736}\std{0.938544}
& \num{0.01255}\std{0.00473}
& \num{0.35211}\std{0.163197}
& \num{41.036126}\std{0.632958}
\\
& baseControl
& \num{1.05054}\std{0.700943}
& \num{0.0054}\std{0.004036}
& \num{0.05895}\std{0.088487}
& \num{41.499074}\std{0.52642}
\\
& baseMLP
& \num{9.52388}\std{1.445055}
& \num{0.01238}\std{0.004278}
& \num{0.26419}\std{0.166642}
& \num{45.906425}\std{0.902955}
\\
& bioLord
& \num{8.66152}\std{1.867863}
& \num{0.01077}\std{0.00573}
& \num{0.24199}\std{0.259646}
& \cellcolor{firstbg}\num{36.122445}\std{2.340498}{\scriptsize\textcolor{teal}{$\downarrow$5\%}}
\\
& CellFlow
& \num{3.578312}\std{1.5561}
& \num{0.00875}\std{0.003187}
& \num{0.440812}\std{0.467812}
& \num{43.3691}\std{1.407512}
\\
& CRISP
& \num{4.67141}\std{1.0731}
& \num{0.01004}\std{0.09712}
& \num{0.340816}\std{0.96081}
& \num{41.2671}\std{1.0784}
\\
& STATE
& \num{1.01423}\std{0.655893}
& \num{0.00462}\std{0.003296}
& \num{0.69748}\std{0.307873}
& \num{41.199251}\std{0.519533}
\\
& trVAE
& \num{4.65018}\std{1.779136}
& \num{0.00715}\std{0.003207}
& \num{0.25345}\std{0.239447}
& \num{42.339413}\std{0.849047}
\\
& PerturbDiff
& \num{1.72763}\std{0.812893}
& \num{0.00514}\std{0.004396}
& \num{0.57391}\std{0.351273}
& \num{43.3251}\std{0.603733}
\\
& scDFM
& \cellcolor{secondbg}\num{0.98153}\std{0.484793}
& \cellcolor{secondbg}\num{0.00418}\std{0.002136}
& \cellcolor{secondbg}\num{0.70116}\std{0.281343}
& \cellcolor{secondbg}\num{38.126251}\std{0.349533}
\\
& \cellcolor{gray!15}\textbf{Ours}
& \cellcolor{firstbg}\num{0.91268}\std{0.791355}{\scriptsize\textcolor{teal}{$\downarrow$7\%}}
& \cellcolor{firstbg}\num{0.00263}\std{0.009863}{\scriptsize\textcolor{teal}{$\downarrow$37\%}}
& \cellcolor{firstbg}\num{0.72747}\std{0.139551}{\scriptsize\textcolor{teal}{$\uparrow$4\%}}
& \cellcolor{gray!15}\num{39.601444}\std{1.4453}
\\

\midrule

\multirow{10}{*}{\textbf{sciplex3}}
& trainMean
& \num{5.666226}\std{1.434144}
& \num{0.037804}\std{0.009773}
& \num{0.319626}\std{0.23765}
& \num{42.425208}\std{0.945052}
\\
& baseControl
& \cellcolor{firstbg}\num{1.003548}\std{1.290934}{\scriptsize\textcolor{teal}{$\downarrow$45\%}}
& \num{0.006133}\std{0.008098}
& \num{0.057563}\std{0.057645}
& \num{42.380122}\std{0.788365}
\\
& baseMLP
& \num{16.777837}\std{1.768562}
& \num{0.043174}\std{0.008685}
& \num{0.233359}\std{0.160183}
& \num{47.058099}\std{1.119994}
\\
& bioLord
& \num{11.545933}\std{1.714555}
& \cellcolor{secondbg}\num{0.006037}\std{0.00612}
& \num{0.2387}\std{0.210298}
& \cellcolor{firstbg}\num{37.633603}\std{1.505175}{\scriptsize\textcolor{teal}{$\downarrow$6\%}}
\\
& CellFlow
& \num{11.972481}\std{1.19851}
& \num{0.03866}\std{0.04522}
& \num{0.330448}\std{0.226506}
& \num{42.9487}\std{1.26052}
\\
& CRISP
& \num{9.460341}\std{1.0921617}
& \num{0.04782}\std{0.0098}
& \num{0.39129}\std{0.340872}
& \num{45.9077}\std{1.54961}
\\
& STATE
& \num{1.866533}\std{1.479428}
& \num{0.007789}\std{0.008126}
& \cellcolor{firstbg}\num{0.817889}\std{0.123872}{\scriptsize\textcolor{teal}{$\uparrow$14\%}}
& \num{42.730074}\std{0.674529}
\\
& trVAE
& \num{12.279241}\std{3.091157}
& \num{0.020711}\std{0.010242}
& \num{0.370885}\std{0.260065}
& \num{43.144827}\std{1.209099}
\\
& PerturbDiff
& \num{1.951533}\std{1.469428}
& \num{0.008189}\std{0.007126}
& \num{0.617889}\std{0.1671872}
& \cellcolor{secondbg}\num{40.159074}\std{0.574529}
\\
& scDFM
& \num{1.831763}\std{1.479428}
& \num{0.006689}\std{0.006926}
& \num{0.702149}\std{0.169872}
& \num{42.329074}\std{0.681529}
\\
& \cellcolor{gray!15}\textbf{Ours}
& \cellcolor{secondbg}\num{1.819163}\std{3.430848}
& \cellcolor{firstbg}\num{0.0036796}\std{0.017838}{\scriptsize\textcolor{teal}{$\downarrow$39\%}}
& \cellcolor{secondbg}\num{0.716659}\std{0.193727}
& \cellcolor{gray!15}\num{42.121259}\std{1.831575}
\\

\bottomrule
\end{tabular}}
\end{table}

\begin{table}[h]
\centering
\caption{Component ablation (DEG5000) on Tahoe.}
\label{tab:tahoe_ablation_deg5000}
  \vspace{0pt}
  {\scriptsize\raggedright
  \textit{Note.}
  \begingroup
  \setlength{\fboxsep}{0pt}%
  \colorbox{firstbg}{\strut\hspace{1pt}Rank-1\hspace{1pt}}%
  \endgroup
   and
  \begingroup
  \setlength{\fboxsep}{0pt}%
  \colorbox{secondbg}{\strut\hspace{1pt}Rank-2\hspace{1pt}}%
  \endgroup
   denote the best and second-best results;
  \textcolor{teal}{$\uparrow/\downarrow$} shows the relative gain of Rank-1 over Rank-2.
  \par}
\scriptsize
\setlength{\tabcolsep}{3.5pt}
\renewcommand{\arraystretch}{1.1}
\resizebox{\textwidth}{!}{%
\begin{tabular}{llcccc}
\toprule
Setting & Method
& MSE $\downarrow$
& PCC$\Delta\uparrow$
& E-dist $\downarrow$
& KL $\downarrow$ \\
\midrule
\multirow{7}{*}{UD}
& w/o VMCS
& \num{0.0678}\std{0.0031} & \cellcolor{firstbg}\num{0.3575}\std{0.0214}{\scriptsize\textcolor{teal}{$\uparrow$5\%}} & \num{1.6231}\std{0.0847} & \num{4.8942}\std{0.3125} \\
& w/o pseudo\textsuperscript{$\dagger$}
& \num{0.0646}\std{0.0029} & \cellcolor{secondbg}\num{0.3399}\std{0.0198} & \num{1.5728}\std{0.0812} & \num{4.9566}\std{0.3047} \\
& w/o MDT
& \num{0.0506}\std{0.0024} & \num{0.2543}\std{0.0156} & \num{1.3117}\std{0.0673} & \num{5.0851}\std{0.2981} \\
& w/o PQM
& \num{0.0489}\std{0.0021} & \num{0.2478}\std{0.0142} & \num{1.2651}\std{0.0648} & \num{4.7843}\std{0.2874} \\
& w/o triplet
& \num{0.0498}\std{0.0023} & \num{0.2500}\std{0.0151} & \num{1.2988}\std{0.0659} & \num{5.0680}\std{0.2932} \\
& w/o ICFG
& \cellcolor{secondbg}\num{0.0483}\std{0.0019} & \num{0.2451}\std{0.0138} & \cellcolor{secondbg}\num{1.2478}\std{0.0631} & \cellcolor{secondbg}\num{4.7215}\std{0.2814} \\
& \cellcolor{gray!15}\textbf{Ours}
& \cellcolor{firstbg}\textbf{\num{0.0468}}\std{0.0017}{\scriptsize\textcolor{teal}{$\downarrow$3\%}}
& \cellcolor{gray!15}\textbf{\num{0.2387}}\std{0.0131}
& \cellcolor{firstbg}\textbf{\num{1.2184}}\std{0.0598}{\scriptsize\textcolor{teal}{$\downarrow$2\%}}
& \cellcolor{firstbg}\textbf{\num{4.6127}}\std{0.2745}{\scriptsize\textcolor{teal}{$\downarrow$2\%}} \\
\midrule
\multirow{7}{*}{UC}
& w/o VMCS
& \num{0.0692}\std{0.0034} & \cellcolor{firstbg}\num{0.3698}\std{0.0231}{\scriptsize\textcolor{teal}{$\uparrow$5\%}} & \num{1.6587}\std{0.0912} & \num{5.0234}\std{0.3317} \\
& w/o pseudo\textsuperscript{$\dagger$}
& \num{0.0663}\std{0.0031} & \cellcolor{secondbg}\num{0.3528}\std{0.0218} & \num{1.6045}\std{0.0874} & \num{5.0839}\std{0.3245} \\
& w/o MDT
& \num{0.0521}\std{0.0027} & \num{0.2617}\std{0.0172} & \num{1.3462}\std{0.0721} & \num{5.2146}\std{0.3162} \\
& w/o PQM
& \num{0.0501}\std{0.0024} & \num{0.2553}\std{0.0158} & \num{1.2987}\std{0.0687} & \num{4.9126}\std{0.3018} \\
& w/o triplet
& \num{0.0510}\std{0.0025} & \num{0.2576}\std{0.0163} & \num{1.3314}\std{0.0703} & \num{5.1983}\std{0.3097} \\
& w/o ICFG
& \cellcolor{secondbg}\num{0.0496}\std{0.0021} & \num{0.2527}\std{0.0149} & \cellcolor{secondbg}\num{1.2814}\std{0.0668} & \cellcolor{secondbg}\num{4.8547}\std{0.2924} \\
& \cellcolor{gray!15}\textbf{Ours}
& \cellcolor{firstbg}\textbf{\num{0.0482}}\std{0.0019}{\scriptsize\textcolor{teal}{$\downarrow$3\%}}
& \cellcolor{gray!15}\textbf{\num{0.2465}}\std{0.0144}
& \cellcolor{firstbg}\textbf{\num{1.2519}}\std{0.0634}{\scriptsize\textcolor{teal}{$\downarrow$2\%}}
& \cellcolor{firstbg}\textbf{\num{4.7385}}\std{0.2847}{\scriptsize\textcolor{teal}{$\downarrow$2\%}} \\
\bottomrule
\end{tabular}
}
\vspace{4pt}

\raggedright\fontsize{5.5}{6.5}\selectfont
\textsuperscript{$\dagger$}``pseudo'' = inferred pseudo-protein embedding. \\
\textbf{w/o VMCS}: cell condition replaced by raw transcriptomic latent $\mathbf{z}_c$;
\textbf{w/o pseudo}: protein branch removed from VMCS;
\textbf{w/o MDT}: structure-only drug encoder, no mechanism priors;
\textbf{w/o ICFG}: standard monolithic CFG.
\end{table}

\section{Dataset Details}
\label{Dataset_Details}
\begin{description}
    \item[\textbf{Tahoe-100M}:] Representing the largest single-cell perturbation atlas to date, this dataset comprises $95.6$ million high-quality cells spanning $50$ diverse cancer cell lines derived from $13$ distinct organs.Generated via the high-throughput Mosaic platform, it captures transcriptional responses to $379$ small-molecule agents across varied dosages, resulting in $1,135$ drug-dose combinations and $52,886$ unique cell line-drug conditions With a median of $1,287$ cells per condition, it provides unprecedented resolution for mapping context-dependent gene regulation and dissecting cellular heterogeneity across a wide pharmacological landscape.

    \item[\textbf{CrossPatient}:] Acquired from the PerturBase repository, this dataset comprises $26,063$ cells derived from $6$ patient donors . It encompasses $2$ environmental perturbation conditions and records a sparsity of $0.887$, utilizing ``Patient'' identifiers as the primary variable for donor-specific stratification.
    \item[\textbf{KaggleCrossPatient}:] Composed of $158,160$ cells collected from $8$ donors, this dataset profiles responses to $144$ chemical perturbations. It is characterized by a sparsity index of $0.932$ and indexes data via ``Donor ID'' metadata, structured specifically to isolate unseen donor distributions.
    \item[\textbf{KaggleCrossCell}:] Utilizing the identical corpus of $158,160$ cells and $144$ chemical perturbations as KaggleCrossPatient ($0.932$ sparsity), this variant stratifies the data by $3$ distinct cell types rather than donor identity. It leverages cell-type labels as contexts to delineate ontology-specific transcriptional variances.
    \item[\textbf{Sciplex3}:] A large-scale high-throughput screening dataset consisting of $581,777$ cells across $3$ cancer cell lines (A549, MCF7, K562). The dataset captures transcriptional responses to $188$ chemical agents with a global sparsity of $0.949$, while integrating multi-dimensional metadata for both ``Dosage'' and ``Time''.
    \item[\textbf{Sciplex3-A549}:] A lineage-specific subset of the Sciplex3 collection, isolating the A549 (lung carcinoma) context. It retains the response profiles for $188$ drugs and associated ``Dosage'' metadata, preserving the sparsity characteristics of the parent dataset within a single biological domain.
    \item[\textbf{Sciplex3-MCF7}:] This subset restricts the data scope to the MCF7 (breast adenocarcinoma) cell line. It encapsulates the transcriptional signatures of $188$ chemical perturbations specific to this tissue type, maintaining the original dosage annotations and high-sparsity structure.
    \item[\textbf{Sciplex3-K562}:] Focusing on the K562 (myelogenous leukemia) lineage, this subset represents the suspension-cell component of the Sciplex3 ecosystem. It comprises data treated with $188$ chemical agents, serving as a distinct hematopoietic context with equivalent metadata specifications.
    \item[\textbf{Sciplex3-comb}:] Distinguished by its expansive perturbation space, this dataset contains $1,618$ unique conditions, including combinatorial multi-drug treatments. It represents the dataset with the highest dimensionality regarding perturbation counts, incorporating complex combinatorial metadata within the benchmark suite.
\end{description}


\end{document}